\documentclass[acmsmall,nonacm]{acmart}
\usepackage{amsmath,amsfonts}
\usepackage[bottom]{footmisc}
\usepackage[most]{tcolorbox}
\usepackage{etoolbox}
\usepackage{afterpage}
\usepackage{subcaption}
\usepackage{capt-of}   %
\usepackage{algorithmic}
\usepackage{graphicx}
\usepackage{listings}
\usepackage{float}      %
\usepackage{placeins}   %
\usepackage{textcomp}
\usepackage{xcolor}
\usepackage[shortlabels]{enumitem}
\usepackage{tikz}
\usepackage{booktabs}
\usepackage{tabularx}
\usepackage[T1]{fontenc}
\usepackage[htt]{hyphenat}
\usepackage{multirow} %
\def\tabref#1{Table~\ref{#1}}
\def\figref#1{Figure~\ref{#1}}

\newcommand{\answer}[2]{
  \begin{tcolorbox}[enhanced, left=3mm,right=3mm,
    colback=gray!10, colframe=gray!80, boxrule=0pt,
    borderline west={4pt}{0pt}{gray!90},
    ]
    \textbf{Answer for RQ#1:}
    #2
    \end{tcolorbox}
}

\begin{document}

\title{Exploring Code Analysis: Zero-Shot Insights on Syntax and Semantics with LLMs}
\thanks{This is the authors' version of the work, accepted for publication in ACM Transactions on Software Engineering and Methodology (TOSEM). The definitive Version of Record will be published by ACM}

\author{Wei Ma}
\authornote{Both authors contributed equally to this research.}
\orcid{0000-0002-0044-466X}
\affiliation{%
  \institution{Singapore Management University}
  \city{Singapore}
  \country{Singapore}
}
\affiliation{%
  \institution{Blekinge Institute of Technology}
  \city{Karlskrona}
  \country{Sweden}
}
\email{wei.ma@bth.se}

\author{Zhihao Lin}
\authornotemark[1]  %
\orcid{0009-0007-7731-6598} %
\affiliation{%
  \institution{Beihang University}
  \city{Beijing}
  \country{China}
}
\email{19241032@buaa.edu.cn}

\author{Shangqing Liu}
\authornote{Corresponding authors.}
\orcid{0000-0002-5598-4006}
\affiliation{%
  \institution{State Key Laboratory of Novel Software Technology, Nanjing University}
  \city{Nanjing}
  \country{China}
}
\email{shangqingliu666@gmail.com}

\author{Qiang Hu}
\authornotemark[2]  
\orcid{0000-0002-8251-1669}
\affiliation{%
  \institution{The University of Tokyo}
  \city{Bunkyo-ku}
  \country{Japan}
}
\email{qianghu0515@gmail.com}

\author{Ye Liu}
\orcid{0000-0001-6709-3721} %
\affiliation{%
  \institution{Singapore Management University}
  \city{Singapore}
  \country{Singapore}
}
\email{li0003ye@e.ntu.edu.sg}

\author{Wenhan Wang}
\orcid{0000-0002-0585-2136} %
\affiliation{%
  \institution{University of Alberta}
  \city{Edmonton}
  \country{Canada}
}
\email{wwhjacob@hotmail.com}

\author{Cen Zhang}
\orcid{0000-0001-5603-1322}
\affiliation{%
  \institution{Nanyang Technological University}
  \city{Singapore}
  \country{Singapore}
}
\email{cen001@e.ntu.edu.sg}

\author{Liming Nie}
\orcid{0000-0002-6058-1586}
\affiliation{%
  \institution{Shenzhen Technology University}
  \city{Shenzhen}
  \country{China}
}
\email{nieliming@sztu.edu.cn}

\author{Li Li}
\orcid{0000-0003-2990-1614}
\affiliation{%
  \institution{Beihang University}
  \city{Beijing}
  \country{China}
}
\email{lilicoding@ieee.org}

\author{Yang Liu}
\orcid{0000-0001-7300-9215}
\affiliation{%
  \institution{Nanyang Technological University}
  \city{Singapore}
  \country{Singapore}
}
\email{yangliu@ntu.edu.sg}

\author{Lingxiao Jiang}
\orcid{0000-0002-4336-8548}
\affiliation{%
  \institution{Singapore Management University}
  \city{Singapore}
  \country{Singapore}
}
\email{lxjiang@smu.edu.sg}

\renewcommand{\shortauthors}{Ma, Lin, Liu and Hu et al.}

\begin{abstract}
Code analysis is a fundamental problem in Software Engineering (SE), playing a crucial role in tasks such as debugging, performance optimization, and security assessment. Human developers typically approach code analysis through syntax parsing, static semantics inference, and dynamic reasoning. Traditional code analysis tools, while effective, are often limited by language specificity, complex configurations, and lack of cross-language generalization. Recently, large language models~(LLMs) have emerged as promising tools for automating various code-related tasks. However, their capabilities for fundamental code analysis remain underexplored. Understanding these capabilities is crucial for assessing and enhancing LLM-based programming tools.

We structure our study around three aspects of code analysis aligned with human practices: syntax parsing, static semantics inference, and dynamic reasoning. We evaluate 21 state-of-the-art LLMs across nine representative tasks spanning multiple programming languages (C, Java, Python, Solidity), including Abstract Syntax Tree (AST) generation, Control Flow Graph (CFG) construction, data dependency analysis, taint analysis, and flaky test reasoning. We apply a rigorous three-layer evaluation protocol (combining automated metrics, expert adjudication, and consistency validation) to 3,124 code samples. The protocol achieves high inter-rater reliability (Cohen's $\kappa = 0.844$--$0.936$) and strong human-machine agreement (Gwet's AC1 = 0.500--0.727, F1 = 0.791--0.882), ensuring methodological rigor. Our results reveal that while best-performing LLMs excel in syntax parsing (AST 90\%+, expression matching 84--100\%) and show initial promise in static analysis, their dynamic reasoning capabilities remain limited ($<$70\%) with high data-shift sensitivity (F1 scores varying 0--1.0 across projects). This capability hierarchy (strong syntax parsing, moderate static analysis, weak dynamic reasoning) appears consistently across different model families and scales, suggesting fundamental rather than transient limitations.

These findings offer new insights into how LLMs can complement traditional code analyzers: LLMs provide cross-language generalization but produce non-deterministic outputs requiring validation, while traditional tools provide deterministic guarantees but require language-specific configuration. We contribute a validated evaluation framework with systematic comparison against traditional analyzers (Tree-sitter, Soot, Joern), task-specific applicability tiers for deployment guidance, and clarify directions for future enhancement of LLM-based code analysis. Our open benchmark is released at \url{https://github.com/mathieu0905/llm_code_analysis.git}.
\end{abstract}



\maketitle

\section{Introduction}

Code analysis is a cornerstone of Software Engineering (SE), enabling developers to ensure code correctness, optimize performance, and identify vulnerabilities. Human developers analyze code by understanding its syntax, constructing representations such as control flow and data dependency graphs, and reasoning about dynamic behaviors under specific inputs. These practices underscore the importance of three fundamental aspects of code analysis: syntax parsing, static semantics inference, and dynamic reasoning.

Traditional code analysis tools, despite their maturity, face practical limitations: they require language-specific configuration and lack cross-language generalization~\cite{10.1109/WETSEB.2019.00008}. This approach leads to a scarcity or immaturity of tools for less commonly used languages.

Large language models (LLMs) offer a compelling alternative. With the success of LLMs in domains like text generation and dialogue systems, both industry and academia have begun to explore how LLMs can streamline software development tasks. For instance, multi-agent systems like MetaGPT~\cite{hong2023metagpt} can autonomously write code based on natural language instructions. LLMs have demonstrated strong capabilities in code-related tasks such as generation~\cite{chen2021evaluating}, repair~\cite{xia2023keep}, and summarization~\cite{tian2023chatgpt}. However, prior studies~\cite{mouselinos2023simple,yang2024robustness,yao2024survey} have revealed robustness challenges: minor input variations can lead to incorrect outputs. Despite these advancements, the application of LLMs to code analysis tasks remains underexplored. This gap motivates our study to evaluate the capabilities of LLMs for code analysis, focusing on their ability to replicate human-like practices in analyzing code syntax, static semantics, and dynamic reasoning.

Following the practice of developers in understanding code through syntax parsing, static semantics inference, and dynamic reasoning, in this work, we ask: \textit{How effectively can LLMs perform code analysis tasks that mirror human developers' practices?} If a language model demonstrates proficiency in code analysis, it inherently showcases strong code comprehension abilities. This capability serves as a cornerstone for building LLM-driven automated programming systems, enabling them to write, test, and refine code.

To study the ability of LLMs on code analysis, we conducted in-depth research to evaluate their code analysis and comprehension capabilities. Our study focuses on three key aspects: (1) parsing program syntax, (2) inferring static semantics, and (3) reasoning about dynamic behaviors. We validate the code analysis capabilities of LLMs in a zero-shot setting, as this approach highlights how LLMs utilize their internal knowledge to perform code analysis without external retrieval, thus establishing their baseline capabilities. We evaluate 21 LLMs spanning diverse architectures and scales. For clarity in presentation, we show representative subsets in main tables and figures, with complete results provided in the appendix. We adopt consistent naming for OpenAI o-series models (e.g., GPT-o1-mini, GPT-o3, GPT-o4-mini). To systematically assess these models, we applied 3,124 code samples spanning four programming languages (C, Java, Python, Solidity) to the following tasks:

\begin{itemize}[leftmargin=*,nosep]
\item 
Syntax parsing (2 tasks): AST generation tests whether LLMs can infer syntactic relationships between tokens in raw code by constructing a tree structure according to predefined syntactic rules. Successfully generating an AST indicates that the model can infer these relationships and the underlying grammar. Expression matching evaluates the model's ability to recognize the syntactic roles of identifiers, such as distinguishing between operators and operands, further showcasing its syntax parsing capability.

\item Static semantics (5 tasks): We evaluate whether LLMs can statically approximate program behavior similar to traditional static analysis tools~\cite{10.1109/WETSEB.2019.00008, 10.1007/978-3-642-33826-7_16}. Control Flow Graph (CFG) generation, Call Graph (CG) generation, and data dependency analysis are fundamental representations of code semantics, requiring LLMs to construct semantic connections between tokens accurately. Pointer analysis and taint analysis demand deeper semantic reasoning capability, as they involve establishing semantic links between variables that are not directly related in the code.

\item Dynamic reasoning (2 tasks): Building on the findings from LLMs' performance in syntax parsing and static analysis, we introduced two more challenging tasks to evaluate their ability to analyze dynamic program behaviors. Equivalent mutant detection evaluates whether LLMs can identify code modifications that preserve a program's behavior, showcasing their ability to recognize functional equivalence despite structural differences. Flaky test reasoning focuses on scenarios where identical code produces inconsistent outcomes due to external factors or nondeterministic behavior, assessing the LLMs' capacity to reason about and predict context-sensitive variations in program execution.
\end{itemize}

Our findings reveal a consistent capability hierarchy across models: (1)~Best-performing LLMs achieve strong syntax parsing capabilities (AST 90\%+, expression matching 84--100\%), producing high‑quality ASTs, though not as deterministic as compiler parsers; (2)~LLMs demonstrate initial competence in static analysis, acting as beginners leveraging internal knowledge, but exhibit 3--5$\times$ performance gaps between top-tier and mid-tier models; (3)~LLMs show limited dynamic reasoning capabilities, with performance plateauing below 70\% and high data-shift sensitivity (F1 scores varying from 0 to 1.0 across projects in dependency and taint analysis); (4)~Within-model ablation studies on dynamic reasoning tasks reveal significant model-strategy interactions: Chain-of-Thought degrades GPT-5-mini on mutant detection ($\Delta$acc$=-0.065$, $p<0.001$) but improves GPT-o1-mini ($\Delta$acc$=+0.230$, $p<10^{-7}$), indicating that prompt strategy must be tuned per model-task combination. This hierarchy, syntax $>$ static $>$ dynamic, holds across different model families, scales, and architectures, suggesting fundamental rather than transient limitations.

Overall, the main contributions of our paper are summarized as follows:
\begin{enumerate}[leftmargin=*,nosep]
    \item \textbf{Comprehensive evaluation framework:} We establish a three-layer evaluation protocol (automated metrics, expert adjudication, consistency validation) with rigorous reliability analysis: \textit{human--human} inter-rater reliability achieves high agreement (Cohen's $\kappa = 0.844$--$0.936$), while \textit{human--machine} agreement using prevalence-corrected metrics (AC1 = 0.500--0.727, F1 = 0.791--0.882) validates automated metrics against expert judgment. The framework spans nine tasks across syntax, static semantics, and dynamic reasoning in four programming languages, assessing 21 models on 3,124 code samples. We release evaluation scripts, standardized prompts, and annotated datasets at \url{https://github.com/mathieu0905/llm_code_analysis.git}.
    
    \item \textbf{Capability hierarchy across models:} Our study suggests that LLMs are capable of comprehending code syntax rules and have certain abilities to infer static semantics but show limited performance on dynamic reasoning tasks. We reveal a consistent capability hierarchy: strong syntactic parsing (90\%+ for best-performing models), initial static analysis competence with 3-5$\times$ performance variance, limited dynamic reasoning ($<$70\%), and high data-shift sensitivity. This pattern holds across model families, scales, and architectures, suggesting fundamental rather than transient limitations.
    
    \item \textbf{Practical deployment guidance:} Through quantitative comparison with traditional tools (Tree-sitter, Soot, Joern), we demonstrate LLMs' complementary role in code analysis: cross-language generalization capability but non-deterministic outputs requiring validation. We identify task-specific applicability tiers and hybrid workflow patterns for integrating LLMs into analysis pipelines, positioning them as complementary assistive tools rather than replacements for deterministic analyzers in safety-critical contexts.
\end{enumerate}

\noindent\textbf{Paper organization.} Section~\ref{sec:approach} describes the study design including nine analysis tasks, their formal definitions, and task selection rationale. Section~\ref{sec:evaluation} presents the experiment setup including the three-layer evaluation protocol, dataset, prompt design, and metrics. Section~\ref{sec:result} reports experimental results across 21 models for syntax parsing, static semantics, and dynamic reasoning tasks. Section~\ref{sec:discussion} synthesizes capability hierarchies, analyzes agreement and validity, compares prompt strategies, and positions LLMs within analysis toolchains. Section~\ref{sec:threats_to_validity} discusses threats to validity. Section~\ref{sec:related_work} surveys related work on code model evaluation and program analysis techniques. Section~\ref{sec:conclusion} concludes with future research directions.

\section{Study Design}
\label{sec:approach}

\subsection{Overview}
\begin{figure}[H]
	\centering
 \scalebox{0.9}{
\includegraphics[width=1.0\textwidth]{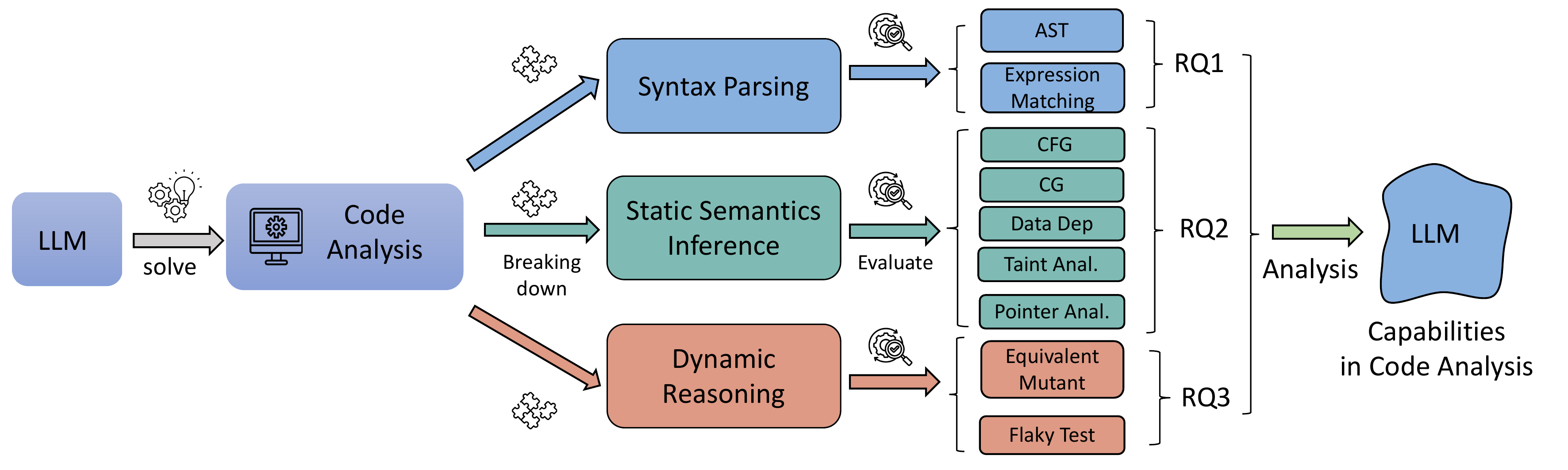}}
	\caption{Overview of our study design.}
	\label{fig:overview}
         \vspace{-1mm}
\end{figure}
\FloatBarrier

We evaluate LLMs on code analysis through three aspects: syntax parsing, static semantics inference, and dynamic reasoning. These mirror how developers analyze code by understanding structure, inferring relationships without execution, and reasoning about runtime behavior.

Our study addresses the following research questions:
\begin{itemize}
    \item \textit{RQ1: Can LLMs parse code syntax (via AST generation and expression matching tasks)?}
    \item \textit{RQ2: Can LLMs infer static semantic structures (CFG, call graph, data/taint dependencies, pointer aliasing)?}
    \item \textit{RQ3: Can LLMs reason about dynamic execution behaviors (mutant equivalence, flaky tests)?}
\end{itemize}

\figref{fig:overview} illustrates our evaluation framework. We employ nine tasks across four programming languages (C, Java, Python, Solidity). Task-specific evaluation metrics and validation protocols are detailed in Section~\ref{sec:evaluation}.

\subsection{Rationale for Task Selection}
\label{sec:task_rationale}

We select nine tasks spanning syntax parsing, static semantics inference, and dynamic reasoning based on three criteria:

\begin{itemize}
    \item \textit{Representativeness}: The task captures fundamental analysis operations used across multiple programming paradigms and tools.
    \item \textit{Verifiability}: Task correctness can be objectively assessed through tool-generated ground truth or expert consensus.
    \item \textit{Practical Relevance}: The task is employed in production tools such as compilers, IDEs, security analyzers, or testing frameworks.
\end{itemize}

Our task suite provides systematic capability profiling through stratified complexity (syntax $\rightarrow$ static $\rightarrow$ dynamic), enabling identification of where LLM performance degrades across abstraction levels.

\paragraph{Syntax Layer (2 tasks)} 
AST Generation and Expression Matching assess structural parsing and syntactic role recognition, operations fundamental to all programming paradigms (representativeness). Both tasks use standardized syntactic representations with deterministic ground truth (verifiability) and underpin practical tools such as compilers, IDEs, and code search/refactoring workflows (practical relevance)~\cite{wang2022tree,10.1145/3510003.3510096}. \textit{AST Generation} evaluates whether LLMs recognize token relationships and hierarchical structure. \textit{Expression Matching} tests whether LLMs can distinguish operators from operands, understand precedence, and match patterns despite variable renaming.

\paragraph{Static Semantics Layer (5 tasks)}
This layer tests whether LLMs infer semantic relationships without executing code. We select five tasks covering core static analysis abstractions used across paradigms (representativeness)~\cite{guo2020graphcodebert,ma2022graphcode2vec}: \textit{CFG} for execution order and branching, \textit{Call Graphs} for function invocation relationships, \textit{Data Dependency} for value flow tracking, \textit{Taint Analysis} for security-critical data propagation, and \textit{Pointer Analysis} for memory aliasing. These tasks leverage production-grade analyzers (Soot, Joern, Slither) for ground truth (verifiability) and directly support security analyzers and optimization tools (practical relevance). They span increasing complexity: CFG and call graph generation require structural inference, data dependency and taint analysis require transitive reasoning, while pointer analysis requires context-sensitive interprocedural reasoning.

\paragraph{Dynamic Reasoning Layer (2 tasks)}
This layer tests whether LLMs reason about runtime behavior without executing code, capabilities essential for testing and debugging across paradigms (representativeness). \textit{Equivalent Mutant Detection} requires determining whether code changes preserve behavior across all possible inputs through symbolic reasoning~\cite{Nica_2012}, validated against established benchmarks (verifiability). \textit{Flaky Test Reasoning} requires inferring non-determinism sources such as async timing, concurrency, randomness, and platform dependencies from static code inspection~\cite{akli2022predicting}, directly supporting testing frameworks (practical relevance).

\paragraph{Language Selection}
We evaluate across four languages covering diverse paradigms, including C (low-level control, explicit memory, pointer arithmetic), Java (object-oriented, static typing, exception handling, polymorphism), Python (dynamic typing, concise syntax, first-class functions), and Solidity (domain-specific blockchain language with transaction semantics and external calls). This diversity tests cross-language generalization by assessing whether models capture language-specific patterns versus transferable principles.

\subsection{Code Syntax Analysis (RQ1)}

Parsing code syntax is needed for LLMs to analyze and manipulate code. Code syntax defines valid symbol combinations in programming languages. These rules matter for tasks ranging from summarization to bug localization. We define two complementary tasks for syntactic evaluation.

\subsubsection{AST Generation} 
\label{sec:ast_eval}

\paragraph{\textit{Motivation}}
Abstract Syntax Trees represent code's hierarchical structure according to language grammar rules. For example,  \colorbox{gray!15}{\texttt{int x = a + b;}} parses into a tree where \texttt{=} forms the root (assignment operation), \texttt{x} is the left child (assignment target), and \texttt{+} is the right child (binary expression) with children \texttt{a} and \texttt{b} (operands). AST generation tests whether LLMs internalize grammar rules to transform textual code into structured representations.

\paragraph{\textit{Task Definition}}
Given source code $C$, the LLM generates an abstract syntax tree $\hat{T} = (V, E, r, \lambda)$, where $V$ represents nodes, $E$ denotes parent-child edges, $r$ is the root node, and $\lambda$ assigns syntactic categories (e.g., \texttt{FunctionDef}, \texttt{IfStatement}). Reference trees are obtained from Tree-sitter for Python/C/Solidity and language-specific parsers for Java. Evaluation metrics are detailed in Section~\ref{sec:ast_metrics}.

\paragraph{\textit{Challenges}}
AST generation requires recognizing language-specific grammar rules, inferring token relationships, and constructing hierarchical structures that preserve semantic information. LLMs must distinguish syntactic categories (statements vs. expressions vs. declarations), maintain correct parent-child relationships, and handle language-specific constructs such as exception handling in Java, pointer declarations in C, and lambda expressions in Python.

\paragraph{\textit{Practical Applications}}
ASTs are fundamental to compilers, IDEs, and learning-based code analysis. Refactoring engines use ASTs to identify code patterns for transformation; syntax highlighters rely on ASTs for token classification; semantic analyzers build upon ASTs for type checking and control-flow analysis. Accurate AST generation demonstrates that LLMs have learned grammar rules essential for code generation, bug detection, and program transformation~\cite{wang2022tree,10.1145/3510003.3510096}.

\subsubsection{Expression Matching} 
\label{sec:exp_eval}

\paragraph{\textit{Motivation}}
Expression matching tests whether LLMs recognize syntactic roles beyond token similarity. Consider finding expressions similar to  \colorbox{gray!15}{\texttt{rate + utilization * factor}}. A text-based matcher might rank \colorbox{gray!15}{\texttt{rate - utilization + factor}} highly (similar names) despite different semantics. A syntax-aware system recognizes  \colorbox{gray!15}{\texttt{r + u * f}} as structurally identical (same operator precedence) despite different identifiers. This capability is essential for code search and pattern matching.

\paragraph{\textit{Task Definition}}
Given target expression $e_{\text{target}}$ and ranked candidates $C = [c_1, c_2, \ldots, c_n]$ from source code, the task retrieves expressions with high structural similarity to $e_{\text{target}}$ in top-$k$ positions. Expression $c_i$ is relevant if it has high structural similarity (measured via semantic embedding distance) and satisfies occurrence constraints (tokens appear in source). Evaluation metrics are detailed in Section~\ref{sec:expression_metrics}.

\paragraph{\textit{Challenges}}
Expression matching requires distinguishing structural similarity from superficial token overlap. LLMs must recognize operator precedence and associativity (e.g.,  \colorbox{gray!15}{\texttt{a + b * c}} differs structurally from  \colorbox{gray!15}{\texttt{(a + b) * c}}), differentiate semantic roles of operators versus operands (operators define computation structure while operands are interchangeable), and handle variable renaming (e.g.,  \colorbox{gray!15}{\texttt{rate + util * factor}} should match  \colorbox{gray!15}{\texttt{r + u * f}}). The task demands abstraction beyond exact string matching to capture computational patterns invariant to identifier names.

\paragraph{\textit{Practical Applications}}
Expression matching enables code search engines to locate computation patterns despite variable renaming, supports refactoring tools in identifying similar code fragments, and assists in code review by finding analogous implementations. This task originated from a blockchain challenge: verifying smart contracts implement reward mechanisms from whitepapers. Our dataset contains 32 reward formulas from four DeFi projects (ALPHA~\cite{AlphaFinanceLab}, BETA~\cite{BETA}, BiFi~\cite{BiFi}, XEN~\cite{XEN}), where LLMs must locate formula implementations in Solidity code despite variable renaming. This requires recognizing syntactic roles: operators define computational structure while operands are placeholders.

\subsection{Code Static Semantics Analysis (RQ2)}

Analyzing static semantics lets LLMs infer code logic without execution. Static semantics capture how constructs relate: control flow, data dependencies, calling relationships. These underpin bug detection~\cite{ayewah2008using}, optimization, and security analysis. We evaluate five tasks spanning structural inference to interprocedural reasoning.

\subsubsection{Control Flow Graph (CFG) Analysis} 
\label{sec:cfg_eval}

\paragraph{\textit{Motivation}}
Control Flow Graphs model statement execution order, capturing control transfer through branches, loops, and exceptions. An \texttt{if-else} creates two edges from the condition (one to \texttt{then}, one to \texttt{else}), both converging at a merge point. A \texttt{while} loop includes a back-edge from body to condition. CFG construction tests whether LLMs understand control flow semantics and can represent execution paths.

\paragraph{\textit{Task Definition}}
Given function $f$ with code $C_f$, the LLM generates control-flow graph $\hat{G} = (B, E, \text{entry}, \text{exit})$, where $B$ represents basic blocks (statement sequences with single entry/exit), $E$ denotes control-flow edges (jumps, branches, fall-throughs), and entry/exit designate start and termination. Reference CFGs are obtained from Joern (C), tree-sitter-graph (Python), and Soot (Java). Evaluation metrics are detailed in Section~\ref{sec:cfg_metrics}.

\paragraph{\textit{Challenges}}
CFG generation requires recognizing control structures (conditionals, loops, exceptions), inferring control transfer semantics (fall-through vs. explicit jumps), and modeling language-specific features such as short-circuit evaluation, exception propagation, and switch statements. LLMs must construct basic blocks, identify entry and exit points, and create correct edges representing all possible execution paths.

\begin{figure}[t]
	\centering
 \scalebox{0.8}{
	\includegraphics[width=0.8\textwidth]{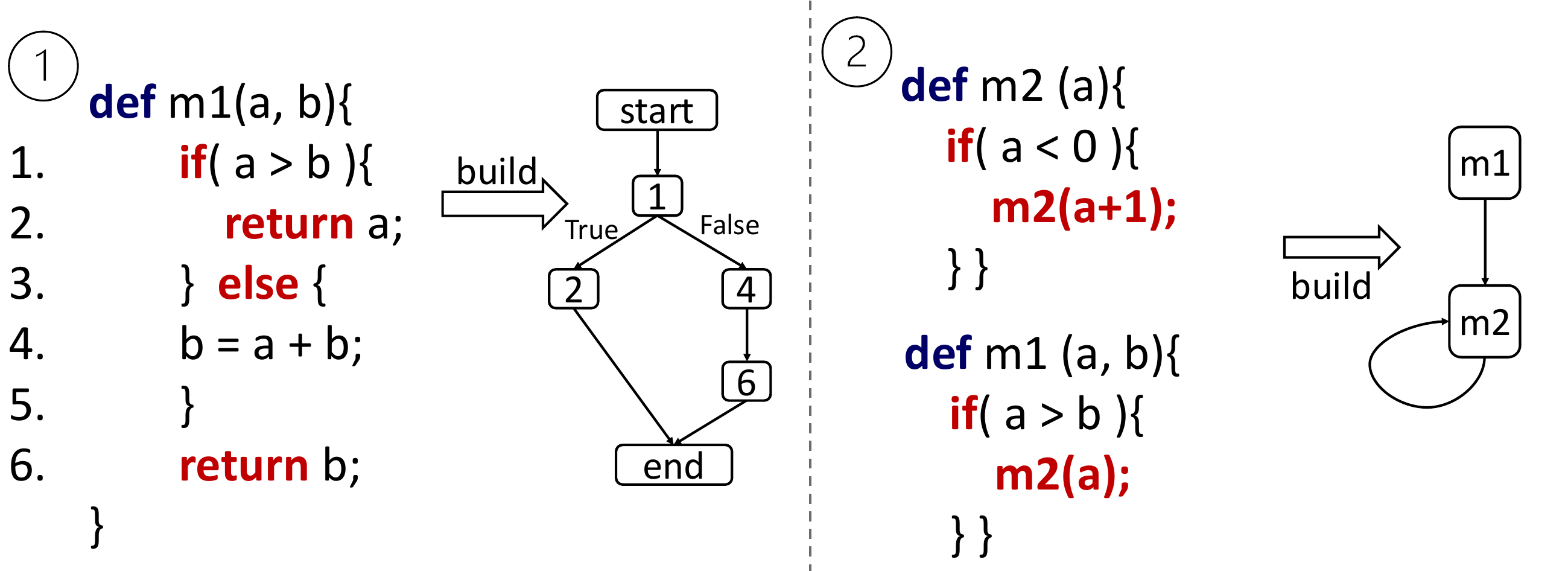}}
	\caption{Examples of \textcircled{1} Control Flow Graph (Python) and \textcircled{2} Call Graph (Python).}
	\label{fig:cfg_cg}
\end{figure}

\paragraph{\textit{Practical Applications}}
CFGs enable path-sensitive analysis, dead code elimination, and vulnerability detection. Testing tools use CFGs to compute execution paths and generate test cases; security analyzers use CFGs to identify unreachable code and detect infinite loops; optimizers use CFGs to enable instruction reordering and dead code elimination~\cite{10.1145/800028.808479}. \figref{fig:cfg_cg}\textcircled{1} illustrates a CFG example.

\subsubsection{Call Graph (CG) Analysis} 
\label{sec:cg_eval}

\paragraph{\textit{Motivation}}
Call graphs represent function invocations for interprocedural analysis. If \texttt{main()} calls \texttt{processData()}, which calls \texttt{validateInput()} and \texttt{computeResult()}, the call graph captures this hierarchy. Call graph construction tests whether LLMs understand function invocation semantics and can resolve complex calling patterns including polymorphism, function pointers, and callbacks.

\paragraph{\textit{Task Definition}}
Given program $P$ with functions $F = \{f_1, f_2, \ldots, f_n\}$, the LLM generates call graph $\hat{G}_c = (F, E_c)$, where nodes represent functions and directed edges $e_{ij} \in E_c$ denote that function $f_i$ invokes $f_j$. Reference call graphs are obtained from Soot (Java), Joern (C/C++), and Slither (Solidity). Evaluation metrics are detailed in Section~\ref{sec:cg_metrics}.

\paragraph{\textit{Challenges}}
Call graph construction requires resolving function invocations across the program, handling polymorphism and dynamic dispatch in object-oriented languages, resolving function pointers in C, and identifying callbacks and lambda expressions. LLMs must trace execution from entry points, distinguish direct calls from indirect invocations, and construct the complete calling hierarchy.

\paragraph{\textit{Practical Applications}}
Call graphs enable impact analysis (determining which functions are affected by changes), vulnerability propagation tracking (identifying whether user input can reach sensitive functions), and optimization (determining which functions can be inlined). Security analyzers use call graphs to trace data flow across function boundaries; refactoring tools use call graphs to assess change impact; program comprehension tools use call graphs to visualize system architecture~\cite{10.1145/279310.279314}. \figref{fig:cfg_cg}\textcircled{2} illustrates the control flow graph and the call graph example that are constructed from the source code.

\subsubsection{Data Dependency} 
\label{sec:dd_eval}

\paragraph{\textit{Motivation}}
Data dependency analysis tracks how values flow between variables through program execution. In the sequence  \colorbox{gray!15}{\texttt{a = input(); b = a * 2; c = b + 1;}}, variable \texttt{c} depends on \texttt{b} (uses its value), \texttt{b} depends on \texttt{a}, and transitively \texttt{c} depends on \texttt{a}. This transitive reasoning is essential for analyzing program semantics and data flow.

\paragraph{\textit{Task Definition}}
Given program $P$ and variable pair $(v_i, v_j)$, the task determines whether $v_j$ data-dependent on $v_i$, that is, whether $v_j$'s value is computed using $v_i$ or a transitively dependent variable. This is formulated as binary classification. Reference labels are obtained from Joern (Java) and Slither (Solidity). Evaluation metrics are detailed in Section~\ref{sec:dd_taint_metrics}.

\paragraph{\textit{Challenges}}
Data dependency analysis requires transitive reasoning across multiple statements and control flow paths. LLMs must trace value flow through assignments, function calls, and control structures, distinguishing direct dependencies (e.g., \texttt{b = a}) from transitive dependencies (e.g., \texttt{c = b} where \texttt{b = a}). The task demands inferring def-use chains, handling conditional assignments where dependencies vary by execution path, and recognizing implicit dependencies through aliasing and pointer indirection.

\begin{figure}[t]
	\centering
 \scalebox{0.6}{
	\includegraphics[width=0.8\textwidth]{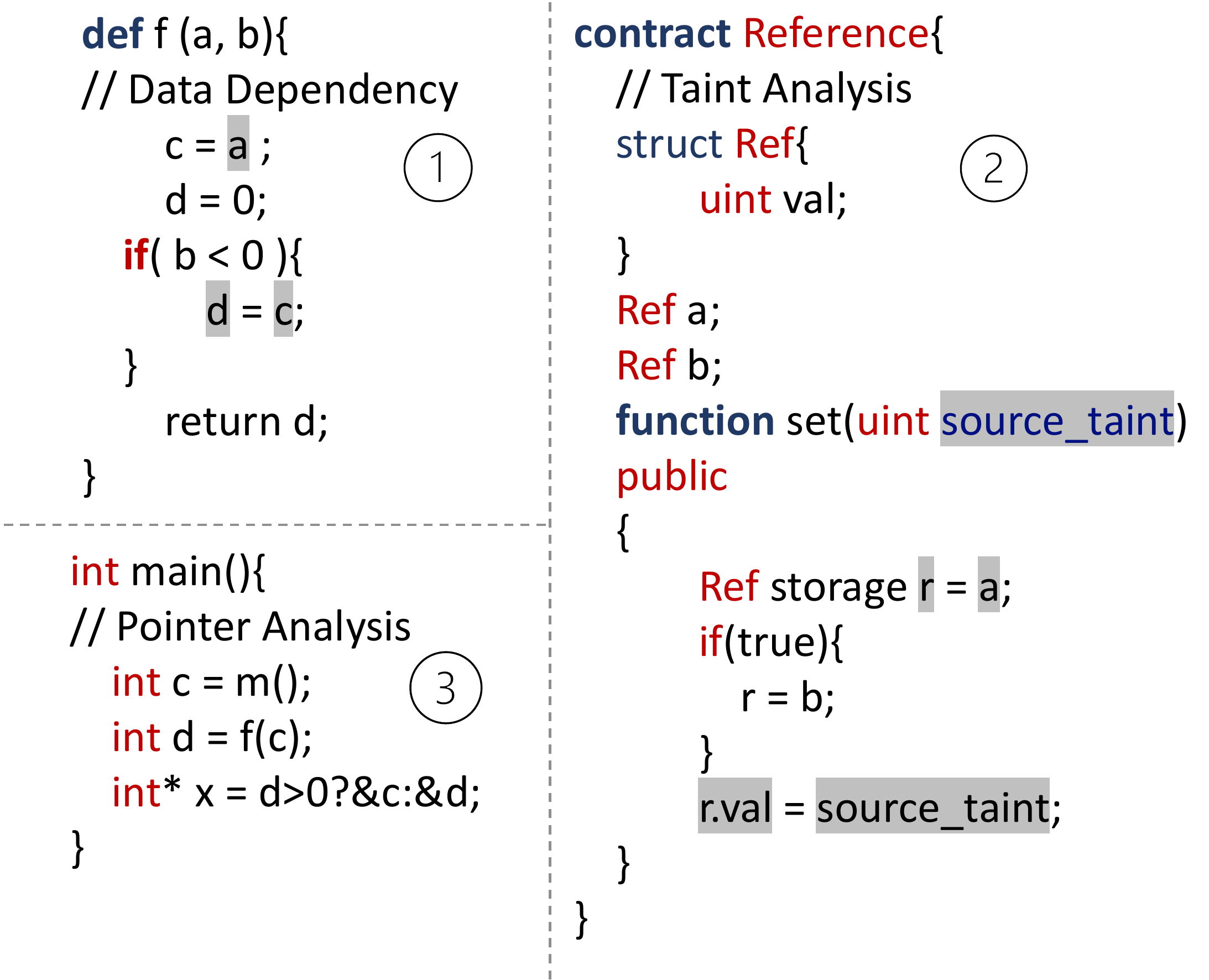}}
	\caption{Analysis examples: \textcircled{1} Data dependency (Python); \textcircled{2} Taint analysis (Solidity); \textcircled{3} Pointer analysis (C).}
	\label{fig:dd_taint_pointer}
\end{figure}

\paragraph{\textit{Practical Applications}}
Data dependency analysis enables program slicing (identifying statements that affect a variable), parallelization (determining whether loops can execute concurrently), and debugging (tracing value origins). Security analyzers use dependency analysis to track sensitive data propagation; optimizers use it to identify dead code and enable instruction reordering~\cite{ferrante1984program,guo2020graphcodebert}. \figref{fig:dd_taint_pointer}\textcircled{1} illustrates variable \texttt{d} depending on \texttt{a}.

\subsubsection{Taint Analysis} 
\label{sec:taint_eval}

\paragraph{\textit{Motivation}}
Taint analysis extends data dependency by tracking external inputs to security-critical operations. If \texttt{userInput} flows through assignments and calls to construct an SQL query, the system is vulnerable to SQL injection. Taint tracking requires recognizing taint sources (external inputs), inferring taint propagation (assignment preserves taint; sanitization removes it), and identifying whether tainted data reaches dangerous sinks.

\paragraph{\textit{Task Definition}}
Given program $P$ with designated taint sources (e.g., \texttt{msg.sender} in Solidity, user input in Java), the task determines for each variable $v$ whether it can be influenced by a taint source through any execution path. This is formulated as binary classification: $\text{tainted}(v) \in \{0, 1\}$, where $1$ indicates potential taint propagation. Reference labels are obtained from Slither (Solidity) and Joern (Java). Evaluation metrics are detailed in Section~\ref{sec:dd_taint_metrics}.

\paragraph{\textit{Challenges}}
Taint analysis extends data dependency with security-specific reasoning. LLMs must identify taint sources (user inputs, network data, external calls), track taint propagation through assignments and function calls while recognizing sanitization operations that remove taint, and determine whether tainted data reaches dangerous sinks (SQL queries, system commands, sensitive storage). The task requires interprocedural analysis to track taint across function boundaries and context-sensitive reasoning to distinguish sanitized from unsanitized paths.

\paragraph{\textit{Practical Applications}}
Taint analysis is essential for security vulnerability detection, particularly for injection attacks (SQL injection, command injection, cross-site scripting). Security analyzers use taint analysis to identify whether untrusted input can reach sensitive operations; code review tools use it to flag potential vulnerabilities; secure coding assistants use it to suggest input sanitization~\cite{kim2014survey}. This requires integrating CFG, call graph, and data dependency analysis. \figref{fig:dd_taint_pointer}\textcircled{2} illustrates variable \texttt{a} tainted via intermediary \texttt{r}.

\subsubsection{Pointer Analysis} 
\label{sec:pointer_eval}

\paragraph{\textit{Motivation}}
Pointer analysis determines which memory locations a pointer may reference. In this code:
\begin{lstlisting}[language=C]
int a, b, *p;
if (condition) p = &a; else p = &b;
\end{lstlisting}
pointer \texttt{p} may point to \texttt{a} or \texttt{b}, depending on runtime \texttt{condition}. Pointer analysis is essential for memory safety verification and alias analysis in C/C++ programs.

\paragraph{\textit{Task Definition}}
Given C program $P$ and pointer variable $p$, the LLM predicts points-to set $\widehat{\text{PT}}(p)$: variables that $p$ may reference during execution. Reference points-to sets are obtained from Frama-C~\cite{10.1007/978-3-642-33826-7_16}, a sound static analyzer for C. Multiple pointers per program are evaluated. Evaluation metrics are detailed in Section~\ref{sec:pointer_metrics}.

\paragraph{\textit{Challenges}}
Pointer analysis is the most challenging static analysis task~\cite{smaragdakis2015pointer}, requiring context-sensitive, flow-sensitive, interprocedural reasoning with aliasing. LLMs must track pointer assignments across control flow paths, handle address-of operations, resolve pointer arithmetic, and propagate aliasing information through function calls.

\paragraph{\textit{Practical Applications}}
Pointer analysis enables detecting null dereferences, buffer overflows, and use-after-free bugs. Memory safety analyzers use pointer analysis to verify safe memory access; optimizers use it to enable aggressive optimizations; program verification tools use it to prove memory safety properties. \figref{fig:dd_taint_pointer}\textcircled{3} illustrates pointer \texttt{x} potentially pointing to \texttt{c} or \texttt{d}.

\subsection{Code Dynamic Reasoning (RQ3)}

Analyzing dynamic semantics requires reasoning about program behavior during execution, especially under uncertainty and state changes. While syntax and static analysis provide foundational inputs, dynamic reasoning tests whether LLMs predict runtime behavior without executing code. This matters for testing, debugging, and quality assurance.

\subsubsection{Equivalent Mutant Detection} 
\label{sec:mutant_eval}

\paragraph{\textit{Motivation}}
Mutation testing introduces code changes to evaluate test quality. Some mutants are equivalent: they preserve semantics despite syntax changes. For example, swapping operands in \colorbox{gray!15}{\texttt{a + b}} yields an equivalent mutant due to commutativity, while changing \texttt{<} to \texttt{<=} is non-equivalent because it alters boundary conditions. Detecting equivalence requires comparing execution semantics across all possible inputs without executing code.

\paragraph{\textit{Task Definition}}
Given original program $P$ and mutated version $P'$ differing by one syntactic change such as operator substitution (\texttt{+} $\rightarrow$ \texttt{-}) or statement reordering, the task determines whether $P$ and $P'$ are semantically equivalent, producing identical outputs for all possible inputs. This is formulated as binary classification. The dataset contains 100 equivalent and 100 non-equivalent mutants from 35 C/Java programs from mutation testing benchmarks. Evaluation metrics are detailed in Section~\ref{sec:mutant_metrics}.

\paragraph{\textit{Challenges}}
Equivalent mutant detection requires symbolic reasoning about program behavior across all possible inputs. LLMs must recognize semantic equivalence rather than syntactic similarity, reason about mathematical properties such as commutativity and associativity, and consider boundary conditions and edge cases. This requires deep semantic reasoning capability beyond static dependency tracking~\cite{Nica_2012}.

\begin{figure}[]
	\centering
 \scalebox{0.8}{
	\includegraphics[width=0.8\textwidth]{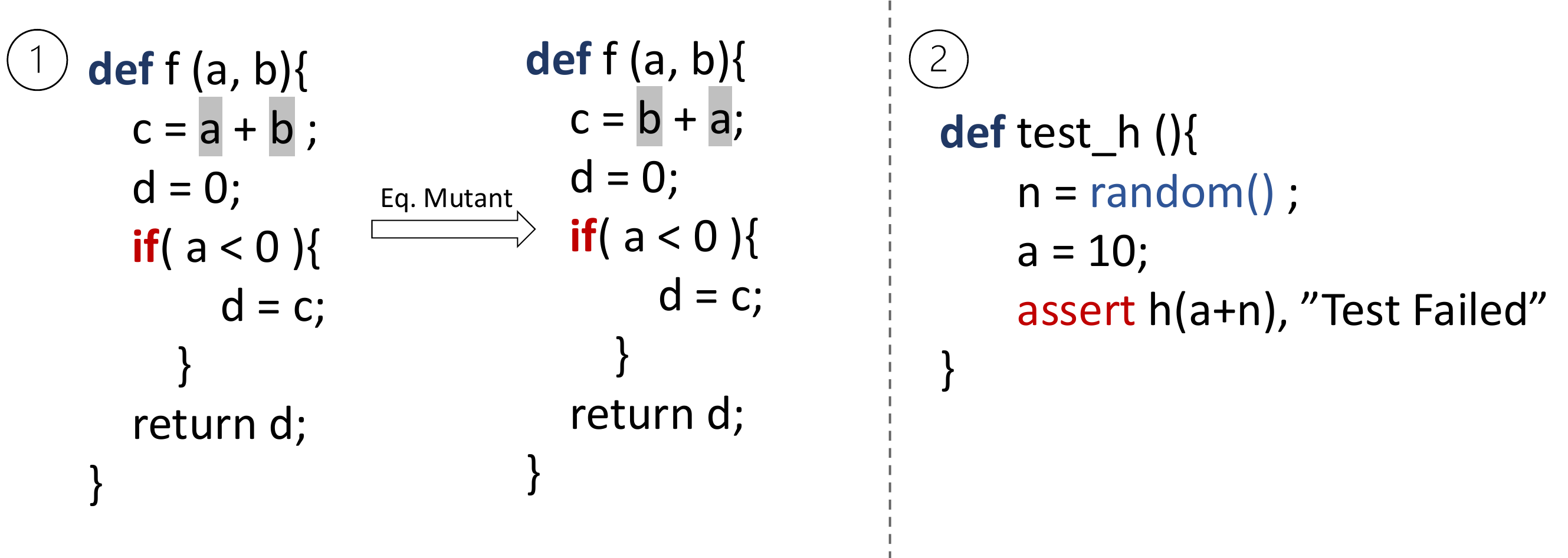}}
	\caption{Dynamic reasoning examples: (1)~Equivalent Mutant and (2)~Flaky Test Root Cause.}
	\label{fig:dynamic}
\end{figure}

\paragraph{\textit{Practical Applications}}
Equivalent mutant detection improves mutation testing efficiency by filtering out equivalent mutants that cannot be killed by any test. Testing tools use this to focus test generation efforts; mutation testing frameworks use it to provide accurate mutation scores; test quality assessment tools use it to avoid false alarms. \figref{fig:dynamic}\textcircled{1} illustrates swapping \texttt{a} and \texttt{b} in addition preserving behavior.

\subsubsection{Flaky Test Reasoning} 
\label{sec:flaky_eval}

\paragraph{\textit{Motivation}}
Flaky tests behave non-deterministically: they pass in some runs but fail in others under identical conditions. Root causes include async timing issues such as insufficient \texttt{Thread.sleep} durations, concurrency problems like race conditions, randomness from non-seeded generators, platform-specific behavior, and I/O variability such as network latency. For example, \texttt{Random()} without a seed introduces randomness, while \texttt{HashMap} iteration order is non-deterministic in Java.

\paragraph{\textit{Task Definition}}
Given a test case exhibiting non-deterministic behavior, the task identifies the root cause from a taxonomy of flaky test categories: \texttt{async\_wait, concurrency, network, randomness, platform\_dependency, time, io, floating\_point, test\_case\_timeout, unordered\_collection}. This is formulated as multi-class classification requiring inference of runtime non-determinism sources from static code. The dataset contains 65 flaky test cases from Java projects with documented root causes~\cite{akli2022predicting}. Evaluation metrics are detailed in Section~\ref{sec:flaky_metrics}.

\paragraph{\textit{Challenges}}
Flaky test diagnosis requires reasoning about runtime uncertainties and execution environment interactions. LLMs must recognize threading models, I/O operations, platform-specific APIs, timing dependencies, and non-deterministic data structures. This requires inferring dynamic behavior from static code inspection without execution.

\paragraph{\textit{Practical Applications}}
Flaky test diagnosis helps developers fix non-deterministic tests efficiently. Testing frameworks use flaky test detection to identify unreliable tests; continuous integration systems use it to filter out flaky failures; test maintenance tools use it to prioritize fixing efforts. Correct root cause diagnosis enables targeted fixes rather than trial-and-error debugging. \figref{fig:dynamic}\textcircled{2} illustrates \texttt{Random()} causing flakiness.

\section{Experiment Setup}
\label{sec:evaluation}

\subsection{Evaluation Challenges and Approach}
\label{sec:eval_challenges}

Evaluating LLMs on structural code analysis presents three fundamental challenges: (1)~\textit{format diversity}, as LLM outputs vary in node naming, structural granularity, and representation schemas; (2)~\textit{semantic equivalence}, as different valid representations exist for the same code, making correctness irreducible to structural matching; (3)~\textit{absent canonical ground truth}, as program analysis tools produce structurally distinct yet valid outputs. Standard metrics such as exact-match, BLEU, and graph isomorphism fail to capture semantic equivalence while penalizing valid representational variations.

We address these challenges through a \textit{three-layer evaluation framework} combining automated metrics, expert assessment, and consistency validation:

\begin{enumerate}
    \item Automated Metrics (Sections~\ref{sec:syntax_metrics}, \ref{sec:static_metrics}, \ref{sec:dynamic_metrics}): Task-specific reconstructability scores validated through rigorous \textit{human--human} inter-rater reliability analysis (Section~\ref{sec:agreement}).
    \item Expert Adjudication (Section~\ref{sec:eval_protocol}): Formal assessment rubrics implementing semantic correctness evaluation through a three-tier classification (pass/minor/fail) that distinguishes structural incompleteness from semantic incorrectness, validated through testable criteria (structural completeness, absence of fabrications, semantic fidelity).
    \item Consistency Validation (Section~\ref{sec:eval_protocol}): \textit{Human--machine} agreement analysis using multiple coefficients (Cohen's $\kappa$, AC1, PABAK, F1) to quantify automated metric validity, accounting for prevalence effects in imbalanced data (Section~\ref{sec:agreement}).
\end{enumerate}

This methodology provides rigorous, reproducible evaluation aligned with practical code analysis requirements. Detailed protocols, metric designs, and implementation strategies are presented in the following subsections.

\subsection{Dataset}
\label{sec:dataset}
Our dataset comprises 291 programs across four programming languages (C, Java, Python, Solidity) with 3,124 extracted samples totaling 206,913 LOC. We adopt differentiated data construction strategies for different capability levels based on their evaluation requirements.

Our evaluation spans three capability levels: (i) \textit{syntax level} assesses parsing capabilities on abstract syntax trees and expression matching; (ii) \textit{static level} evaluates structural graph generation (CFG, call graphs) and semantic analysis (data dependencies, taint flows, pointer aliasing); (iii) \textit{dynamic level} examines reasoning about program behavior (equivalent mutant detection, flaky test diagnosis). Task-specific sample counts, data sources, and evaluation baselines are detailed in Table~\ref{tab:task_datasets}.

\begin{table*}[]
	\centering
	\caption{Tasks and Datasets used in this study.}
	\vspace{-2mm}
	\label{tab:task_datasets}
	\scalebox{0.6}{
		\large
		\begin{tabular}{c|l|c|c|c|c|c}
			\toprule
			Task        & Level         & Language            & Ground-Truth Tool / Evaluation           & Programs & Extracted Data Size & LoC \\ \hline
			AST                 & \multirow{2}{*}{syntax}                      & Python, Java, C, Solidity & Tree-Sitter       & 75       & 75    &  1,059   \\ %
			Expression Matching &                                      &     Solidity                      & Top-5/10/20             &    4      &      32   &  4,238 \\ \hline
			CFG                 & \multicolumn{1}{c|}{\multirow{5}{*}{static}} & Python, Java, C, Solidity & Tree-Sitter / Soot / Joern / Slither & 75       & 75    &  1,059  \\ %
			CG                  & \multicolumn{1}{c|}{}                        & Python, Java, C, Solidity & Tree-Sitter / Soot / Joern / Slither & 75       & 75     &  2,356   \\ %
			Data Dependence     & \multicolumn{1}{c|}{}                        & Solidity, Java                  & Slither / Joern           & 38       & 1,210     &  100,947  \\ %
			Taint Analysis      & \multicolumn{1}{c|}{}                        & Solidity, Java                  & Slither / Joern           & 38       & 1,050     &  78,244  \\ %
			Pointer Analysis    & \multicolumn{1}{c|}{}                        & C                         & Frama-C           & 40       & 342     &  2,726   \\ \hline
			Flaky Test Reasoning      & \multirow{2}{*}{dynamic}                     & Java                      & Expert Evaluation & 13       & 65     & 1,615    \\ %
			Equivalent Mutant Detection       &                                              & C, Java                   &  Scripted Patches & 35       & 200    &   15,728   \\ \hline
			Total & \multicolumn{3}{l}{} & \multicolumn{1}{c}{291} & \multicolumn{1}{c}{3,124}  & 206,913 \\ \bottomrule
	\end{tabular}}
\end{table*}

\paragraph{Syntax Level (AST, Expression Matching).} For AST evaluation, we manually constructed 75 programs (1,059 LOC) covering fundamental language constructs (Table~\ref{tab:syntax_cheatsheet}) across Python, Java, C, and Solidity, with ground truth via Tree-sitter. For expression matching, we collected 32 samples from 4 Solidity DeFi projects (ALPHA, BETA, BiFi, XEN; 4,238 LOC)~\cite{AlphaFinanceLab,BETA,BiFi,XEN,etherscan} to evaluate matching mathematical reward equations from whitepapers with contract implementations, addressing blockchain reward mechanism verification~\cite{braem2021richspecificationsethereumsmart}, evaluated via Top-5/10/20 retrieval using CodeBERT embeddings (Section~\ref{sec:expression_metrics}).

\paragraph{Static Level.} This level comprises two task categories: structural tasks evaluating graph generation capabilities and analysis tasks assessing semantic reasoning. For \textit{structural tasks (CFG, Call Graph)}, we manually constructed 75 programs (1,059 LOC, shared with AST) for CFG evaluation and 75 programs (2,356 LOC) for call graph evaluation across Python, Java, C, and Solidity, with ground-truth structures generated via Soot (Java), Joern (C), and Slither (Solidity) for automated Layer~1 comparison. For \textit{analysis tasks (Data Dependence, Taint, Pointer)}, we collected 38 real-world projects, including 15 Solidity DeFi protocols from Etherscan and 23 Java projects from mature, widely adopted libraries (mainly Apache Commons series, together with representative JSON processing, networking/I/O, testing, and program-analysis utilities), extracting 1,210 data-dependence pairs and 1,050 taint-analysis samples from these codebases (100,947 LOC for data dependence, 78,244 LOC for taint, analyzed via Slither for Solidity and Joern for Java~\cite{joern,slither}). This selection captures diverse yet practical coding patterns across smart-contract finance and general-purpose software through well-maintained, extensively used codebases. For pointer analysis, we use 40 C programs from SVF benchmark (2,726 LOC, 342 cases) with ground truth via Frama-C. Manual construction for structural tasks enables contamination control for zero-shot evaluation and supports consistent expert validation, while real-world projects for analysis tasks ensure representativeness of production code patterns.

\paragraph{Dynamic Level (Mutant Detection, Flaky Test).} We use established benchmarks: 200 equivalent-mutant pairs from 35 C/Java programs (15,728 LOC)~\cite{9440157} evaluated via scripts that apply patches; and 65 flaky test cases~\cite{akli2022predicting} from 13 Java projects (1,615 LOC) with documented root causes evaluated by expert assessment.

\begin{table*}[t]
	\centering
	\caption{Syntax Cheatsheet for C, Python, Java and Solidity}
	\vspace{-2mm}
	\label{tab:syntax_cheatsheet}
	\scalebox{0.6}{
		\begin{tabular}{c|c|c|c}
			\hline
			C                                    & Python                               & Java                              & Solidity \\ \hline
			Arrays                               & Array-like                           & Arrays                            & Arrays \\ 
			Basic Input and Output               & Casting                              & Collections                       & Assemble \\ 
			Binary Operators                     & Conditional Statements (Branching)   & Conditional Statements (Branching)& Assert \\ 
			Common Library Functions             & Constructors                         & Enumeration (enum)                & Conditional Statements (Branching) \\ 
			Conditional Statements (Branching)   & Dictionary                           & Exception Handling                & Do While Loop \\
			Console Input/Output                 & Dynamic Type                         & Generics                          & Enumeration \\
			Data Types                           & File Handling                        & Inheritance                       & For Loop \\ 
			Enumeration (enum)                   & Formatting                           & Input-output (I/O operations)     & Library \\ 
			File Handling                        & Functions                            & Lambda                            & Mapping \\ 
			Format Specifiers                    & Generators                           & Loops                             & Modifier \\ 
			Functions                            & Exception Handling                   & Maths Class                       & Struct \\ 
			Header Reserved Keywords             & Inheritance                          & Methods                           & While Loop \\
			Heap Space                           & Lambda                               & Operators                         &  \\ 
			Loops                                & Lists                                & Polymorphism                      &  \\ 
			Number Literals                      & Loops                                & Redirection Piping                &  \\ 
			Operators                            & Multi Threading                      & Special Class                     &  \\ 
			Pointers                             & Overriding                           & Strings                           &  \\ 
			Preprocessor Directives              & Polymorphism                         & Typecasting                       &  \\ 
			Strings                              & Range                                &                                   &  \\ 
			Structures                           & repr() method                        &                                   &  \\ 
			Ternary \& Assignment Operators      & Strings                              &                                   &  \\ 
			Unary Operators                      & Super() Function                     &                                   &  \\ 
			Union                                & Tuple                                &                                   &  \\ \hline
	\end{tabular}}
\end{table*}

\subsubsection{Evaluated LLMs}
We evaluate 21 LLMs spanning commercial and open-source offerings. Models are queried in a zero-shot setting with deterministic or low-temperature decoding to reduce randomness, assessing their internal knowledge without external retrieval. The evaluated set comprises GPT-5-nano, GPT-5-mini, GPT-o1-mini, GPT-5-codex, GPT-OSS-20B, GPT-5, GPT-o4-mini, GPT-o3-mini, GPT-o3, GPT-4o-mini, GPT-4o, Claude-sonnet-4, Gemini-2.5-Flash-09-2025, Gemini-2.5-Pro, Deepseekchat-v3, Qwen3-next-80b-a3b-instruct, Qwen3-coder-plus, Kimi-K2, StarChat, CodeLlama-70b, and CodeLlama-13b. We select (1) GPT series to provide strong baselines across families (GPT-4o, GPT-o1/o3/o4 lines); (2) strong recent proprietary models (Claude-sonnet-4, Gemini-2.5-Pro/Flash, Deepseekchat-v3, Kimi-K2, and Qwen family) on programming and reasoning; and (3) open-source baselines (CodeLlama-13b/70b, StarChat, GPT-OSS-20B).

\subsection{Prompt Design and Optimization}
\label{sec:prompt}
LLMs utilize the prompt-based learning paradigm~\cite{liu2023pre}. The design of prompts significantly impacts model performance. We follow an iterative optimization process. Starting from initial prompts drafted based on task requirements, we generate multiple variations using GPT-4 as a meta-optimizer with the instruction ``Act as a prompt optimizer and refine the following prompt for [task], [draft]''. Each candidate is evaluated on a pilot set per task, measuring task-specific metrics such as reconstructability for AST and F1 for data dependence. We select the best-performing template and validate it on a validation set. If validation performance drops compared to pilot results, we iterate with refined prompts informed by failure analysis. Otherwise, we finalize the template. This iterative process yielded two prompt categories: role-based for single-function tasks and instruction-based for multi-faceted tasks. 

We operationalize prompt design through three concrete principles. \textit{Role assignment} specifies one of six roles (AST parser, CFG analyzer, call graph analyzer, code static analyzer, pointer analyzer, expression matcher) for focused task framing. \textit{Task decomposition} breaks complex tasks into explicit subtasks, such as identifying entry points, tracing execution paths, and constructing graphs. \textit{Format constraints} mandate structured outputs like JSON with specific keys to enable automated evaluation.

Role-based prompts assign a specific function to guide focused output generation. The generic template is:
\begin{tcolorbox}[colback=gray!5,colframe=gray!40]
\small\texttt{You are [ROLE] for [LANG]. [TASK DESCRIPTION]. [OUTPUT FORMAT]. The input is [INPUT].}
\end{tcolorbox}

Placeholder definitions:
\begin{itemize}[leftmargin=*,itemsep=2pt]
\item \text{[ROLE]}: One of six specialized roles—AST parser, expression tree matcher, CFG analyzer, call graph analyzer, code static analyzer, or pointer analyzer.
\item \text{[LANG]}: Target programming language—Python, Java, C, or Solidity.
\item \text{[TASK DESCRIPTION]}: Task-specific instruction describing the expected analysis output.
\item \text{[OUTPUT FORMAT]}: Structured specification for automated evaluation, such as JSON schema or indented text format.
\item \text{[INPUT]}: Source code under analysis.
\end{itemize}

\noindent\textit{Concrete instantiation for AST generation (Python):} ``You are a Python Abstract Syntax Tree (AST) parser. I will give you a Python code file. You give me its AST in Json format. Each AST node only has three attributes, children, type and value. The input file is: [code].'' (See Appendix~\ref{appendix:prompt_ast} for complete template.)

Instruction-based prompts provide explicit commands without role assignment, suitable for tasks requiring multiple analytical perspectives. The instruction template is:
\begin{tcolorbox}[colback=gray!5,colframe=gray!40]
\small\texttt{Please analyze [LANG]. [DOMAIN KNOWLEDGE]. Please identify if [TASK DESCRIPTION]. [OUTPUT FORMAT]. The input is [INPUT].}
\end{tcolorbox}

Placeholder definitions:
\begin{itemize}[leftmargin=*,itemsep=2pt]
\item \text{[DOMAIN KNOWLEDGE]}: Task-relevant background knowledge providing context for the analysis.
\item \text{[LANG]}, \text{[TASK DESCRIPTION]}, \text{[OUTPUT FORMAT]}, \text{[INPUT]}: Same as role-based prompts.
\end{itemize}

\noindent\textit{Concrete instantiation for equivalent mutant detection:} ``Please analyze the two following provided code files in C or Java. Identify if they are semantically equal. `Semantically equal' means two codes have the same meaning, that they have the same output given the same input. [Example pair provided...] Please identify if the two following codes are semantically equal. Please only answer `yes' or `no'. Input: [code].'' (See Appendix~\ref{appendix:prompt_mutant} for complete template with examples.)

We employ role-based prompts for RQ1 and RQ2. For tasks in RQ3, we use instruction prompts as they work better during our trials, since these tasks require multiple analytical perspectives rather than a single role. All prompts and template variants are available in our online repository~\cite{mywebsite}. Complete prompt templates for all nine tasks are provided in Appendix~\ref{appendix:prompts}.

\paragraph{\textbf{Prompt Strategy Ablation.}} To assess prompt robustness and understand how reasoning scaffolds affect performance across different model architectures, we conduct within-model ablation studies on dynamic tasks (RQ3: Equivalent Mutant Detection and Flaky Test Reasoning). We compare four prompt strategies: \textit{normal} (baseline instruction), \textit{CoT} (chain-of-thought prompting with explicit reasoning steps), \textit{role} (role-based prompting assigning specific personas), and \textit{role+CoT} (combining both approaches). Each strategy is evaluated on the same model using identical test sets, with statistical significance assessed via McNemar's test for binary classification tasks and paired bootstrap confidence intervals for multi-class tasks. Detailed results and analysis are presented in Section~\ref{sec:prompt_strategy}.

\subsection{Three-Layer Evaluation Protocol}
\label{sec:eval_protocol}

This section details the implementation of our three-layer evaluation framework introduced in Section~\ref{sec:eval_challenges}. Layer~1 handles format diversity through normalization pipelines that assess semantic reconstructability. Layer~2 employs expert judgment with formal assessment criteria to evaluate semantic correctness. Layer~3 validates both human consistency through inter-rater reliability analysis and automated metric reliability against expert consensus. The protocol establishes formal assessment criteria, quantifies human consistency through multiple agreement coefficients (Cohen's $\kappa$, Gwet's AC1, PABAK), and provides statistical significance testing for all key comparisons.

\paragraph{\textbf{Layer 1: Automated Metrics}} 
We compare LLM outputs with references produced by program analysis tools when available (e.g., Tree-sitter, Slither, Frama-C) and compute task-specific reconstructability scores. Layer~1 automated comparison addresses representational heterogeneity between LLM outputs and static analysis tool references through task-specific normalization pipelines designed to assess \textit{semantic reconstructability}, defined as the degree to which outputs preserve program structure sufficient for developer reasoning, rather than syntactic fidelity to tool-specific representations.

Large variations in output formats make direct alignment challenging. Different LLMs may generate structures that differ in node types, naming conventions, and granularity. Even with automated scripts, extensive post-processing is required, and complete elimination of ambiguities remains challenging. To address these challenges, we apply three categories of transformations to enable fair comparison while preserving semantic equivalence:
\begin{itemize}
\item[(1)]  AST Alignment addresses node proliferation, nesting depth inconsistency, and semantic granularity mismatch through node filtering (removing compiler artifacts), type coarsening (mapping to unified semantic categories), and backbone extraction (retaining key control-flow construct types), thereby reducing tree depth while preserving structure; 

\item[(2)]  CFG Alignment (using Joern for C, tree-sitter-graph for Python, Soot for Java) handles basic-block granularity mismatch, node type naming variation, and abstraction level differences via node-type normalization (mapping to coarse categories including backbone types), graph canonicalization (collapsing auxiliary node chains when graphs are large), and backbone-focused metrics (computing coverage over backbone types with binary existence flags); 

\item[(3)]  CG Alignment (using Joern for C, tree-sitter for Python, Soot for Java, Slither for Solidity) manages qualifier inconsistency, synonym variation (e.g., \texttt{print}/\texttt{println}/\texttt{printf}), optional library granularity, and anonymous variants (e.g., \texttt{Lambda\$1.apply}) through identifier tokenization, synonym expansion (IO/initialization/collection query families), fuzzy matching with tuned similarity thresholds achieving high agreement, wrapper collapse (yielding size reduction), and variant-key normalization (deduplicating lambda/inner-class nodes). 
\end{itemize}

 Complete strategies are detailed in Appendices~\ref{appendix:ast_norm}, \ref{appendix:cfg_norm}, and \ref{appendix:cg_norm}.

Quantitative thresholds across all pipelines were calibrated through iterative refinement on a held-out validation subset (15 cases per task) until automated decisions met a pre-set agreement threshold with expert judgments, balancing precision (avoiding false alignments) with recall (accepting valid representational variations). Calibration methodology is described in Appendix~\ref{appendix:threshold_calibration}. However, semantic equivalence complexity necessitates expert validation (Layer~2) for cases where structural metrics provide incomplete assessment.

\paragraph{\textbf{Layer 2: Expert Adjudication}}
We visualize ASTs, CFGs, and CGs produced by LLMs and conduct structured expert evaluation through formal assessment protocols. Two domain experts (with backgrounds in programming languages, compilers, and program analysis) independently assess each output following task-specific reconstructability criteria. We employ a three-tier classification that distinguishes structural incompleteness from semantic incorrectness, ensuring that missing information and false information receive appropriately different treatment:

\begin{itemize}
    \item \textit{Pass (reasonable).} Output is structurally complete and semantically correct. All critical constructs are present, relationships conform to program semantics, and no fabrications exist. Minor simplifications (e.g., merging sequential blocks, omitting trivial leaf nodes) are permitted if they preserve reconstructability.
    \item \textit{Minor (reasonable with incompleteness).} Output is semantically correct in what it represents but structurally incomplete. Core execution paths and main constructs are preserved, but secondary details (e.g., utility function calls, empty branches, peripheral nodes) are missing. Crucially, \textit{no incorrect information is present}; the output is a valid subset of the complete structure.
    \item \textit{Fail (unreasonable due to incorrectness or severe incompleteness).} Output contains semantic incorrectness (fabricated nodes/edges, wrong control flow, incorrect caller-callee relationships) or severe structural incompleteness (missing critical constructs like loop back-edges, main execution paths, entry points) that prevents meaningful program understanding.
\end{itemize}

This classification distinguishes false information (hallucinations, contradictions) from missing information (omissions), recognizing that incorrect outputs mislead downstream analysis while incomplete outputs merely limit scope. Evaluators assess whether the visualized structure preserves sufficient semantic information to understand the original program's control flow, data dependencies, and call relationships without consulting the source code. If disagreement occurs, a third expert adjudicates and the reviewers discuss to reach consensus. We compute inter-rater reliability (Cohen's $\kappa$, Gwet's AC1) on pre-adjudication labels.

Complexity of semantic consistency evaluation requires expert judgment. Our main focus is whether the structures generated by the LLM maintain semantic consistency with the original program, rather than simply matching nodes and edges. For instance, a slightly simplified CFG may still correctly represent the control flow, whereas incorrectly decomposing complex nested structures can undermine the overall semantics. Expert judgment from domain specialists helps identify these subtle yet critical semantic deviations, enhancing both credibility and interpretability. To ensure consistent and reproducible assessment, we establish formal evaluation criteria for each structural task that implement the three-tier classification:

\begin{itemize}
    \item \textit{AST.} An AST is labeled \texttt{pass} if it satisfies completeness (all critical constructs present), structural correctness (parent-child relationships conform to grammar), syntactic validity (required delimiters present), and reconstructability (tree is parsable and preserves sufficient program structure for understanding control flow and data relationships). It is labeled \texttt{minor} if core structure is preserved but secondary details are missing (e.g., access modifiers, type annotations, trivial leaf nodes) with no incorrect information. It is labeled \texttt{fail} if it contains incorrectness (fabricated nodes, grammar-violating attachments), severe incompleteness (critical statements missing), or non-parsability.
    
    \item \textit{CFG.} A CFG is labeled \texttt{pass} if it satisfies structural completeness (entry/exit nodes exist), edge correctness (branch/loop/exception edges correct), absence of hallucination (no fabricated edges), loop modeling (back-edges present), and semantic fidelity (correctly modeling language-specific constructs such as short-circuit operators, ternary expressions, and exception handling with \texttt{try}/\texttt{catch}/\texttt{finally}). It is labeled \texttt{minor} if main execution paths are preserved but peripheral elements are missing (empty branches, utility calls) with no incorrect edges. It is labeled \texttt{fail} if it contains incorrectness (fabricated edges, wrong control structure), severe incompleteness (critical control edges missing), or structural violations (unreachable regions, missing entry/exit).
    
    \item \textit{CG.} A CG is labeled \texttt{pass} if it satisfies completeness (core application calls complete), precision (no significant fabricated cross-module calls), and coverage (entry and core functions covered). It is labeled \texttt{minor} if main execution paths from entry points are preserved but peripheral calls are missing (utility functions, library wrappers) with no incorrect edges. It is labeled \texttt{fail} if it contains incorrectness (fabricated calls, wrong receiver resolution), severe incompleteness (core application calls missing), or structural violations (missing entry points, fundamentally wrong caller-callee semantics).
\end{itemize}

\paragraph{\textbf{Layer 3: Consistency Validation}}
We conduct rigorous inter-rater reliability analysis and human-machine agreement assessment to validate evaluation consistency. This layer addresses a fundamental challenge: the absence of ground truth for semantic correctness requires establishing both the reliability of human expert judgment and the validity of automated approximations.

\textit{Inter-Rater Reliability.} We quantify agreement between independent reviewers (r1, r2) using multiple coefficients because single metrics may be sensitive to class imbalance or marginal distribution artifacts. We compute: Cohen's $\kappa$ for chance-corrected agreement (standard baseline); Scott's $\pi$ for pooled marginal distributions; Gwet's AC1 (robust to prevalence effects, critical when pass rates are highly skewed); and PABAK (Prevalence-Adjusted Bias-Adjusted Kappa, providing prevalence-independent interpretation). We interpret coefficients using Landis \& Koch thresholds: $\kappa < 0.20$ (slight), $0.20 \leq \kappa < 0.41$ (fair), $0.41 \leq \kappa < 0.61$ (moderate), $0.61 \leq \kappa < 0.81$ (substantial), $\kappa \geq 0.81$ (almost perfect). For disagreements, we report pre-adjudication metrics (quantifying initial annotator variance) and post-consensus outcomes (establishing final labels) separately.

\textit{Human–Machine Agreement.} We compare automated Layer~1 judgments against human consensus (formed via majority vote from r1, r2 with ties excluded) using the same coefficient suite. We report confusion matrices (where human consensus serves as gold standard) and derive precision, recall, F1, and accuracy. This dual-layer validation quantifies both the consistency of human judgment (establishing the upper bound on achievable agreement) and the alignment between automated metrics and expert assessment (validating whether Layer~1 normalization successfully operationalizes semantic reconstructability). Detailed error taxonomies (miss, fabrication, structural violation) are maintained per task to enable qualitative error analysis.

To validate the evaluation protocol, we conduct consistency analysis on a representative subset of models across structural tasks (AST, CFG, CG). Two independent reviewers assess outputs using the formal rubrics, with disagreements adjudicated by a third expert through consensus discussion. We compute multiple agreement coefficients (Cohen's $\kappa$, Gwet's AC1, PABAK) for inter-rater reliability and human-machine agreement to establish evaluation reliability bounds. Detailed validation results and analysis are reported in Section~\ref{sec:agreement}.

The three-layer protocol is designed to be model-agnostic: new LLMs require only API configuration without modifying evaluation logic, prompts, or assessment criteria. The rubrics, once validated through inter-rater reliability analysis, can assess future models without re-validation as they operationalize fundamental semantic properties rather than model-specific characteristics. We release evaluation scripts, reference implementations, standardized prompts, and datasets (3,124 samples with ground truth annotations) to support community assessment of emerging LLMs.

\subsection{Evaluation Metrics for Syntax Tasks (RQ1)}
\label{sec:syntax_metrics}

\subsubsection{\textbf{AST Generation Metrics}}
\label{sec:ast_metrics}
\paragraph{Matching Challenge.} LLM outputs exhibit systematic format drift from parser outputs, including naming variance (e.g., \texttt{if\_statement} vs. \texttt{IfStatement}), depth differences (parsers produce 7-8 layers; LLMs generate 3-4 flattened layers), and noise nodes (compiler internals like \texttt{TypedefDecl}). Exact structural matching would penalize semantically correct but representationally different outputs.

\paragraph{Metric Design.} 
The LLM-generated AST often encounters three dominant error types: (1) missing statement tokens, where important statement kinds (assignments, calls, branches, loops, exceptions) are absent or severely under-represented; (2) missing syntax keywords, where language keywords (e.g., visibility modifiers and control keywords) do not appear in the tree even though they exist in the code; and (3) wrong structure, where control blocks (e.g., \texttt{if} / loops) are missing, split, or attached in the wrong place. We therefore define three normalized error indicators, which we refer to as Statement Gap, Keyword Gap, and Structure Gap. Intuitively, Statement Gap measures ``how many of the core statement types and their rough counts are missing'' by comparing which of a small set of normalized statement categories (ASSIGN/CALL/RETURN/BRANCH/LOOP/TRY/EXCEPT) appear in the model tree versus the reference, with extra tolerance for non-critical categories but no forgiveness for missing branches/loops/returns. Keyword Gap measures ``how many keyword families are missing'': we group language keywords into a few coarse families (branch, loop, exception, jump) based on what appears in the gold AST, and compute the fraction of these families that do not appear in the model AST. Structure Gap measures ``how different the control-flow skeleton is'' by checking whether the main control nodes still connect in the same way and whether typical shapes (condition plus body, with roughly the same order) are preserved. All three indicators are scaled to $[0,1]$ (higher is worse) and directly mirror the observed failure modes; we use them both as diagnostics and as the basis for the pass/fail classification below.

\paragraph{Pass/Fail Classification.} We classify each output using three error thresholds: (1) Statement Gap $\leq$ 0.30 (missing core statements or strong semantic tokens), (2) Keyword Gap $\leq$ 0.35 (missing language keywords), and (3) Structure Gap $\leq$ 0.45 (wrong parent-child attachments or grammar violations). An output passes if all three error rates are below their respective thresholds. 
For C programs, thresholds are relaxed to 0.35, 0.40, and 0.50 respectively to accommodate greater tool-induced representational variance across parsers.

\textit{Threshold Calibration.} These thresholds were calibrated on a held-out validation set (15 cases per language) through iterative refinement with human expert validation. The calibration process balances two objectives: (1) avoiding false passes for outputs with obvious structural errors, and (2) avoiding false failures for outputs with minor but consumable deviations. For example, the 0.30 threshold for statement incompleteness allows for common identifier placeholders and minor truncations while rejecting outputs missing critical control structures. The 0.45 threshold for structural incorrectness tolerates slight attachment or ordering deviations while catching severe grammar violations. The C-specific relaxation reflects that C parsers exhibit greater naming and nesting variance (e.g., Tree-sitter vs Clang AST representations) while preserving semantic equivalence. Thresholds were validated by computing human-machine agreement (Cohen's $\kappa$, AC1) between automated classifications and expert consensus labels on the validation set, achieving inter-rater reliability $\kappa = 0.936$ and AC1 $= 0.975$ (Table~\ref{tab:consistency_struct}), confirming near-perfect alignment with expert judgment.

\subsubsection{\textbf{Expression Matching Metrics}}
\label{sec:expression_metrics}
	\paragraph{Matching Challenge.} Direct string matching fails to capture semantic equivalence, as expressions like \colorbox{gray!15}{\texttt{rate + ur * s1}} and \colorbox{gray!15}{\texttt{base\_rate + utilization * slope1}} are semantically similar despite lexical differences. Traditional edit distance conflates structural similarity (operator order) with superficial similarity (variable naming).
	\paragraph{Metric Design.} We employ CodeBERT~\cite{feng2020codebert} embeddings to compute cosine similarity between expressions, measuring Top-$k$ hit rate as 
	\begin{equation}
		\text{Hit@k} = \frac{1}{|E|} \sum_{e \in E} \mathbb{1}\left[\exists i \leq k: \mathrm{sim}(c_i, e) \geq \theta \land \mathrm{occurs}(c_i)\right]
	\end{equation}
	where $E$ is the set of target expressions, $c_i$ is the $i$-th candidate in the LLM-generated ranked list, $\mathrm{sim}(\cdot, \cdot)$ denotes cosine similarity between CodeBERT embeddings, $\theta$ is the similarity threshold calibrated through pilot validation, and $\mathrm{occurs}(c_i)$ verifies that candidate $c_i$ appears in the source code or has sufficient token overlap. We report $k \in \{5, 10, 20\}$ to assess precision-oriented (top-5) and recall-oriented (top-20) retrieval.
	\paragraph{Motivation.} The metric reflects practical code search scenarios where developers need relevant expressions ranked highly. CodeBERT embeddings capture both syntactic structure (operator precedence, nesting) and semantic roles, generalizing across variable renaming and formatting differences. The occurrence constraint prevents spurious matches from semantically unrelated code regions, ensuring that high-similarity candidates are contextually valid. The similarity threshold balances precision (avoiding false matches) with recall (accepting valid renamings). Multi-$k$ reporting reveals retrieval quality trade-offs, with top-5 for precision-critical tasks and top-20 for recall-oriented exploration.

\subsection{Evaluation Metrics for Static Semantics Tasks (RQ2)}
\label{sec:static_metrics}

\subsubsection{\textbf{CFG Analysis Metrics}}
\label{sec:cfg_metrics}
\paragraph{Matching Challenge.} CFG representations exhibit tool-dependent variance, as Soot generates fine-grained basic blocks while Clang produces coarse-grained control nodes. LLM outputs may use semantically valid alternative factorizations (splitting/merging blocks, short-circuit evaluations). Exact graph isomorphism would penalize valid representations.

\paragraph{Metric Design.} 
The LLM-generated CFG often encounters three dominant error types: (1) redundancy, where extra control-flow jumps are added that do not change which code is reachable; (2) fabrication, where entirely spurious control edges create impossible execution paths; and (3) wrong structure, where the main control-flow skeleton (branches, loops, entry/exit connectivity) is incomplete or miswired. We capture these with three normalized indicators, which we refer to as Jump Redundancy, Jump Fabrication, and CFG Shape Error. Jump Redundancy measures ``how many of the model’s jump edges are unnecessary'' by comparing model with gold jump edges and treating extra jumps as errors. Jump Fabrication measures ``how many kinds of jump behavior only exist in the model'' by checking which jump categories (e.g., branch/loop variants, excluding calls) appear in the gold CFG and counting categories that appear only in the model. CFG Shape Error measures ``how much of the core control-flow shape is lost'' by looking at the coverage of core CFG nodes (entry/exit, branch, loop) and how many gold jump paths remain reachable in the model; if either the backbone or the reachable jump paths are badly degraded, this error becomes large. All three indicators lie in $[0,1]$ (higher is worse) and serve both as diagnostics and as the basis for the pass/fail classification below; we report their distribution and the induced pass rate as our primary CFG metrics. 

\paragraph{Pass/Fail Classification.} We classify each output using three error thresholds: (1) Jump Redundancy $\leq$ 0.40 (excessive nodes/edges that do not alter semantics), (2) Jump Fabrication $\leq$ 0.25 (spurious control edges that misrepresent program flow), and (3) CFG Shape Error $\leq$ 0.45 (wrong control structure, missing critical paths). An output passes only if all three error rates are below their respective thresholds.

\textit{Threshold Calibration.} These thresholds were calibrated on a validation set through human expert validation. The calibration reflects the differential impact of error types: redundancy errors (0.40 threshold) have lower semantic impact as extra nodes/edges do not mislead downstream analysis; fabrication errors (0.25 threshold) receive stricter treatment as spurious control edges actively misrepresent program flow; structural incorrectness (0.45 threshold) ensures that outputs missing critical paths are rejected. The thresholds balance tolerance for minor deviations (e.g., different basic-block granularities) against rejection of semantically incorrect control flow, achieving inter-rater reliability $\kappa = 0.858$ and AC1 $= 0.920$ (Table~\ref{tab:consistency_struct}), demonstrating substantial agreement with expert judgment.

\subsubsection{\textbf{Call Graph Analysis Metrics}}
\label{sec:cg_metrics}
\paragraph{Matching Challenge.} Call graphs exhibit structural hierarchy, where entry-point functions and their immediate callees form the execution backbone, while library leaf nodes provide secondary detail. Treating all edges equally would over-penalize missing utility calls while under-rewarding correct main execution paths. Dynamic dispatch (polymorphism, reflection, lambdas) introduces resolution ambiguity where analyzers may produce valid alternative target sets.

\paragraph{Metric Design.} 
The LLM-generated call graph often encounters three dominant error types: (1) redundancy, where extra call nodes or edges are introduced that do not correspond to essential application-level calls; (2) fabrication, where non-existent interprocedural paths are created; and (3) missing calls, where critical calls are omitted. We therefore design three normalized indicators, which we refer to as Call Redundancy, Call Fabrication, and Missed Calls. Call Redundancy measures ``how many extra functions or call edges the model invents'' by checking, among core application functions and their calls, what fraction of predicted elements cannot be matched to any reference node or edge. Call Fabrication focuses even more narrowly on spurious core call sites and measures ``how many predicted core functions never appear in the reference graph'' after normalizing away implementation variants and library-only nodes. Missed Calls measures ``how many real calls are missing'' by collapsing both graphs to a variant-level call graph, matching gold call edges to possible model edges (with allowances for dynamic dispatch), and computing the fraction of gold call edges that remain uncovered. All three indicators are bounded in $[0,1]$ (higher is worse) and directly mirror the observed failure modes; we report their distribution and the induced pass rate as our primary call-graph metrics.

\paragraph{Pass/Fail Classification.} We employ a strict pass criterion requiring three error rates below their thresholds: Call Redundancy $\leq$ 0.22, Call Fabrication $\leq$ 0.12, and Missed Calls $\leq$ 0.25. An output passes only if all three indicators fall below these limits.

\textit{Threshold Calibration.} Call graph thresholds are stricter than AST/CFG thresholds, calibrated on a validation set through expert review with slight relaxation (0.22/0.12/0.25 vs initial 0.20/0.10/0.20) to accommodate dynamic dispatch resolution ambiguity where multiple valid call targets exist. The rationale: interprocedural call errors propagate more widely in downstream analyses (vulnerability propagation, impact analysis). Redundancy errors (0.22 threshold, stricter than CFG's 0.40) reflect that extra call edges more easily pollute downstream analysis. Fabrication errors (0.12 threshold, strictest across all tasks) reflect the highest risk of spurious calls misleading security analysis. Missing call errors (0.25 threshold) balance tolerance for minor omissions with rejection of outputs that lose essential call paths. The final thresholds achieve inter-rater reliability $\kappa = 0.844$ and AC1 $= 0.914$ (Table~\ref{tab:consistency_struct}), demonstrating substantial agreement despite the inherent complexity of interprocedural analysis.

\subsubsection{\textbf{Data Dependency and Taint Analysis Metrics}}
\label{sec:dd_taint_metrics}
	\paragraph{Matching Challenge.} Data dependency and taint analysis datasets exhibit class imbalance, as in real-world programs, most variable pairs are \textit{not} data-dependent, and most variables are \textit{not} tainted. Accuracy would be misleadingly high by predicting all negatives. Precision and recall alone provide incomplete views, where high precision with low recall suggests overly conservative predictions, while high recall with low precision suggests over-tagging.
	
	\paragraph{Metric Design.} We employ F1 score, the harmonic mean of precision and recall ($F1 = 2 \cdot \frac{P \cdot R}{P + R}$), which balances both concerns and is robust to class imbalance. We report \textit{per-project average F1} to account for project-level variance, as different codebases exhibit different dependency/taint patterns (e.g., security-critical smart contracts have dense taint propagation; utility libraries have sparse dependencies). Aggregating across projects prevents large projects from dominating the metric. We parse LLM free-text responses into ternary labels (\texttt{yes}/\texttt{no}/\texttt{unknown}), treating \texttt{unknown} as negative to penalize uncertainty in safety-critical contexts.
	
	\paragraph{Motivation.} F1 aligns with practical requirements, as in vulnerability detection, false negatives (missed tainted variables) create security risks, while false positives (flagging clean variables) increase manual review burden. Balanced F1 reflects this dual cost. Per-project averaging prevents dataset composition bias (e.g., one large project skewing results). The ternary-to-binary mapping (\texttt{unknown} $\rightarrow$ negative) enforces a ``fail-safe'' stance, where uncertain predictions are treated as analysis failures, appropriate for security-sensitive tasks. F1 is standard in information security and software testing literature, enabling direct comparison with prior work.

\subsubsection{\textbf{Pointer Analysis Metrics}}
\label{sec:pointer_metrics}
	\paragraph{Matching Challenge.} Pointer analysis outputs are \textit{sets} of variables (not ordered sequences or binary labels). Traditional precision/recall metrics are sensitive to set size and require threshold tuning. Exact set matching is too strict, as missing one alias out of five yields zero credit despite 80\% correctness.
	
	\paragraph{Metric Design.} We employ the Jaccard index $J(A, B) = \frac{|A \cap B|}{|A \cup B|}$, which naturally handles variable-sized sets, ranges from 0 (disjoint) to 1 (identical), and balances over-prediction (spurious aliases) with under-prediction (missed aliases) without threshold parameters. We compute \textit{per-pointer Jaccard}, then aggregate via \textit{per-program average} as $\text{Jaccard}_{\text{program}} = \frac{1}{|P_{\text{program}}|} \sum_{p \in P_{\text{program}}} J(\widehat{\text{PT}}(p), \text{PT}(p))$, where $P_{\text{program}}$ is the set of pointers in a program, $\text{PT}(p)$ is the ground truth pointer set, $\widehat{\text{PT}}(p)$ is the predicted pointer set. We report the mean across programs to account for program-level complexity variance (small programs with 2-3 pointers vs. larger programs with 10+ pointers).
	
	\paragraph{Motivation.} Jaccard is the standard set-similarity metric in information retrieval and data mining, widely used for comparing entity resolutions and clustering results. In pointer analysis, partial correctness is meaningful, as identifying 4 out of 5 actual pointees provides useful information for downstream analyses (e.g., aliasing-aware optimization, buffer overflow detection). Per-program averaging prevents large programs from dominating the metric, reflecting that analysis difficulty varies with program structure rather than just pointer count. Jaccard's threshold-free nature avoids arbitrary cutoffs, making results reproducible and comparable across studies.

\subsection{Evaluation Metrics for Dynamic Reasoning Tasks (RQ3)}
\label{sec:dynamic_metrics}

\subsubsection{\textbf{Equivalent Mutant Detection Metrics}}
\label{sec:mutant_metrics}
	\paragraph{Matching Challenge.} Unlike structural tasks (AST/CFG/CG) with partial correctness gradations, equivalence is a \textit{binary semantic property}, where mutants are either behaviorally equivalent or not. There is no ``partial equivalence''; a mutant that differs on even one input is non-equivalent.
	
	\paragraph{Metric Design.} We employ accuracy, which directly measures classification correctness as $\text{Accuracy} = \frac{\text{correct predictions}}{\text{total mutants}}$. For this balanced dataset (100 equivalent, 100 non-equivalent), accuracy provides an unbiased assessment without class-imbalance artifacts.
	
	\paragraph{Motivation.} Accuracy is standard for balanced binary classification tasks in software testing and mutation analysis literature. Equivalence detection is a diagnostic task, where practitioners need to know ``is this mutant equivalent?'' with boolean certainty, not probabilistic scores. The 50/50 class balance prevents accuracy inflation from majority-class bias. Alternative metrics (F1, MCC) would provide similar insights but add complexity without changing interpretation for balanced data. Accuracy's simplicity enables direct comparison with prior mutation testing studies and human expert performance baselines.

\subsubsection{\textbf{Flaky Test Reasoning Metrics}}
\label{sec:flaky_metrics}
	\paragraph{Matching Challenge.} Flaky test categories are \textit{mutually exclusive}, where a test fails non-deterministically for one primary reason (though multiple factors may contribute, the dominant cause defines the class). Unlike structural tasks, there is no notion of ``partial category correctness''; predicting \texttt{async\_wait} when the true cause is \texttt{randomness} provides no diagnostic value.
	
	\paragraph{Metric Design.} We employ accuracy, which measures $\text{Accuracy} = \frac{\text{correct category predictions}}{\text{total flaky tests}}$. To account for potential class imbalance in flaky test datasets (some categories like \texttt{async\_wait} are more common than \texttt{floating\_point}), we also report macro-averaged F1 as $\text{Macro-}F1 = \frac{1}{K} \sum_{k=1}^{K} F_{1,k}$, where $K$ is the number of categories and $F_{1,k}$ is the F1 score for category $k$. Macro-F1 treats all categories equally, preventing common categories from dominating the metric.
	
	\paragraph{Motivation.} Accuracy reflects practical debugging scenarios, where developers need the correct root cause diagnosis to fix flaky tests efficiently; wrong diagnoses lead to wasted effort. Macro-F1 gives rare but important categories (e.g., \texttt{floating\_point} precision issues in scientific computing) equal weight in evaluation, preventing the metric from being skewed by frequent categories. The combination of accuracy (overall correctness) and macro-F1 (category-balanced correctness) provides complementary views, where high accuracy with low macro-F1 indicates bias toward common categories, while high macro-F1 with moderate accuracy indicates balanced but imperfect performance. Both metrics are standard in multi-class classification literature and enable comparison with automated flaky test detection tools.

\section{Experimental Results}
\label{sec:result}
\subsection{Code Syntax Analysis~(RQ1)}
\subsubsection{AST Generation}
\label{sec:results_ast}

\begin{figure}[h]
     \centering
     \includegraphics[width=\textwidth]{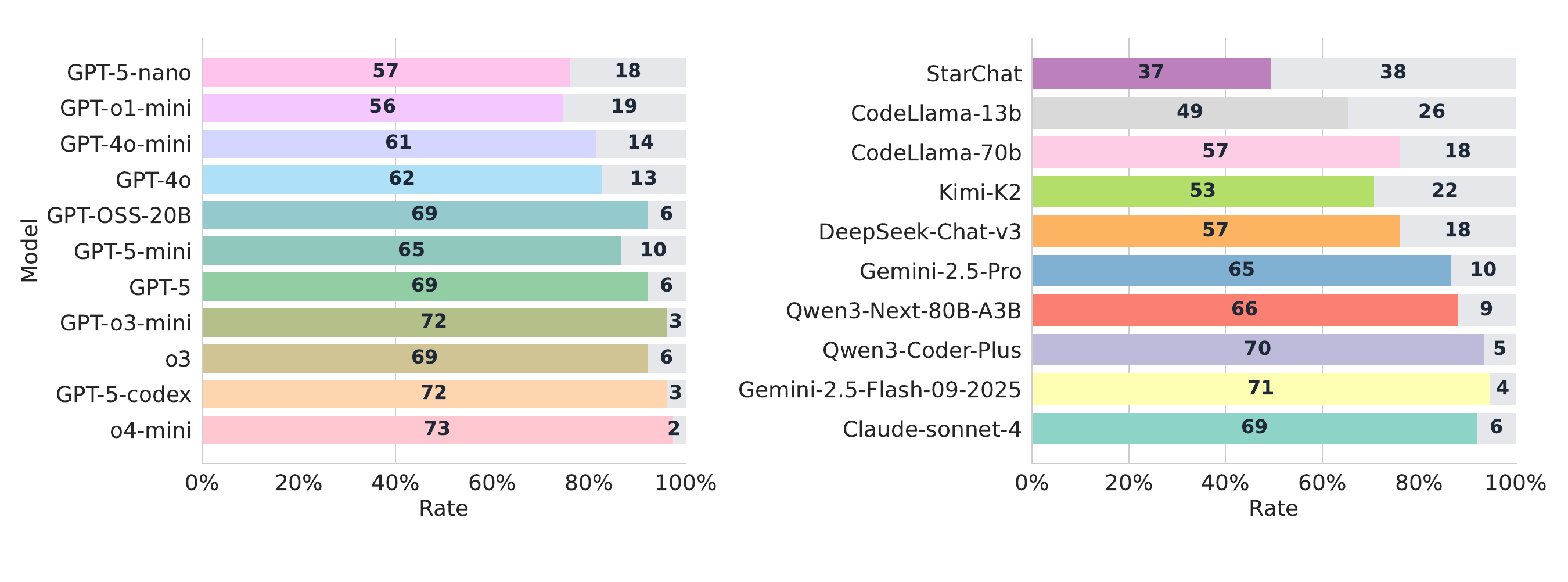}
     \caption{Automated pass/fail rates for AST across models (Layer 1).}
     \label{fig:ast_auto}
\end{figure}

\paragraph{Layer 1 (Automated Metrics)} Following the criteria defined in Section~\ref{sec:ast_metrics}, we compare LLM-generated ASTs against tool baselines and use structural indicators to decide pass/fail shown in \figref{fig:ast_auto}. Under this setup, most recent models achieve similarly high pass rates, with only small differences among the top group, whereas earlier open-source baselines such as CodeLlama-13B and StarChat are noticeably weaker (AST pass rates around 50–65\%). High-performing models rarely fail, and when they do it is mostly due to mild token or keyword omissions; mid-tier models show a larger tail of cases mixing such omissions with structural mistakes; and the weak models accumulate many failures dominated by dropped statements or entire blocks and more frequent structural errors that break the code skeleton.
Across model families, AST accuracy also tracks model scale. Within the GPT-5 family, pass rates increase from 76\% (GPT-5-nano) to 86.7\% (GPT-5-mini) and 92\% (GPT-5), and within the CodeLlama family from 65.3\%
(CodeLlama-13B) to 76\% (CodeLlama-70B). At the same time, model type matters: code-specialized or reasoning-oriented models such as GPT-5-codex, Qwen3-coder-plus, Gemini-2.5-Pro, and Claude-sonnet-4 consistently occupy the top, while
earlier dense open-source chat models (e.g., CodeLlama-13B, StarChat) lag behind both in overall pass rate and in robustness, frequently omitting statements or distorting the global AST shape.

\begin{figure}[]
     \centering
     \includegraphics[width=0.7\textwidth]{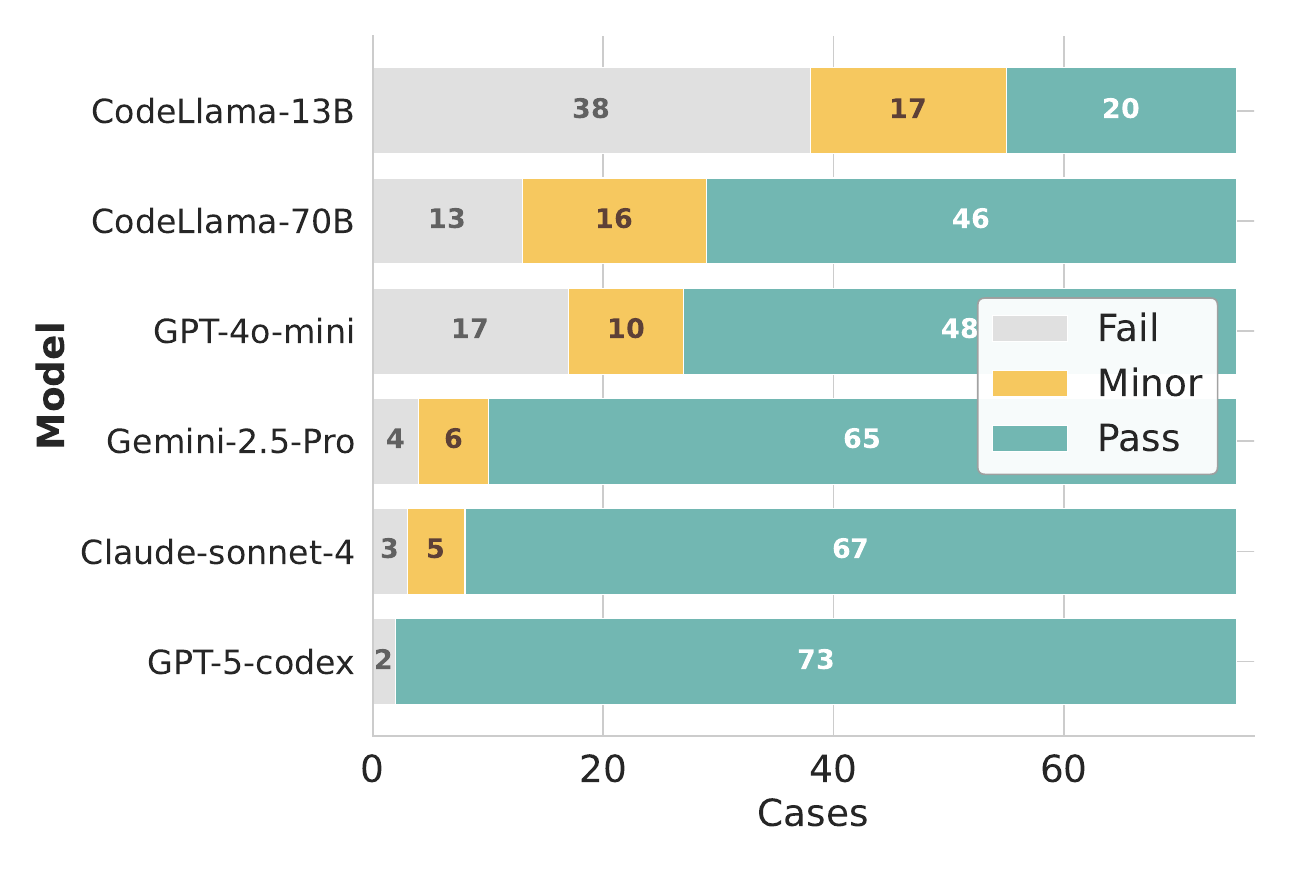}
     \caption{AST structural evaluation: human pass/minor/fail.}
     \label{fig:ast_count}
\end{figure}

\paragraph{Layer 2 (Expert Adjudication)} We present human-labeled structural quality (pass/minor/fail) and issue-category breakdowns under the same case set. \figref{fig:ast_count} summarizes pass/minor/fail from human evaluation. Consistent with automated trends, top commercial models (\textit{GPT-5-codex}, \textit{Claude-sonnet-4}, and \textit{Gemini-2.5} variants) achieve very strong structural quality with few severe errors. \textit{GPT-4o-mini}, although earlier and smaller, performs comparably to the open-source \textit{CodeLlama-70b}. Besides, \textit{CodeLlama-70b} clearly outperforms \textit{CodeLlama-13b}.

\begin{figure}[]
     \centering
     \includegraphics[width=0.7\textwidth]{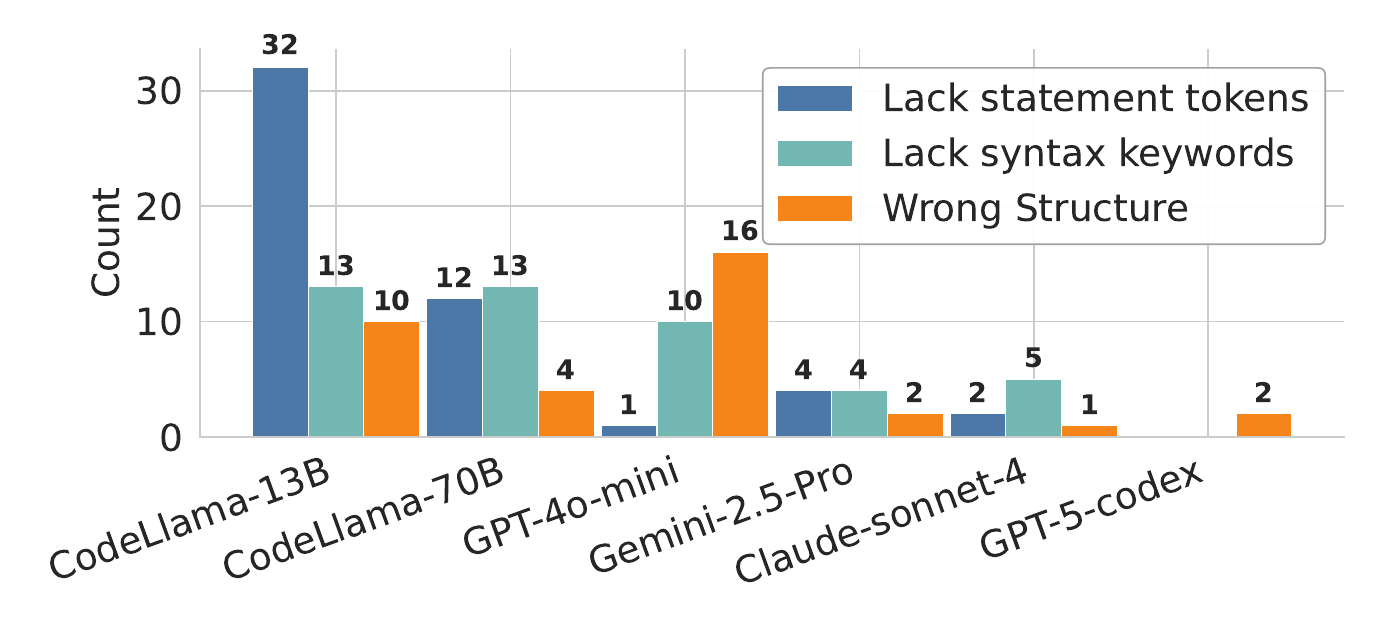}
     \caption{AST issue categories by model.}
     \label{fig:ast_issues}
\end{figure}

\begin{figure}
   \centering
   \scalebox{0.9}{
   \includegraphics[width=0.9\textwidth]{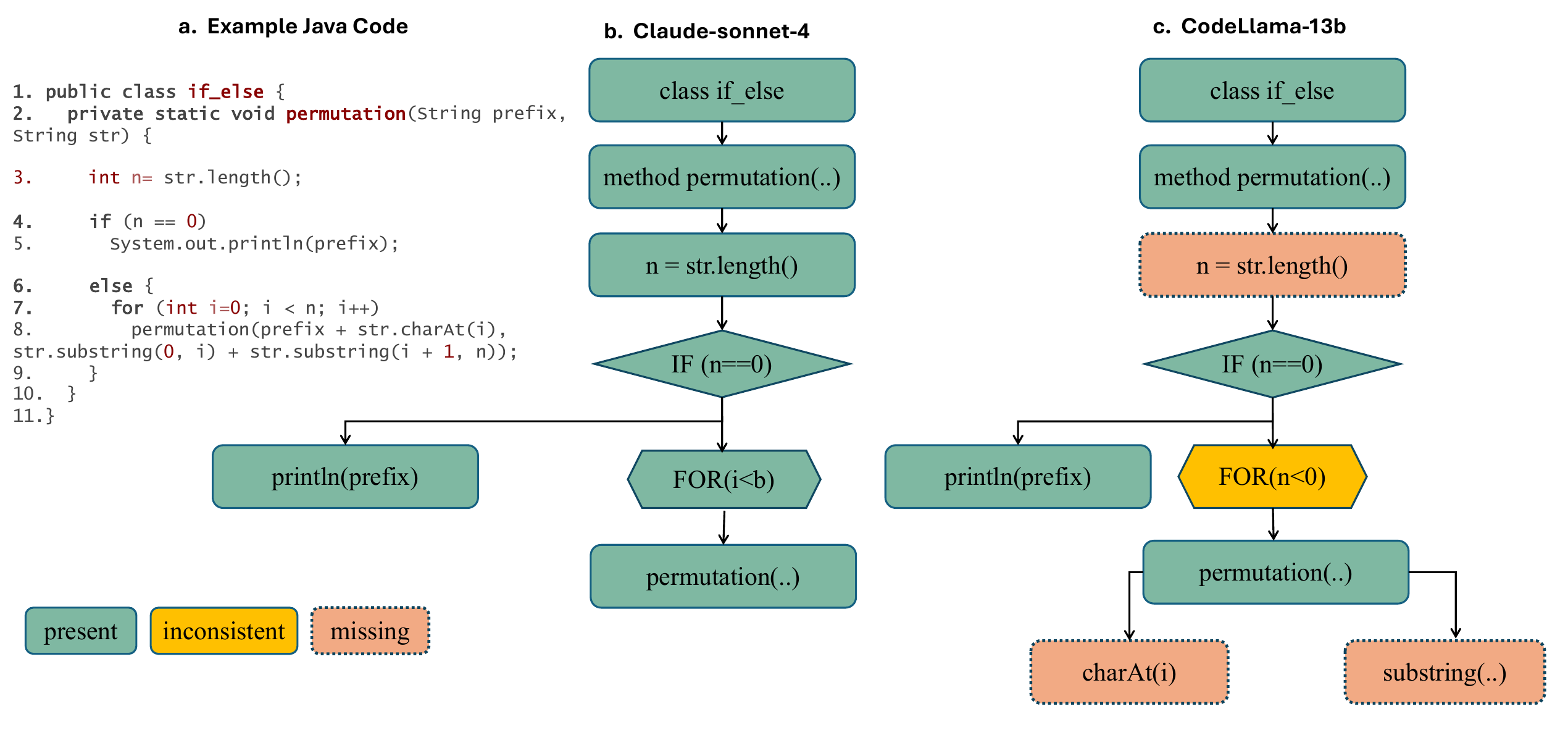}}
    \caption{AST examples. Left: reasonable; Right: unreasonable.}
   \label{fig:ast_example}
\end{figure}

We further analyze issue categories in \figref{fig:ast_issues}. One AST can exhibit multiple issues, and even reasonable ASTs may contain minor defects. We group issues into: missing statement tokens, missing syntax tokens, and wrong structure. Missing-statement-tokens reflect omissions within statements (e.g., missing ``System'' in ``System.out.print(a)''); missing-syntax-tokens reflect omissions of syntax markers (e.g., access modifiers); wrong-structure indicates faulty syntax structure (e.g., malformed if-else). In our evaluation, wrong-structure is the most severe category. Reasonable ASTs with minor issues typically involve missing statement or trivial syntax tokens (e.g., return type, access modifier).
Taken together, \figref{fig:ast_auto}, \figref{fig:ast_count}, and \figref{fig:ast_issues} suggest a capacity effect: larger or better-instructed models tend to parse AST syntax more reliably. %
\figref{fig:ast_example} shows an example: the reasonable AST preserves the class/method skeleton and matches the if/for structures. The unreasonable AST misses a key declaration/assignment, mis-specifies the for-loop initializer and condition, and omits the update clause, contradicting the source skeleton.

\paragraph{Layer 3 (Consistency Validation)} Inter-rater reliability validates evaluation consistency, with Cohen's $\kappa = 0.936$ and Gwet's AC1 = 0.975. Human-machine agreement shows substantial concordance (AC1 = 0.746) despite Cohen's $\kappa = 0.339$ being deflated by prevalence effects (83\% pass rate). Detailed multi-metric analysis and model pass rates under human consensus are provided in Section~\ref{sec:agreement}.

\subsubsection{Expression Matching} 
\tabref{tab:expression_tab} summarizes the latest results. With continued model iterations, most current commercial models achieve strong expression matching performance, often near-saturated hit rates (31-32/32). Many other recent models fall within a narrow range (27-30/32) with only small differences. Performance drops more clearly beyond this range: older open-source baselines are weakest (CodeLlama-70b at 11/32; StarChat and CodeLlama-13b at 0/32).

\begin{table*}[t]
\centering
\caption{Expression Matching across 21 models.}
\label{tab:expression_tab}
\scalebox{0.9}{
\begin{tabular}{l|c|c|c|c}
\toprule
Model & Expr cases & Expr@5 & Expr@10 & Expr@20 \\
\midrule
GPT-5 & 32 & 31/32 (0.969) & 31/32 (0.969) & 31/32 (0.969) \\
GPT-o4-mini & 32 & 30/32 (0.938) & 30/32 (0.938) & 30/32 (0.938) \\
GPT-o3-mini & 32 & 30/32 (0.938) & 30/32 (0.938) & 30/32 (0.938) \\
GPT-o3 & 32 & 30/32 (0.938) & 30/32 (0.938) & 30/32 (0.938) \\
GPT-5-nano & 32 & 30/32 (0.938) & 30/32 (0.938) & 30/32 (0.938) \\
GPT-5-mini & 32 & 29/32 (0.906) & 29/32 (0.906) & 29/32 (0.906) \\
GPT-o1-mini & 32 & 29/32 (0.906) & 29/32 (0.906) & 29/32 (0.906) \\
GPT-5-codex & 32 & 29/32 (0.906) & 29/32 (0.906) & 29/32 (0.906) \\
GPT-4o & 32 & 27/32 (0.844) & 27/32 (0.844) & 27/32 (0.844) \\
GPT-OSS-20B & 32 & 22/32 (0.688) & 22/32 (0.688) & 22/32 (0.688) \\
GPT-4o-mini & 32 & 18/32 (0.562) & 18/32 (0.562) & 18/32 (0.562) \\
Gemini-2.5-Pro & 32 & 32/32 (1.000) & 32/32 (1.000) & 32/32 (1.000) \\
Gemini-2.5-Flash-09-2025 & 32 & 31/32 (0.969) & 31/32 (0.969) & 31/32 (0.969) \\
Kimi-K2 & 32 & 31/32 (0.969) & 31/32 (0.969) & 31/32 (0.969) \\
Claude-sonnet-4 & 32 & 30/32 (0.938) & 30/32 (0.938) & 30/32 (0.938) \\
Deepseekchat-v3 & 32 & 27/32 (0.844) & 27/32 (0.844) & 27/32 (0.844) \\
Qwen3-coder-plus & 32 & 27/32 (0.844) & 27/32 (0.844) & 27/32 (0.844) \\
Qwen3-next-80b-a3b-instruct & 32 & 21/32 (0.656) & 21/32 (0.656) & 21/32 (0.656) \\
CodeLlama-70b & 32 & 11/32 (0.344) & 11/32 (0.344) & 11/32 (0.344) \\
CodeLlama-13b & 32 & 0/32 (0.000) & 0/32 (0.000) & 0/32 (0.000) \\
StarChat & 32 & 0/32 (0.000) & 0/32 (0.000) & 0/32 (0.000) \\
\bottomrule
\end{tabular}}
\end{table*}

\begin{figure}[]
	\centering
       \scalebox{0.7}{
\includegraphics[width=0.8\textwidth]{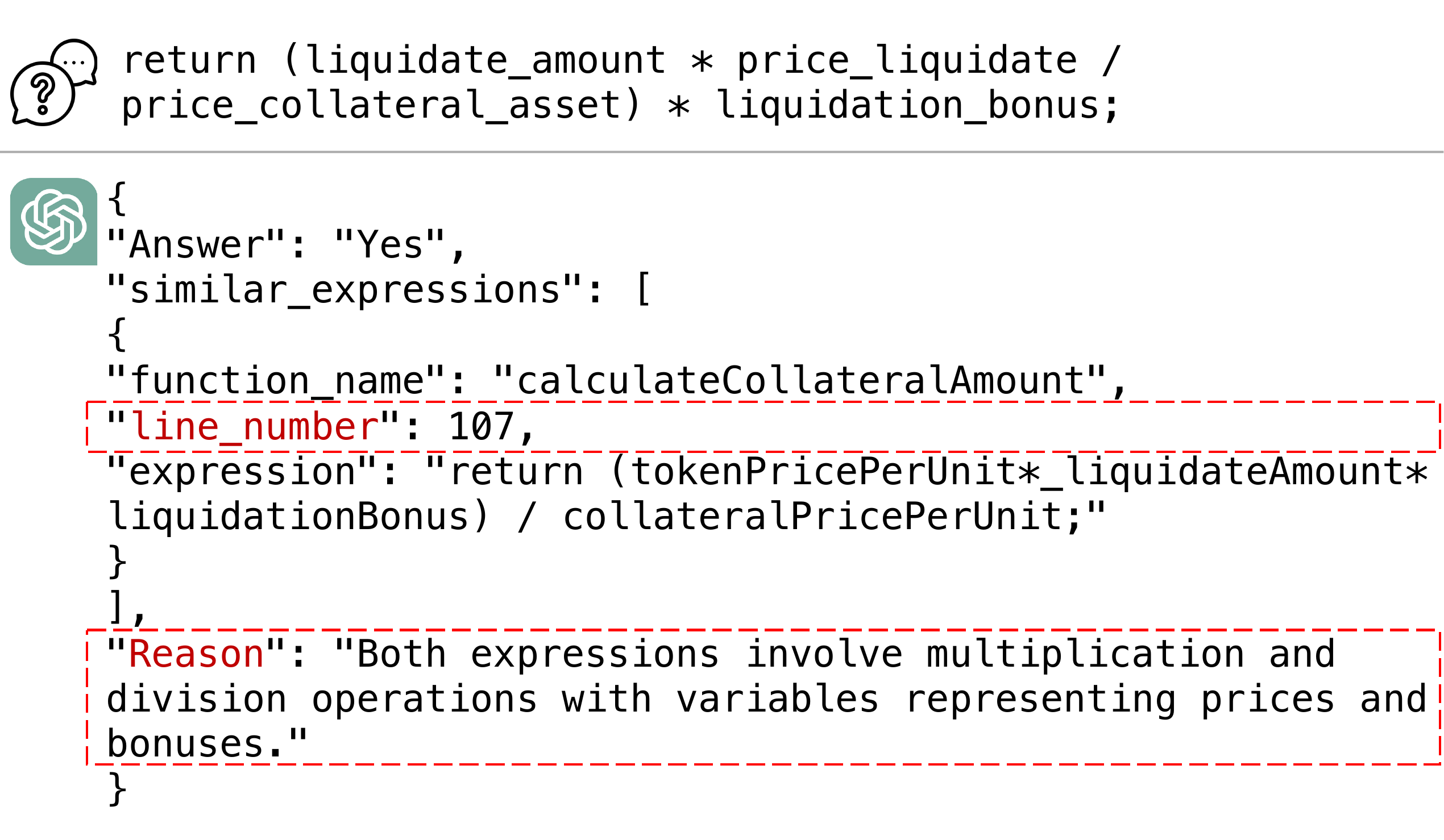}}
	\caption{One Response Example about Expression Matching.}
	\label{fig:expression_matching_sample}
\end{figure}

We find that models judge expression similarity mainly by the structure of operators, the roles of operands, and evaluation order. \figref{fig:expression_matching_sample} illustrates this behavior. Line-number references in raw outputs are unreliable and excluded from evaluation. Larger capacity helps (e.g., CodeLlama-70b > 13b), and instruction tuning plus training data likely underpin the strong results of recent commercial models.

\noindent
\answer{1}{Recent SOTA LLMs perform well on syntax‑oriented tasks: they often produce high‑quality ASTs and accurately match expression structures, indicating a strong grasp of programming‑language syntax. However, compared with compiler parsers, their outputs are less precise and less stable (e.g., minor omissions or formatting variants), so human validation remains necessary. We also observe a clear size effect: larger models tend to achieve stronger syntactic performance.}

\subsection{Code Static Analysis~(RQ2)}
\subsubsection{CFG Generation}
\label{sec:results_cfg}

\begin{figure}[]
   \centering
   \includegraphics[width=\textwidth]{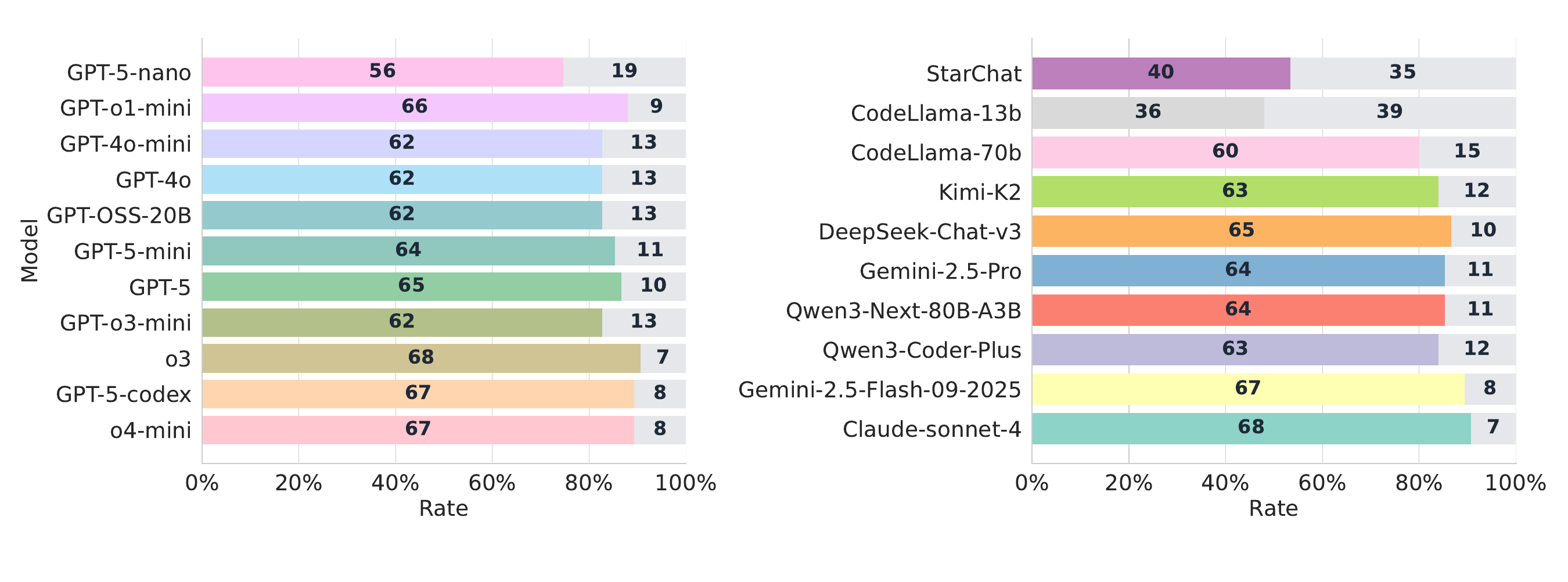}
   \caption{Automated pass/fail rates for CFG across models (Layer 1).}
   \label{fig:cfg_passfail}
\end{figure}

\paragraph{Layer 1 (Automated Metrics)} Following the human-defined reasonableness criteria, we compare LLM-generated CFGs against static-analysis baselines and use structural metrics to decide pass/fail; see the automated summary in \figref{fig:cfg_passfail}. Under this setup, recent commercial models generally achieve mid-to-high pass counts (roughly 62-68/75), with the best reaching 68/75. Many other current models fall within a narrow band, while earlier open-source baselines are clearly weaker: CodeLlama-70b reaches 60/75, and StarChat/CodeLlama-13b drop to 40/75 and 36/75. CFG remains more challenging than AST, with more frequent structural errors and occasional fabrications. 
To better understand these gaps, we examine failed CFGs and observe that high-performing models mostly exhibit subtle structural glitches (e.g., slightly mis-joined branches or misplaced basic blocks), mid-tier models show more fabricated edges and redundant control-flow introduced around error handling or logging, and the weakest baselines frequently hallucinate entire blocks or loop structures and merge unrelated branches, which aligns with their much lower pass counts. Across model families, CFG accuracy improves with scale. For example, within the GPT-5 and CodeLlama families, Code-specialized or reasoning-oriented models (such as GPT-5-codex, Claude-sonnet-4, and Gemini-2.5-Pro) consistently outperform comparably sized general-purpose chat models, indicating that architectural and training choices matter as much as parameter count for capturing detailed control flow.

\begin{figure}[]
    \centering
    \includegraphics[width=0.7\textwidth]{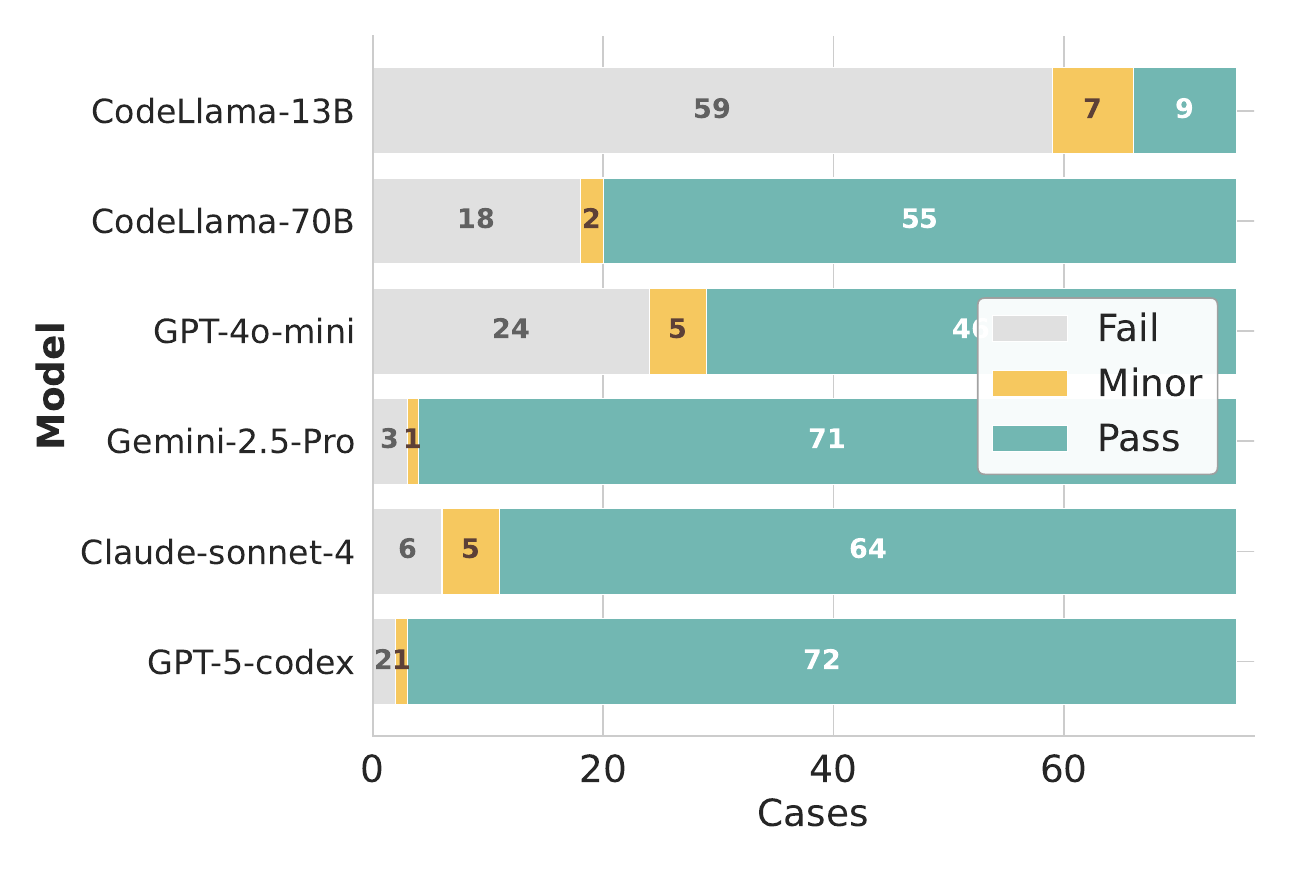}
    \caption{CFG structural evaluation: human pass/minor/fail.}
    \label{fig:cfg_count}
\end{figure}

\begin{figure}[]
    \centering
    \includegraphics[width=0.7\textwidth]{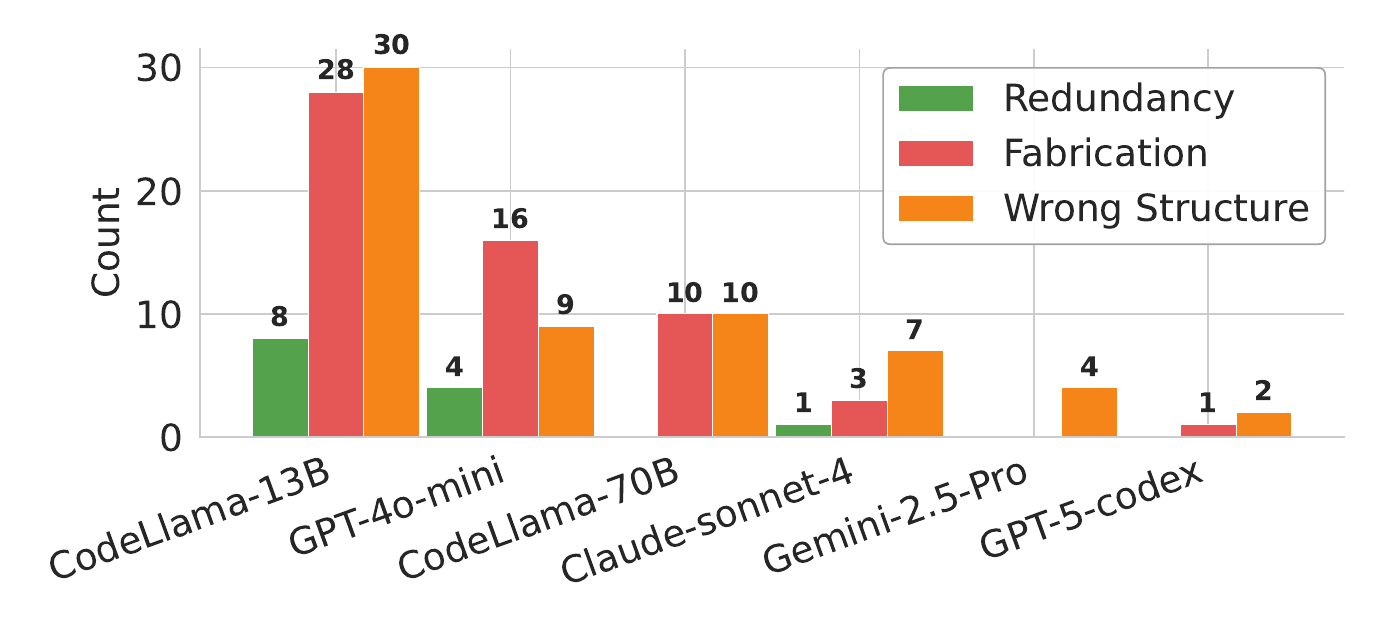}
    \caption{CFG issue categories by model.}
    \label{fig:cfg_issues}
\end{figure}

\begin{figure}
   \centering
   \scalebox{0.9}{
   \includegraphics[width=0.9\textwidth]{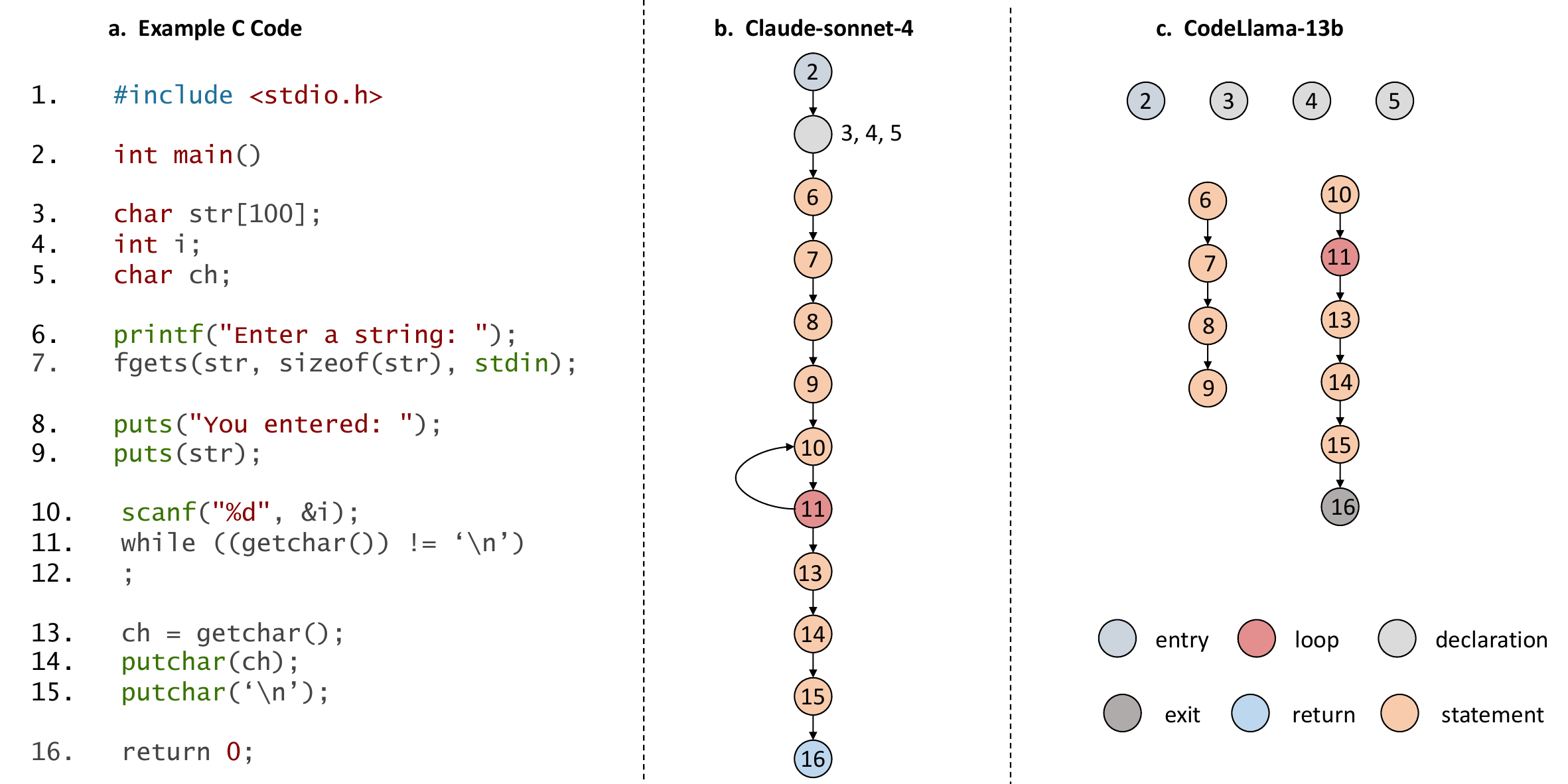}}
    \caption{CFG examples. Left: reasonable; Right: unreasonable.}
   \label{fig:cfg_example}
\end{figure}

\paragraph{Layer 2 (Expert Adjudication)} We next report human-labeled structural quality (pass/minor/fail) and issue-category breakdowns for the same case set. \figref{fig:cfg_count} summarizes the distribution of reasonable and unreasonable CFGs. \figref{fig:cfg_issues} details issue categories by model. We categorize issues as redundancy (minor), fabrication, and wrong-structure (severe). Redundancy includes meaningless nodes (e.g., null nodes); fabrication refers to non-existent nodes or statements; wrong-structure denotes misrepresented control flow (e.g., loops, if-else). Consistent with Layer 1, within the six manually evaluated models, recent commercial models place more weight in pass with fewer severe errors. In contrast, the open-source baselines show higher rates of wrong-structure and occasional fabrication. We also observe a capacity effect: larger or better-instructed models tend to reduce these errors.
\figref{fig:cfg_example} shows an example: The reasonable CFG preserves sequential order and models the empty-body while loop via a self-loop and exit. The unreasonable CFG lacks a while loop back-edge, and loop semantics.

\paragraph{Layer 3 (Consistency Validation)} Inter-rater reliability shows $\kappa = 0.858$ and AC1 = 0.920. Human-machine agreement is substantial (AC1 = 0.706, PABAK = 0.611), validating automated metric effectiveness despite Cohen's $\kappa = 0.427$ being suppressed by class imbalance. Comprehensive analysis in Section~\ref{sec:agreement}.

\subsubsection{Call Graph Generation}
\label{sec:results_cg}

\begin{figure}[]
   \centering
   \includegraphics[width=\textwidth]{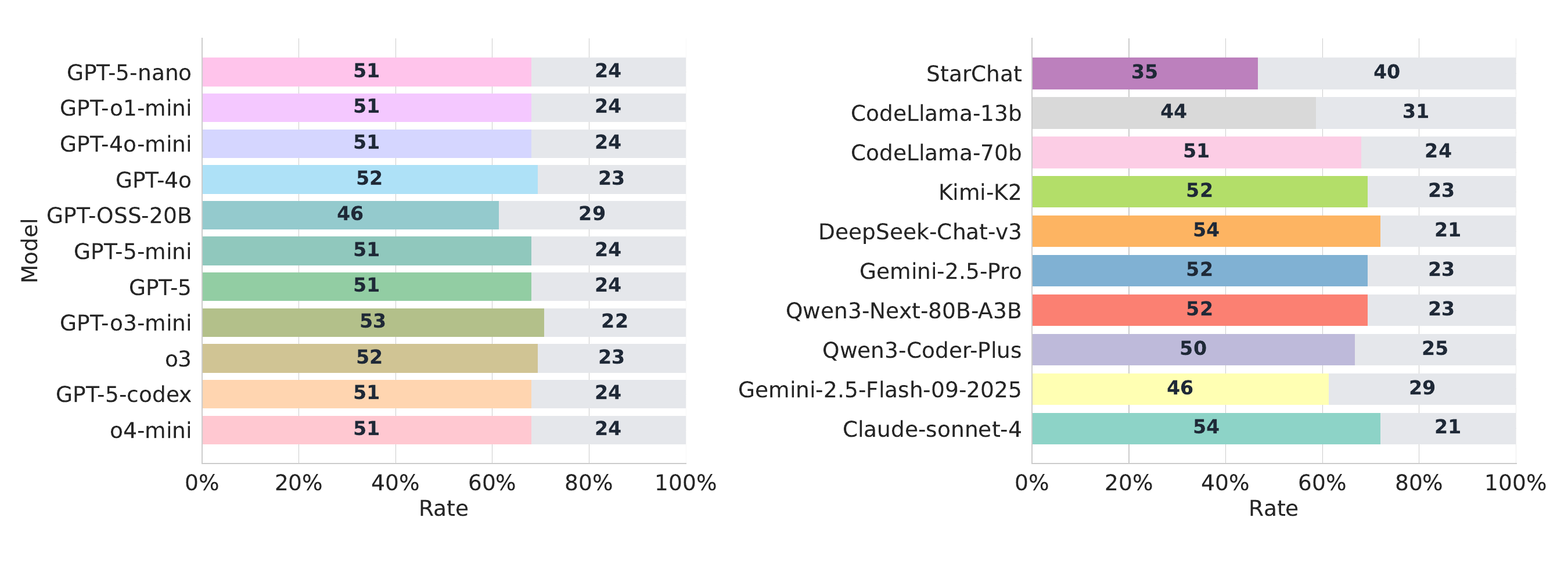}
   \caption{Automated pass/fail rates for CG across models (Layer 1).}
   \label{fig:cg_passfail}
\end{figure}

\paragraph{Layer 1 (Automated Metrics)} Following the human-defined reasonableness criteria, we compare LLM-generated CGs with static-analysis baselines and use structural indicators (coverage and edge consistency) to decide pass/fail; see Figure~\ref{fig:cg_passfail}. Under this setup, many recent models cluster in the low: 51–54 over 75 cases, while older open-source baselines drop further into the 30–40. CG remains more challenging than AST/CFG: when we inspect failed CGs, we find that problems are dominated by missing calls (especially through wrappers, callbacks, and virtual dispatch) and spurious edges to unrelated helpers or overly generic library routines, with redundancy playing a smaller role. Strong models mostly miss a few non-critical calls but rarely hallucinate large call edges, whereas mid-tier models both miss deeper callee chains and introduce extra edges in recursive or event-driven code; weaker baselines exhibit many missing and fabricated edges at the same time, often losing the core call skeleton entirely. As with CFG, CG pass rates improve with model size within a family (e.g., CodeLlama-13B→70B, GPT-5-nano→GPT-5), but code-focused or reasoning-optimized models (GPT-5-codex, Qwen3-coder-plus, Gemini-2.5-Pro, Claude-sonnet-4) cluster near the top of this band (around 50–54/75), while earlier dense open-source chat models remain in the lower 30–40/75 range.

\begin{figure}[h]
    \centering
    \includegraphics[width=0.7\textwidth]{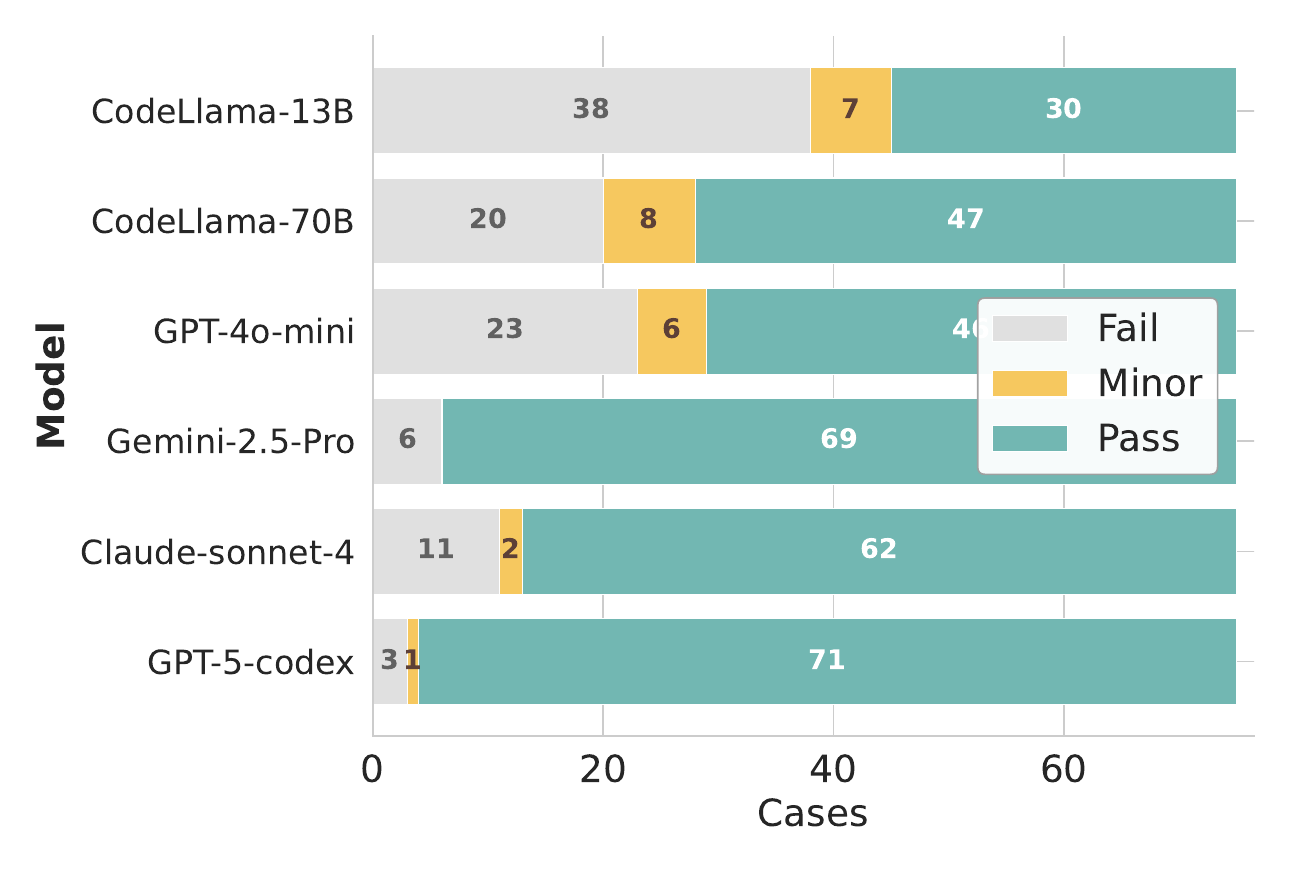}
    \caption{CG structural evaluation: human pass/minor/fail.}
    \label{fig:cg_count}
\end{figure}

\begin{figure}[h]
    \centering
\includegraphics[width=0.7\textwidth]{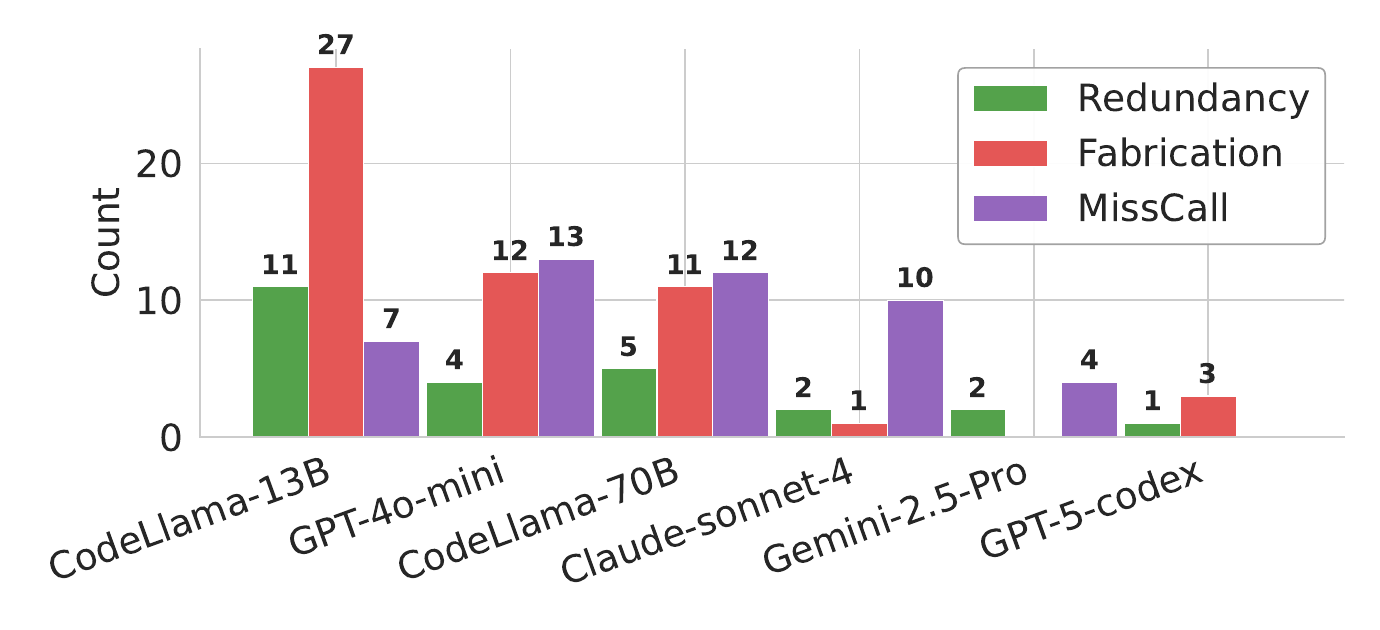}
    \caption{CG issue categories by model.}
    \label{fig:cg_issues}
\end{figure}

\begin{figure}[]
 \centering
 \includegraphics[width=\textwidth]{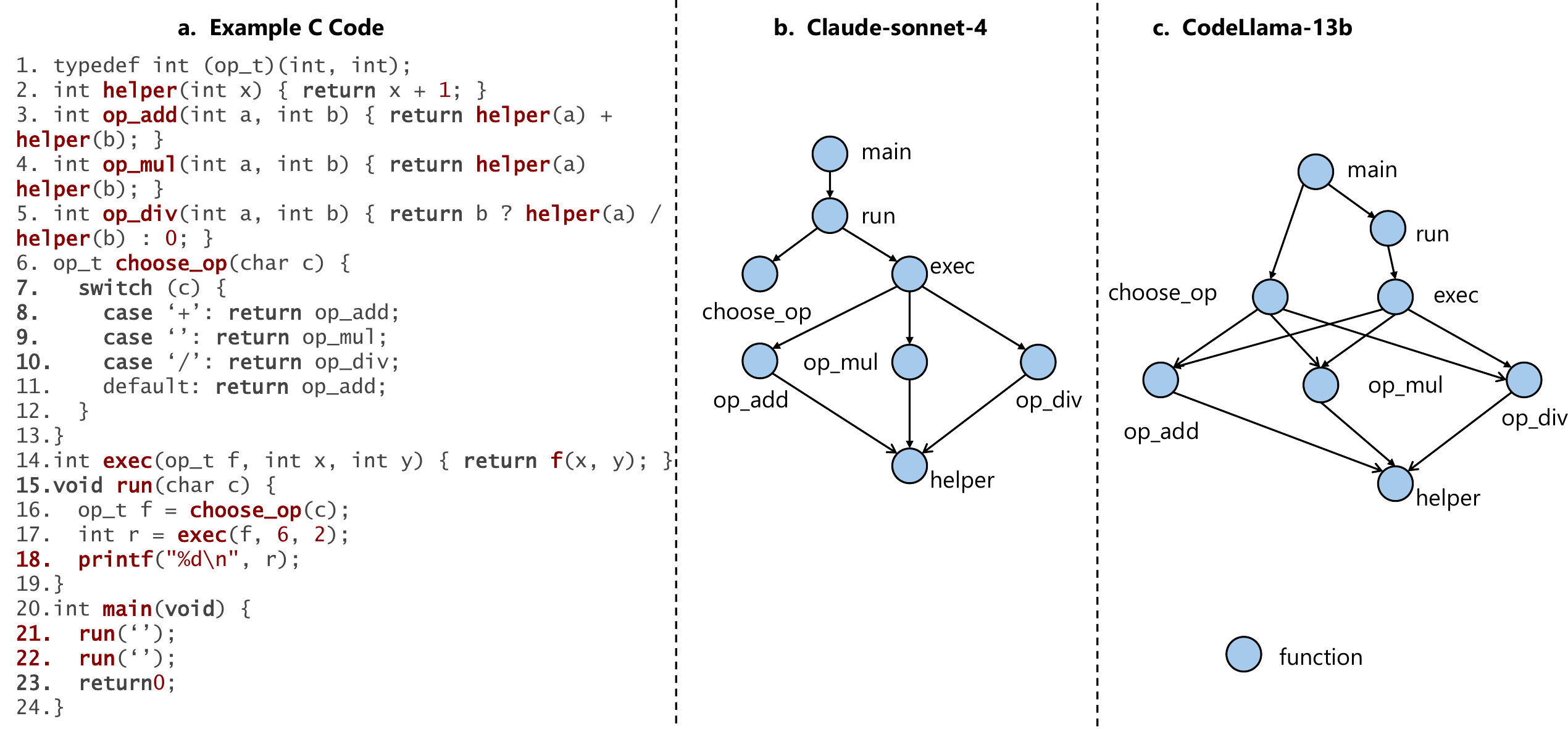}
 \caption{CG examples. Left: reasonable; Right: unreasonable.}
 \label{fig:cg_example}
\end{figure}

\paragraph{Layer 2 (Expert Adjudication)} We report human-labeled structural quality (pass/minor/fail) and issue-category breakdowns on the same case set. \figref{fig:cg_count} summarizes pass/minor/fail from human evaluation, and \figref{fig:cg_issues} details issue categories per model. We group issues into redundancy (minor), fabrication, and missing call (severe when coverage is affected). Across the six manually reviewed models, stronger models place more weight in pass, while weaker baselines exhibit higher rates of missing calls and occasional fabrications.
Figure~\ref{fig:cg_example} shows that the reasonable CG captures main→run and run→choose\_op/exec, with exec→op\_* and op\_*→helper. The unreasonable CG fabricates choose\_op→op\_* edges, omits run→choose\_op, and adds an extra main→choose\_op, altering caller–callee semantics.


\paragraph{Layer 3 (Consistency Validation)} Inter-rater reliability shows $\kappa = 0.844$ and AC1 = 0.914. Human-machine agreement is moderate (AC1 = 0.515), reflecting task complexity where multiple valid call graph factorizations exist. Full analysis including model pass rates under human consensus in Section~\ref{sec:agreement}.

\subsubsection{Data Dependency and Taint Analysis}
\begin{figure}[t]
   \centering
\begin{subfigure}[b]{0.32\textwidth}\centering\includegraphics[width=\textwidth]{sections/figures/dp_taint/gpt-5-nano__dp_taint_f1_dual_centered_paper.pdf}\caption*{\small{GPT-5-nano}}\end{subfigure}%
\begin{subfigure}[b]{0.32\textwidth}\centering\includegraphics[width=\textwidth]{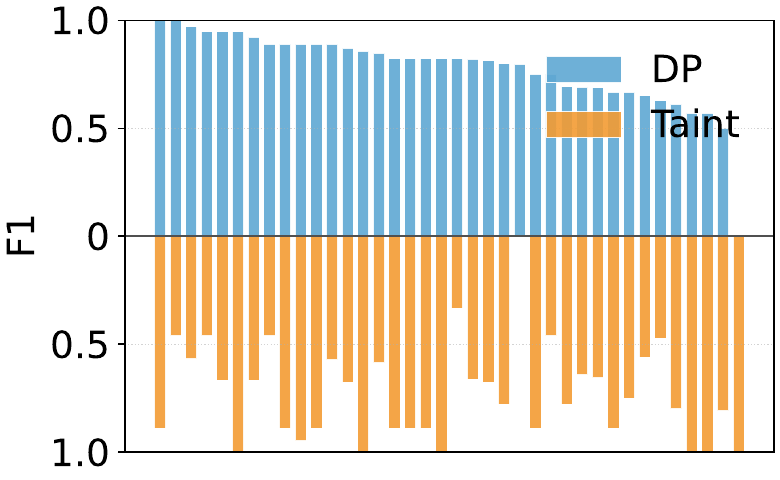}\caption*{\small{Claude-sonnet-4}}\end{subfigure}%
\begin{subfigure}[b]{0.32\textwidth}\centering\includegraphics[width=\textwidth]{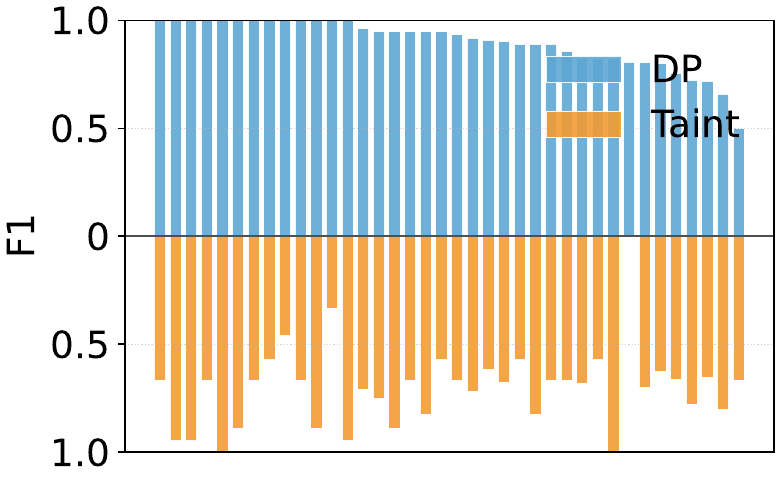}\caption*{\small{Gemini-2.5-Flash}}\end{subfigure}%

\begin{subfigure}[b]{0.32\textwidth}\centering\includegraphics[width=\textwidth]{sections/figures/dp_taint/gpt-oss-20b__dp_taint_f1_dual_centered_paper.pdf}\caption*{\small{GPT-OSS-20B}}\end{subfigure}%
\begin{subfigure}[b]{0.32\textwidth}\centering\includegraphics[width=\textwidth]{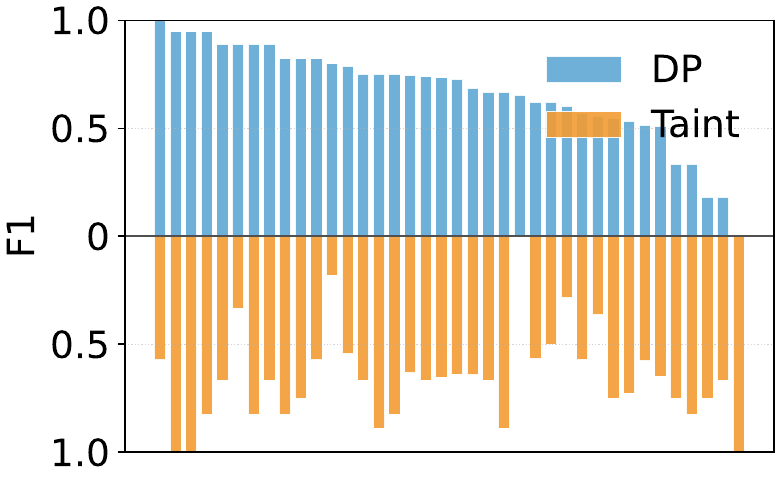}\caption*{\small{CodeLlama-70b}}\end{subfigure}%
\begin{subfigure}[b]{0.32\textwidth}\centering\includegraphics[width=\textwidth]{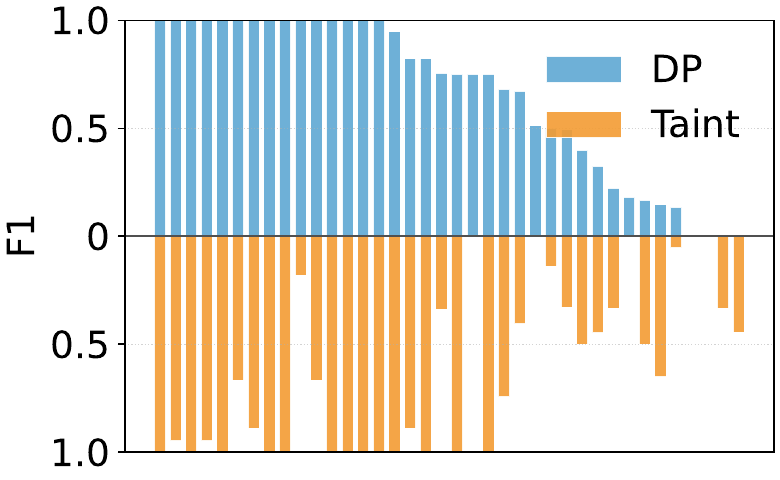}\caption*{\small{CodeLlama-13b}}\end{subfigure}%

   \caption{F1 of each project on Data Dependency (blue) and Taint Analysis (orange) for representative models. Complete results for all 21 models are provided in Appendix~\ref{appendix:dp_taint_full}.}
   \label{fig:data_dep_taint}
\end{figure}

As shown in Table~\ref{tab:dp_taint_f1_all}, for data dependence the best-performing models reach macro averaged F1 of $0.898$ and $0.887$ (Gemini-2.5-Flash and Gemini-2.5-Pro), followed by a compact group around $0.84$ to $0.83$ (e.g., GPT-5-nano and GPT-5-mini). Mid-tier and open-source entries cluster around $0.66$ to $0.75$ (e.g., CodeLlama 13b $0.659$, CodeLlama 70b $0.669$, and GPT OSS 20B $0.710$). StarChat surpasses CodeLlama 13b on this task while still trailing most commercial models. The taint column in the same table is clearly lower even with a per-project balanced set; representative strong values are $0.761$ for GPT-5-nano, $0.725$ for Claude-sonnet-4, and $0.700$ for Gemini-2.5-Flash, while several models sit at or below the random guess baseline of $0.50$ (e.g., $0.473$ and $0.388$ for GPT-4o and GPT-o4-mini, and $0.364$ and $0.281$ for GPT-o3-mini and Deepseekchat-v3). 
We evaluate these tasks on thirty-eight projects using analyzer derived facts from Slither for Solidity and Joern for Java with balanced per project sampling and report macro averaged F1 across projects. Across model families, data dependence macro F1 typically lies in $0.66$ to $0.90$, while taint spans $0.28$ to $0.76$. As shown in Figure~\ref{fig:data_dep_taint}, per project dispersion is large, especially for taint. The upper section for data dependence ranges from approximately $0$ to $1.0$, and the lower section for taint ranges from $0$ to approximately $0.8$, confirming substantial variance. Variability is pronounced on Solidity projects with external calls and transaction driven flows, whereas Java libraries are steadier on shallow patterns such as ``parameter $\to$ use'' and ``getter $\to$ variable'' but degrade on intermodule reasoning, aliasing, and collection semantics. Typical errors include, for data dependence, missing multistep or interprocedural chains and mistaking co-occurrence for dependence, and for taint, overtagging interface parameters as user controlled, weak recognition of sanitization and guards, and incomplete propagation through framework or platform specific interfaces.

\begin{table*}[t]
\centering
\caption{Prediction Performance (F1) on Data Dependency and Taint across 21 models.}
\label{tab:dp_taint_f1_all}
\scalebox{0.9}{
\begin{tabular}{lcc|lcc}
\toprule
Model & DP F1 & Taint F1 & Model & DP F1 & Taint F1 \\
\midrule
GPT-5-nano & 0.846 & 0.761 & GPT-5-mini & 0.826 & 0.592 \\
GPT-o1-mini & 0.748 & 0.435 & GPT-5-codex & 0.695 & 0.606 \\
GPT-OSS-20B & 0.710 & 0.624 & GPT-5 & 0.700 & 0.545 \\
GPT-o4-mini & 0.705 & 0.388 & GPT-o3-mini & 0.734 & 0.364 \\
GPT-o3 & 0.710 & 0.672 & GPT-4o-mini & 0.535 & 0.678 \\
GPT-4o & 0.692 & 0.473 & Claude-sonnet-4 & 0.777 & 0.725 \\
Gemini-2.5-Flash-09-2025 & 0.898 & 0.700 & Gemini-2.5-Pro & 0.887 & 0.664 \\
Deepseekchat-v3 & 0.741 & 0.281 & Qwen3-next-80b-a3b-instruct & 0.559 & 0.630 \\
Qwen3-coder-plus & 0.607 & 0.503 & Kimi-K2 & 0.746 & 0.614 \\
StarChat & 0.699 & 0.481 & CodeLlama-70b & 0.669 & 0.655 \\
CodeLlama-13b & 0.659 & 0.616 &  &  &  \\
\bottomrule
\end{tabular}}
\end{table*}

\subsubsection{Pointer Analysis}
\begin{figure}[t]
   \centering
   \includegraphics[width=0.9\textwidth]{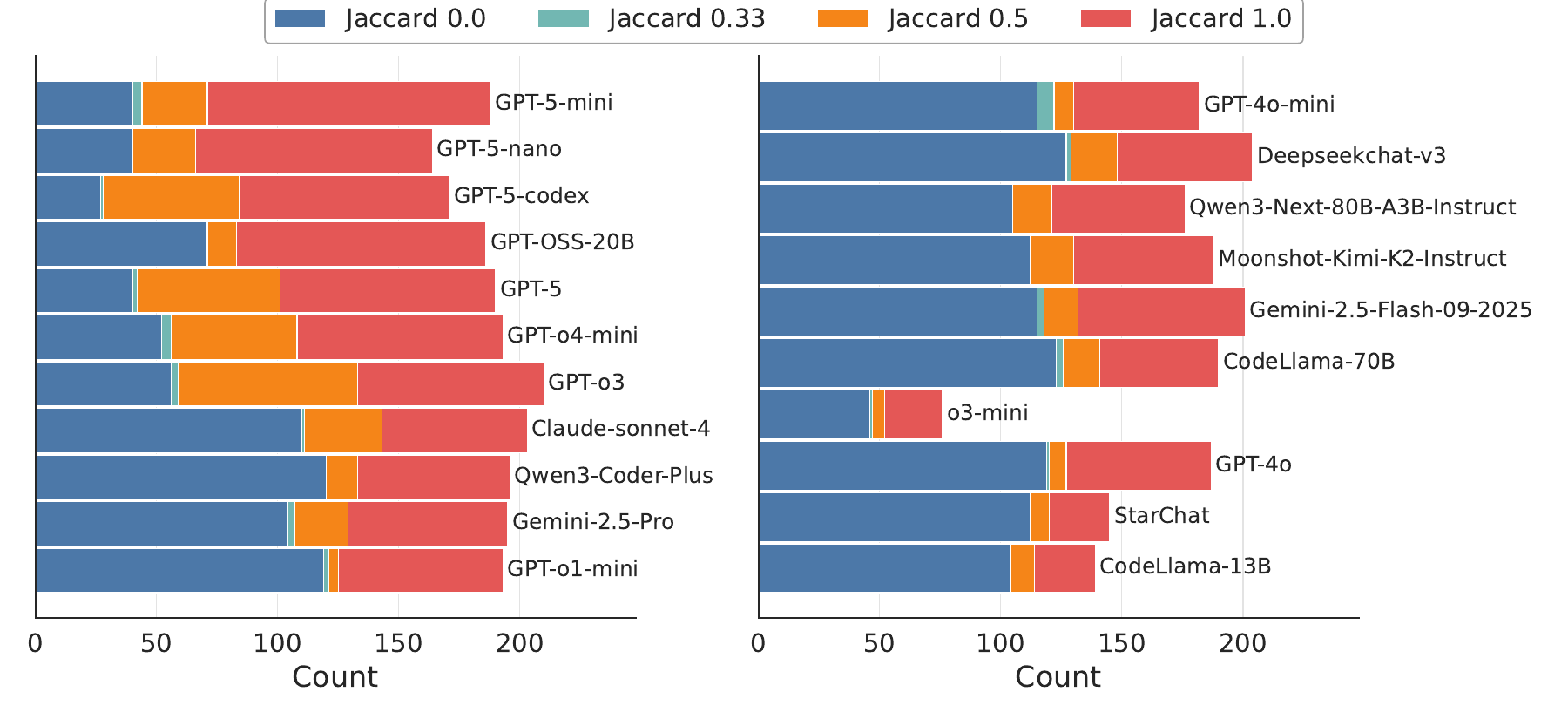}
   \caption{Jaccard Index of Each Pointer}
   \label{fig:jacrd_cor_datapoint_each_pointer}
\end{figure}

\begin{figure}[t]
   \centering
   \includegraphics[width=0.9\textwidth]{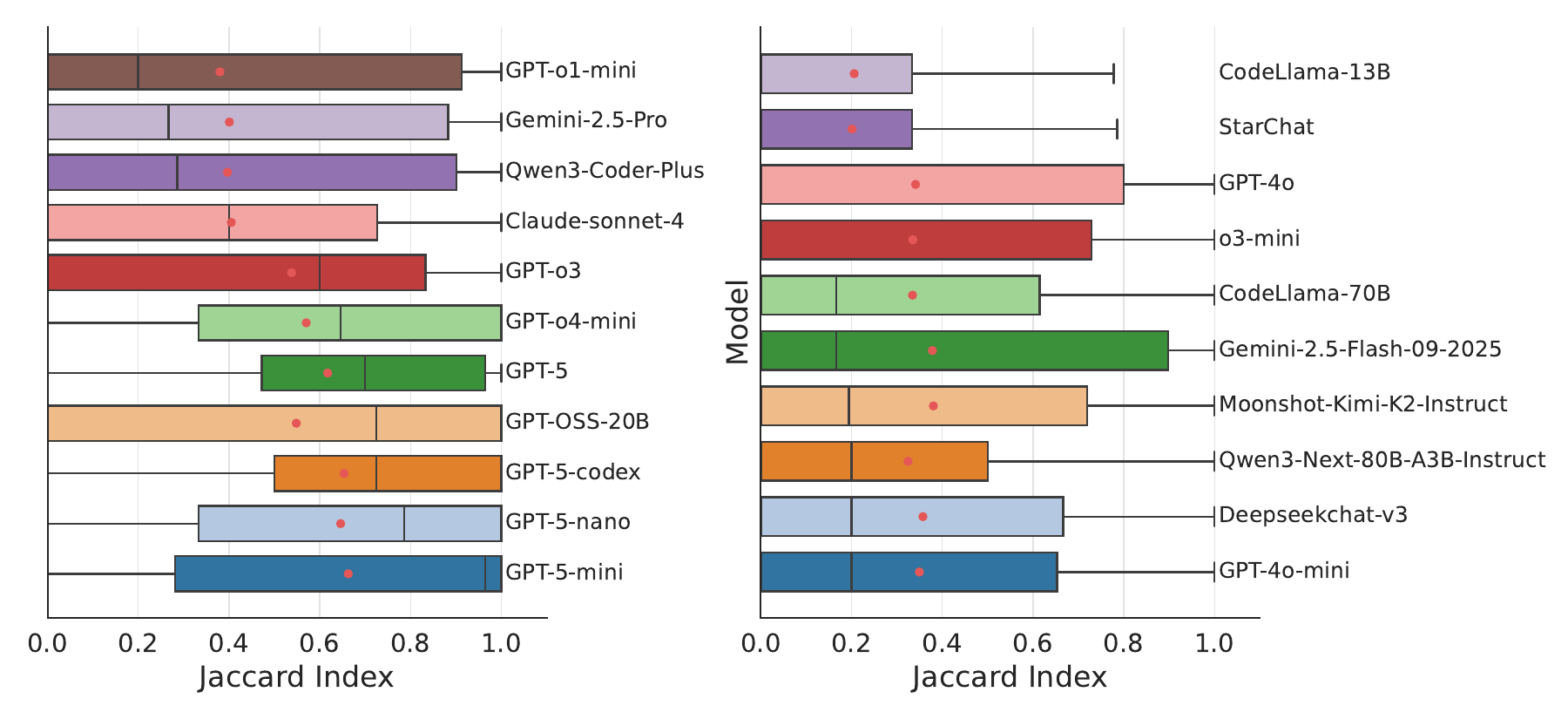}
   \caption{Jaccard Index of Each Program}
   \label{fig:jacrd_cor_datapoint}
\end{figure}

We summarize pointer analysis with Jaccard similarity between predicted and ground-truth points-to sets. Across 21 models, the best scores are still modest: \textit{GPT-5-mini} (0.552), \textit{GPT-5} (0.516), \textit{GPT-5-codex} (0.511), and \textit{GPT-o3} (0.499) cluster around 0.5; many recent models sit in the 0.29-0.36 band (e.g., \textit{GPT-4o-mini} 0.290, \textit{Deepseekchat-v3} 0.320, \textit{Qwen3-next-80b} 0.297), while weaker baselines fall below 0.2 (e.g., \textit{StarChat} 0.150, \textit{CodeLlama-13b} 0.136). Open-source \textit{CodeLlama-70b} (0.276) outperforms \textit{CodeLlama-13b} markedly but still trails the top cluster.
Per-pointer distributions in \figref{fig:jacrd_cor_datapoint_each_pointer} reveal a polarized pattern across models: many predictions are either entirely correct (Jaccard=1) or entirely incorrect (Jaccard=0), with relatively fewer partial matches. This polarization, together with the moderate top-line scores, indicates current LLM outputs are not yet reliable replacements for pointer analysis. To further assess data-shift effects, we compute the program-level mean Jaccard (\figref{fig:jacrd_cor_datapoint}); the wide spread across projects underscores sensitivity to codebase characteristics and motivates hybrid workflows (tool-extracted facts + LLM synthesis + tool re-checks).


\noindent
\answer{2}{Larger and more recent models achieve higher structural correctness on syntax and static analysis tasks, whereas smaller open models are weaker. The evaluated models still show fabrication errors, including nonexistent nodes, edges, or statements, indicating that hallucination remains a pervasive failure mode in structural code analysis. Performance varies by program and language, reflecting sensitivity to control constructs and data abstractions present in the dataset. Reducing structural hallucinations and stabilizing behavior over diverse codebases should be a priority for future methods and evaluations.}

\subsection{Code Dynamic Analysis~(RQ3)}
\subsubsection{Equivalent Mutant Detection}
Table~\ref{tab:performance_eq} reports few-shot and zero-shot F1. Top few-shot scores cluster around 0.90--0.93: \textit{GPT-5-mini} (0.925), \textit{GPT-o3-mini} (0.920), \textit{GPT-o4-mini} (0.920), and \textit{Gemini-2.5-Flash} (0.905). \textit{GPT-5} and \textit{GPT-5-codex} are close ($\approx0.890$), and \textit{Gemini-2.5-Pro} reaches 0.855. Zero-shot is generally comparable for strong models (e.g., \textit{Gemini-2.5-Flash} 0.925, \textit{GPT-o4-mini} 0.920). Weaker or earlier models perform substantially worse: \textit{GPT-4o-mini} $\approx0.50$, \textit{CodeLlama-13b} $\approx0.50$; \textit{StarChat} shows uneven behavior (0.417 few-shot vs 0.591 zero-shot).
Dynamic equivalence remains challenging but benefits from targeted prompting on some models. Performance does not strictly follow model strength: instruction tuning and task familiarity matter. Few-shot helps certain models (e.g., \textit{GPT-o1-mini}), but can be neutral or negative on others. Within-model prompt strategy comparisons with McNemar significance testing (Section~\ref{sec:prompt_strategy}) reveal model-dependent effects: CoT/role improve GPT-o1-mini ($\Delta$acc$=+0.23$-$+0.26$, $p<10^{-7}$) but degrade GPT-5-mini ($\Delta$acc$=-0.04$-$-0.07$, $p<0.01$), suggesting that reasoning scaffolds interact with model architectures in task-specific ways.

\begin{table*}[t]
\centering
\caption{Performance (F1) about Equivalent Mutant Detection across 21 models.}
\label{tab:performance_eq}
\scalebox{0.9}{
\begin{tabular}{lcc|lcc}
\toprule
Model & Few-shot & Zero-shot & Model & Few-shot & Zero-shot \\
\midrule
GPT-5-nano & 0.895 & 0.875 & GPT-5-mini & 0.925 & 0.885 \\
GPT-o1-mini & 0.720 & 0.645 & GPT-5-codex & 0.890 & 0.880 \\
GPT-OSS-20B & 0.834 & 0.875 & GPT-5 & 0.895 & 0.885 \\
GPT-o4-mini & 0.920 & 0.920 & GPT-o3-mini & 0.920 & 0.880 \\
GPT-o3 & 0.880 & 0.860 & GPT-4o-mini & 0.500 & 0.495 \\
GPT-4o & 0.600 & 0.540 & Claude-sonnet-4 & 0.700 & 0.660 \\
Gemini-2.5-Flash-09-2025 & 0.905 & 0.925 & Gemini-2.5-Pro & 0.855 & 0.755 \\
Deepseekchat-v3 & 0.570 & 0.505 & Qwen3-next-80b-a3b-instruct & 0.550 & 0.530 \\
Qwen3-coder-plus & 0.555 & 0.520 & Kimi-K2 & 0.515 & 0.510 \\
StarChat & 0.417 & 0.591 & CodeLlama-70b & 0.531 & 0.533 \\
CodeLlama-13b & 0.500 & 0.500 &  &  &  \\
\bottomrule
\end{tabular}}
\end{table*}

\subsubsection{Flaky Test Reasoning} 
\begin{figure}[!h]
	\centering
	\begin{subfigure}[b]{0.32\textwidth}\centering\includegraphics[width=\textwidth]{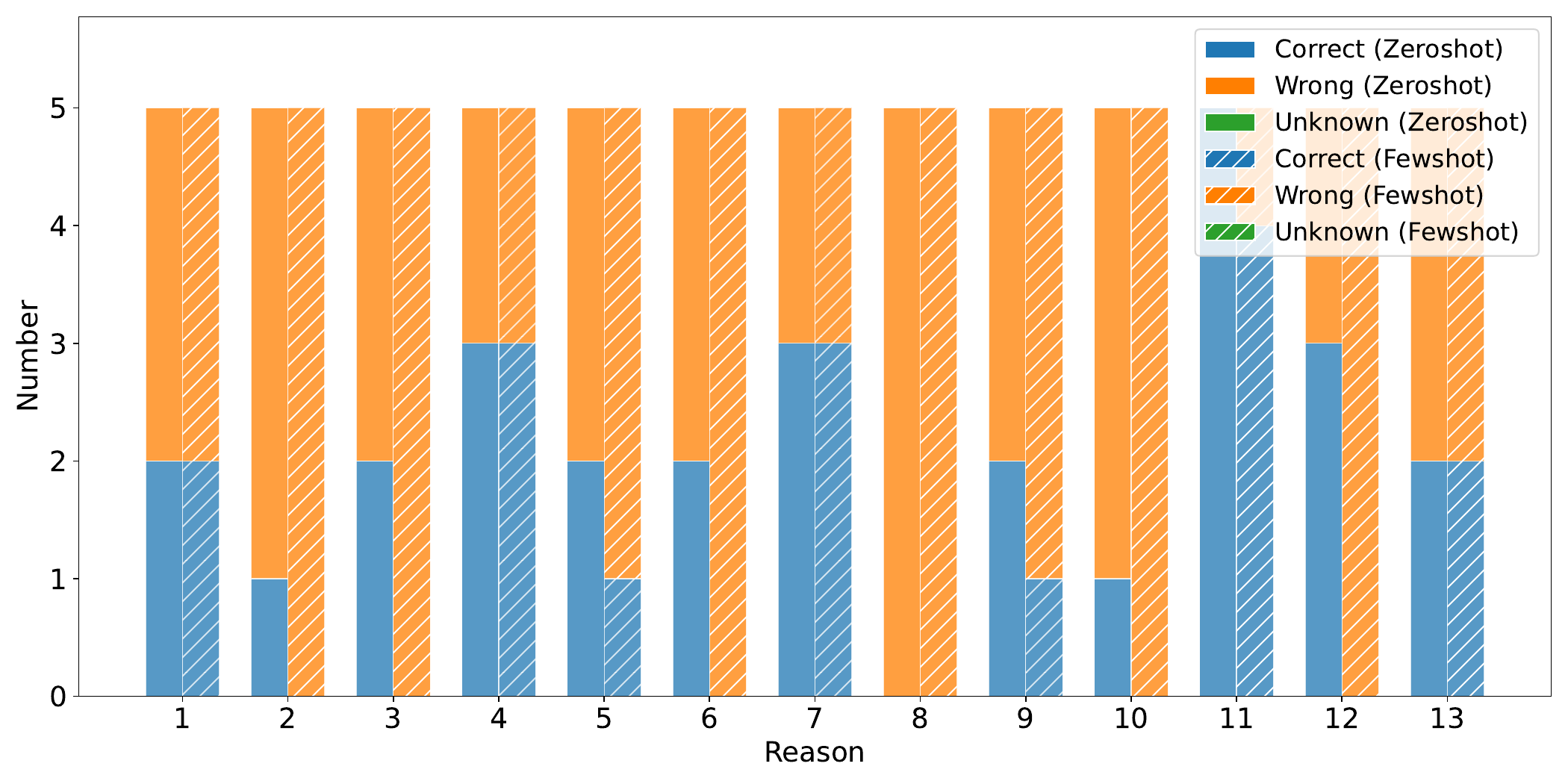}\caption*{\small{Claude-sonnet-4}}\end{subfigure}%
	\begin{subfigure}[b]{0.32\textwidth}\centering\includegraphics[width=\textwidth]{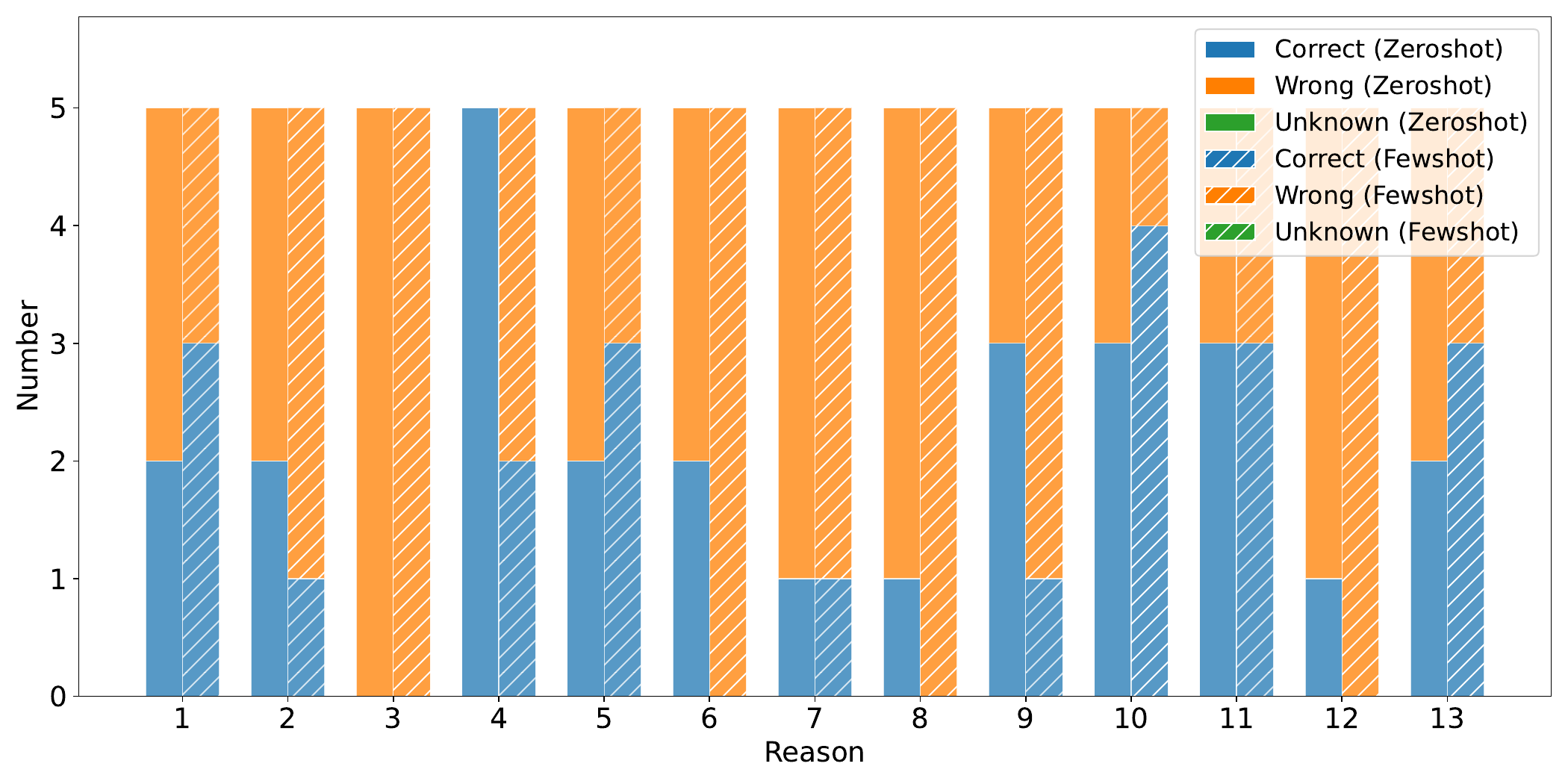}\caption*{\small{GPT-o3}}\end{subfigure}%
	\begin{subfigure}[b]{0.32\textwidth}\centering\includegraphics[width=\textwidth]{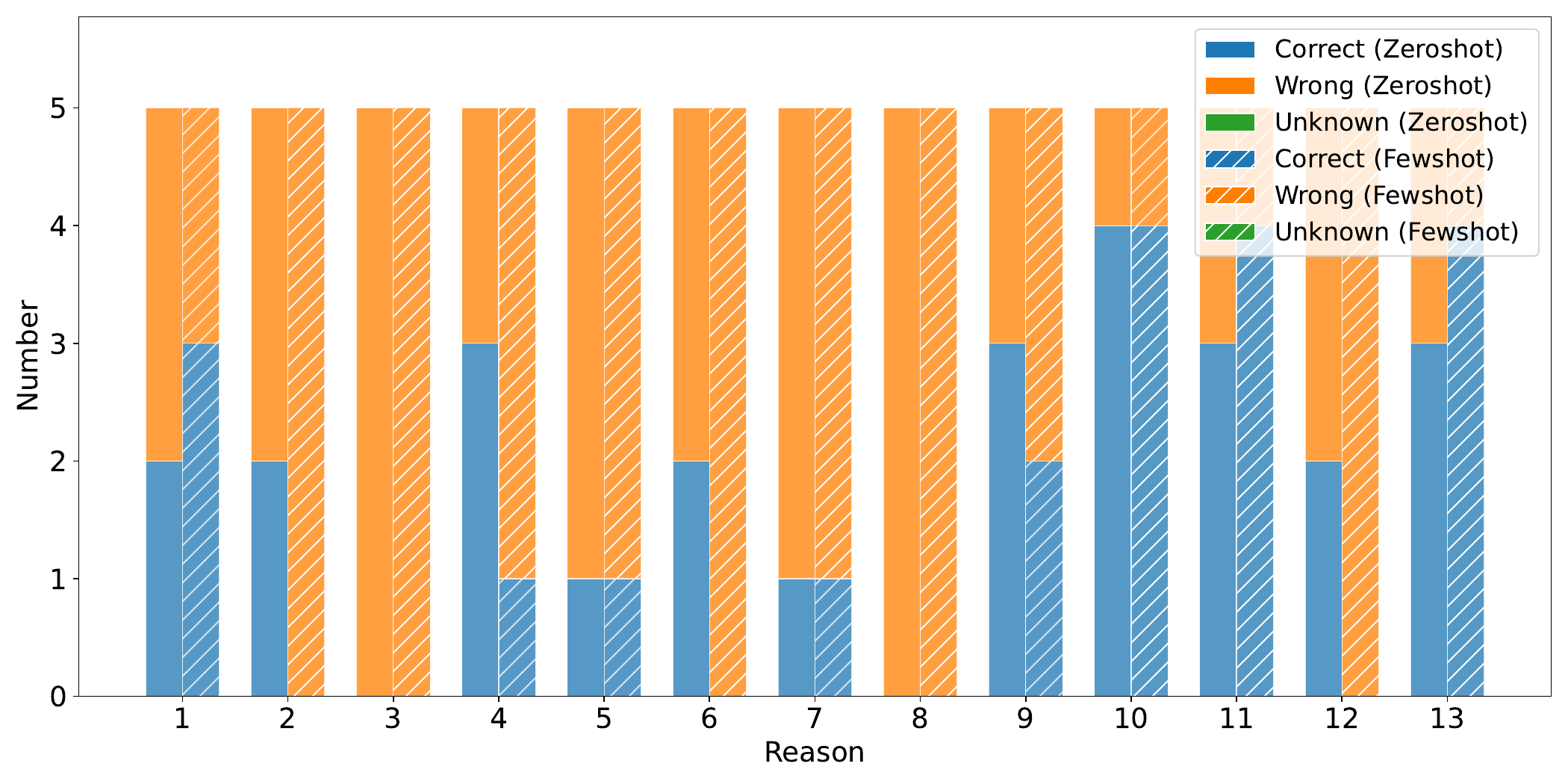}\caption*{\small{GPT-5}}\end{subfigure}%
	
	\begin{subfigure}[b]{0.32\textwidth}\centering\includegraphics[width=\textwidth]{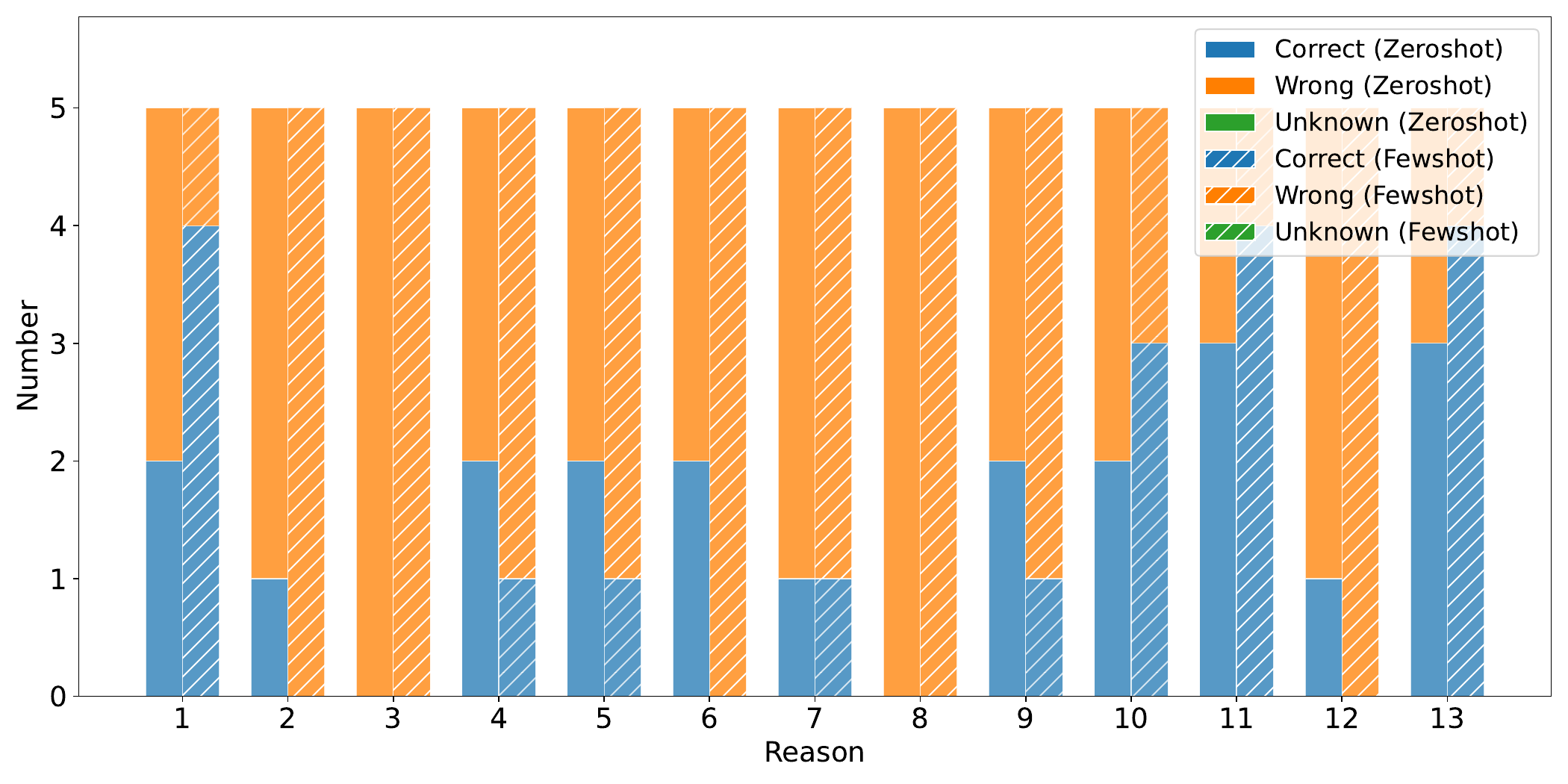}\caption*{\small{GPT-5-mini}}\end{subfigure}%
	\begin{subfigure}[b]{0.32\textwidth}\centering\includegraphics[width=\textwidth]{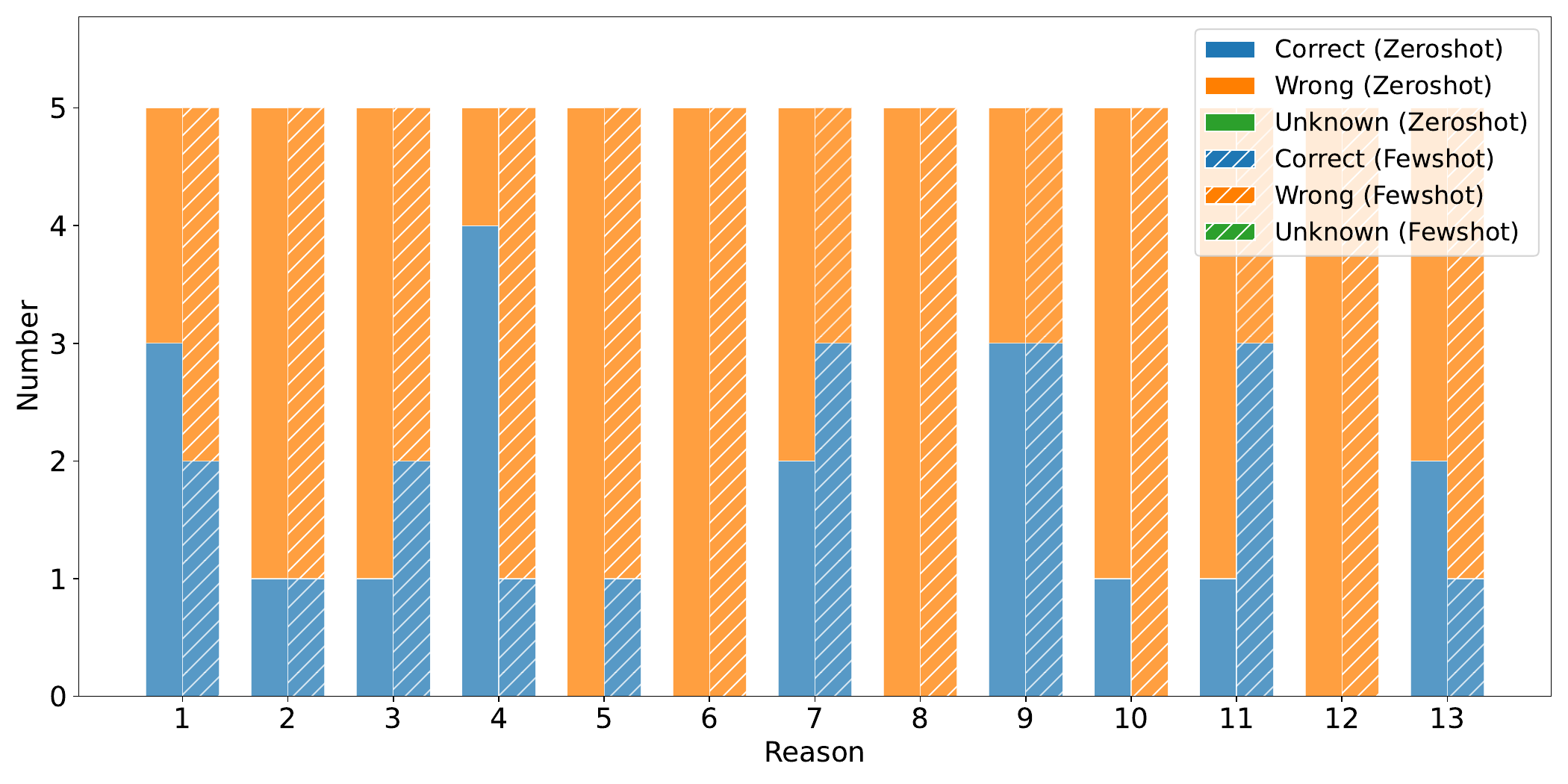}\caption*{\small{CodeLlama-70b}}\end{subfigure}%
	\begin{subfigure}[b]{0.32\textwidth}\centering\includegraphics[width=\textwidth]{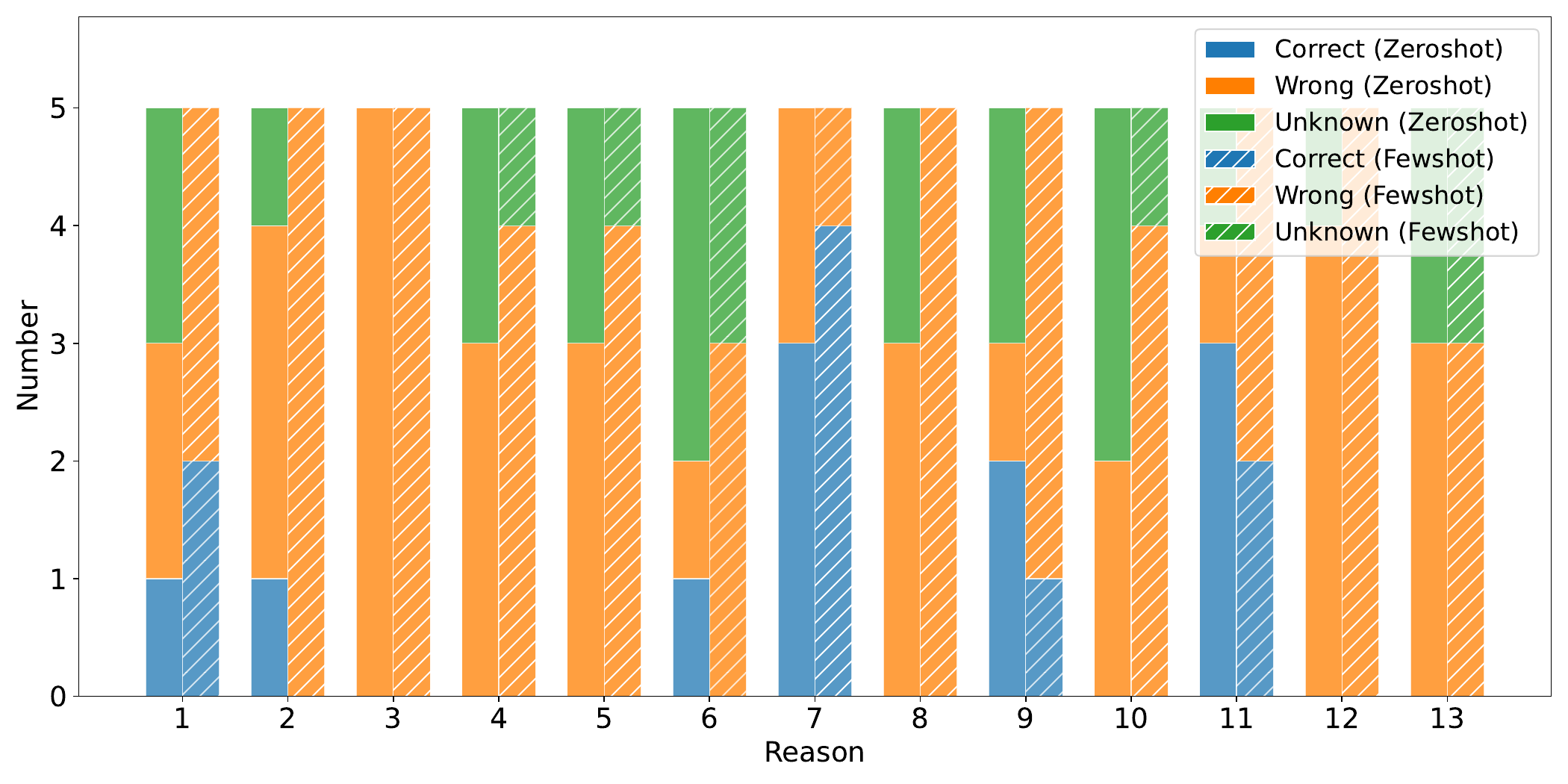}\caption*{\small{CodeLlama-13b}}\end{subfigure}%
	
	\caption{Predictions of LLMs for Flaky Test Reasoning (representative models). Each subplot compares few-shot (left bar) and zero-shot (right bar) performance across 13 flaky test categories. For readability, the x-axis uses compact numeric codes (1--13). Mapping: 1=Async~Wait, 2=Concurrency, 3=Floating, 4=IO, 5=Network, 6=Platform~Dependency, 7=Randomness, 8=Resource~Leak, 9=Test~Timeout, 10=Test~Order, 11=Time, 12=Range, 13=Unordered~Collections. Complete results for all 21 models are provided in Appendix~\ref{appendix:flaky_full}.}
	\label{fig:flaky_test}
\end{figure}

\begin{figure}[]
	\centering
	\includegraphics[width=0.9\textwidth]{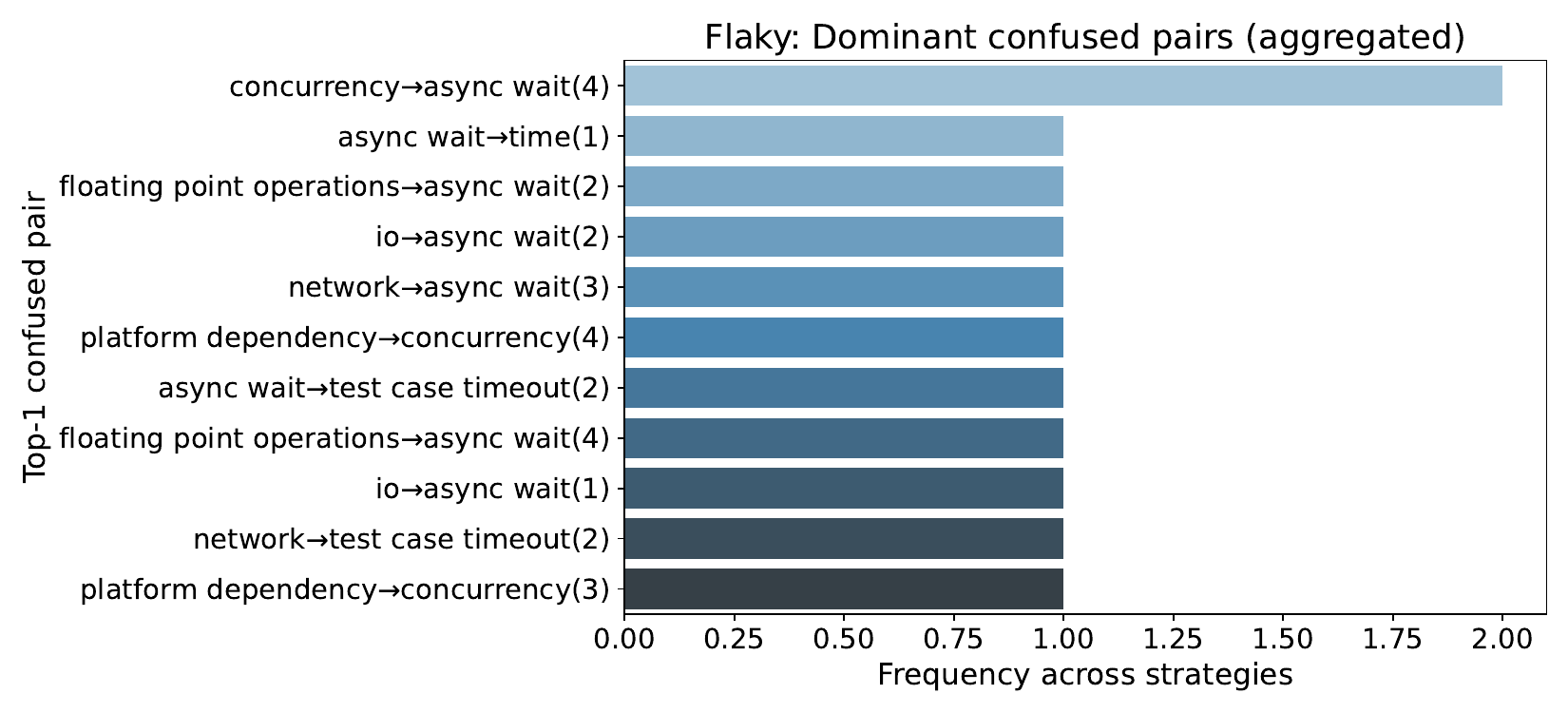}
	\caption{Flaky reasoning: aggregated dominant confused pairs across strategies.}
	\label{fig:flaky_top_confusions}
\end{figure}

\begin{figure}[t]
	\centering
	\includegraphics[width=0.7\textwidth]{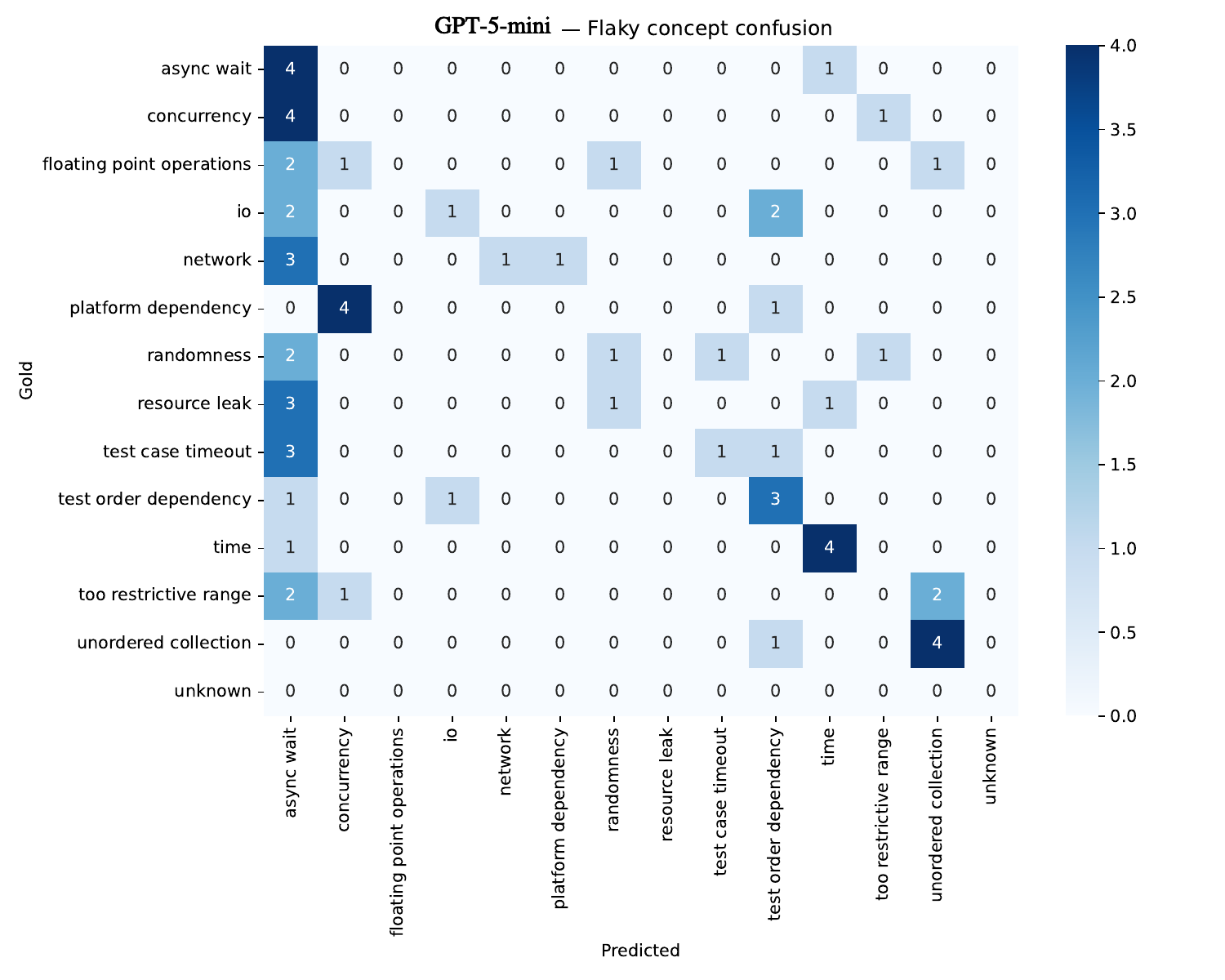}
	\caption{Flaky reasoning: confusion matrix (concept), representative model (GPT-5-mini). Full matrices for all models are included in the artifact.}
	\label{fig:flaky_confmat_gpt5mini_concept}
\end{figure}

We evaluate 13 classes with five samples each under zero/few-shot settings. Accuracy remains low across most models: on the summary split, the best-performing models reach only $\approx0.40$-$0.43$ (e.g., \textit{Claude-sonnet-4} 0.431, \textit{GPT-o3} 0.415, \textit{GPT-5} 0.400), while many models are in the 0.25-0.35 range; the concept split is even lower (top $\approx0.33$). Few-shot helps some models but can be neutral or negative for others, reflecting sensitivity to example selection.
Figure~\ref{fig:flaky_test} visualizes per-class predictions for representative models (few-shot vs zero-shot). To go beyond counts, Figure~\ref{fig:flaky_top_confusions} aggregates dominant confused pairs, and Figure~\ref{fig:flaky_confmat_gpt5mini_concept} shows a representative confusion matrix. Confusion matrices for additional models (GPT-o1-mini, GPT-4o, CodeLlama-70b, Gemini-2.5-Pro, Qwen3-coder-plus) and complete results for all 21 models are provided in Appendix~\ref{appendix:flaky_full}.

\textbf{Class-wise Error Pattern Analysis.} Systematic examination of confusion matrices reveals three dominant error patterns that explain why certain flaky test categories consistently fail:

\textit{(1) Over-attribution to ``async wait'':} This category acts as a prediction attractor, receiving mis-classifications from multiple unrelated categories. Models collapse diverse non-determinism sources into async wait, failing to distinguish three distinct types: \textit{temporal non-determinism} (timing delays, sleep operations), \textit{I/O non-determinism} (network latency, external system responses), and \textit{computational non-determinism} (floating-point precision instability). This suggests models anchor on asynchronous keywords (``async'', ``await'', ``thread'') without deeper execution-semantics reasoning.

\textit{(2) Time-adjacent confusion cluster:} Categories \texttt{time}, \texttt{test\_case\_timeout}, and \texttt{async\_wait} form a tight confusion cluster, with models frequently misattributing among them. The pattern differs by model: GPT-5-mini frequently confuses async wait $\leftrightarrow$ time (e.g., async wait$\rightarrow$time: 1 instance in concept split), while GPT-o1-mini confuses async wait $\leftrightarrow$ test case timeout (async wait$\rightarrow$test case timeout: 2 instances). This indicates difficulty in distinguishing \textit{execution-level timeout semantics} (test infrastructure killing long-running tests) from \textit{application-level timing operations} (explicit sleep/wait calls) and \textit{implicit scheduling delays} (asynchronous I/O completion).

\textit{(3) Low-recall categories:} \texttt{concurrency}, \texttt{io}, and \texttt{floating\_point\_operations} suffer from scattered predictions, with recall often near zero. Concurrency errors disperse into async wait and platform dependency predictions, revealing difficulty in recognizing multi-threading synchronization issues as distinct from asynchronous I/O patterns. Floating-point errors consistently misattribute to timing issues (async wait or time), conflating numerical precision instability with temporal non-determinism. These low-recall patterns reflect fundamental gaps in execution-semantics reasoning capability.

Statistical analysis of prompt strategy effects (Section~\ref{sec:prompt_strategy}) reveals that improvements are often modest and inconsistent. For instance, CoT on GPT-5-mini shows small accuracy gains ($\Delta$acc$=+0.062$) and F1 improvements ($\Delta$F1$=+0.073$) on concept split, but with wide confidence intervals indicating high uncertainty. Confusion matrix analysis reveals that prompt strategies primarily redistribute errors rather than resolving them: floating point$\rightarrow$async wait increases by +2 instances while concurrency$\rightarrow$async wait decreases by $-1$, demonstrating systematic but incomplete label-shift patterns without addressing fundamental semantic confusion. These patterns indicate models rely on surface lexical correlations rather than deep execution-semantics reasoning required to distinguish root causes of non-deterministic test behavior.

\noindent
\answer{3}{LLMs continue to exhibit notable shortcomings in reasoning about dynamic semantics and remain heavily affected by data distribution shifts. To fully harness their potential in real-world applications, they must leverage richer runtime information, employ more robust prompting strategies, and adopt hybrid approaches that integrate traditional tools, thereby mitigating blind spots in real-world reasoning contexts.}

\section{Discussion}
\label{sec:discussion}

We discuss our experimental results and conclusions across several aspects: (1) synthesis of capability hierarchy across tasks, (2) key findings and interpretation, (3) within-model prompt strategy comparison for dynamic tasks (RQ3), (4) agreement and validity of assessments, (5) positioning LLMs in the analysis toolchain, and (6) implications for research and practice.

\subsection{Synthesis: Capability Hierarchy Across Tasks}
\label{sec:capability_hierarchy}

Synthesizing results across all three research questions reveals a consistent \textbf{capability hierarchy}: LLMs demonstrate strong proficiency in syntax parsing (RQ1), moderate competence in static semantics analysis (RQ2), and mixed performance in dynamic reasoning (RQ3), with strong results in equivalence detection but notable challenges in execution-semantics reasoning. This pattern holds across model families, scales, and architectures, suggesting fundamental differences in how LLMs process syntactic structure versus execution semantics.

\paragraph{\textbf{Error Pattern Evolution Across RQs}}
Error characteristics shift qualitatively as task complexity increases:
\begin{itemize}
    \item RQ1 errors: Predominantly \textit{omission-based} (missing leaf nodes, trivial tokens, or formatting inconsistencies). Structural integrity is generally preserved. Errors are \textit{local} (within single statements) and \textit{recoverable} (downstream tools can infer missing details).
    \item RQ2 errors: Transition to \textit{fabrication and misalignment} (generating non-existent edges in CFG/CG, incorrect alias sets in pointer analysis, spurious dependencies in taint analysis). Errors become \textit{non-local} (affecting graph connectivity) and \textit{non-recoverable} (require human/tool verification to detect).
    \item RQ3 errors: Dominated by \textit{reasoning failures}. For flaky tests, models conflate conceptually distinct non-determinism sources, with accuracy ranging 0.25-0.43 due to severe class confusions. Errors are \textit{semantic} (not structural) and reveal gaps in execution modeling rather than syntax/graph analysis.
\end{itemize}

\paragraph{\textbf{Scale and Architecture Effects}}
Capability hierarchy persists despite model improvements:
\begin{itemize}
    \item Scale helps, but unevenly: Increasing parameters from CodeLlama-13B to CodeLlama-70B, or from GPT-4o-mini to GPT-5, yields consistent gains for RQ1 ($+10$--$15\%$ pass rate) and moderate gains for RQ2 ($+5$--$10\%$). RQ3 improvements are negligible or inconsistent ($-2\%$ to $+5\%$), suggesting limitations in current model design and training beyond parameter count.
    \item Reasoning-centric training: Models with reinforcement learning for reasoning (e.g., o1/o3/o4-series) show marginal RQ3 improvements over similarly sized baselines but do not close the gap with deterministic tools. This suggests dynamic semantics require external grounding (execution traces, symbolic constraints) beyond self-supervised text training.
    \item Prompt strategy effects on dynamic reasoning: Our within-model ablation studies on RQ3 tasks (Section~\ref{sec:prompt_strategy}) reveal significant model-strategy interactions. Chain-of-Thought degrades GPT-5-mini on mutant detection ($\Delta$acc$=-0.065$, $p<0.001$) but improves GPT-o1-mini ($\Delta$acc$=+0.230$, $p<10^{-7}$). Similarly, role-based prompting shows opposite effects across models for flaky test reasoning. These heterogeneous effects indicate that prompt strategy must be tuned per model-task combination, and that models lack stable internal representations of dynamic semantics requiring external scaffolding whose effectiveness varies by architecture.
\end{itemize}

\subsection{Key Findings and Interpretation}
\paragraph{\textbf{Model-Specific Observations}}
Deepseekchat-v3 and Claude Sonnet 4 deliver strong results on structured code analysis (e.g., AST and CFG), reflecting the benefits of scale and instruction tuning. Within the OpenAI family, newer generations (e.g., GPT-4o, GPT-5, and the o-series such as o1/o3/o4 variants) generally surpass smaller predecessors on parsing and analysis tasks, indicating gains beyond mere parameter count due to refined training and reasoning procedures. Among open-source models, CodeLlama-70B outperforms the 13B variant and earlier baselines, though the gap narrows on certain dynamic tasks where reasoning-centric training becomes more decisive. Smaller reasoning-centric models (e.g., compact o1-style variants) can be competitive on dynamic tasks against larger general models, suggesting that training objectives and reasoning procedures, beyond parameter count, play a key role for execution-semantics-oriented analysis.

\paragraph{\textbf{Discrepancy Between Static Analysis and Dynamic Reasoning}}
LLMs demonstrate strong performance on syntax parsing (90\%+ pass rates) and reasonable competence on structural graph generation (CFG 80-90\%, CG 68-72\% for best-performing models), but exhibit mixed results on execution-semantics reasoning. Equivalent mutant detection achieves high accuracy for best-performing models (F1 0.90-0.93), though weaker models remain near random (0.50). In contrast, flaky test reasoning remains challenging across all models (accuracy 0.25-0.43), even with Chain-of-Thought prompting~\cite{mywebsite}. Our RQ3 experiments further show that prompt strategy interacts with model architecture: Chain-of-Thought can stabilize certain dynamic judgments for some models while harming others (Section~\ref{sec:prompt_strategy}), reinforcing that robust dynamic analysis will likely require both reasoning scaffolds and external program evidence.

\paragraph{\textbf{Data Shift}}
In pointer analysis and taint analysis, noticeable discrepancies in F1 scores occur across different projects (or distinct data distributions), suggesting LLMs experience significant performance drops when lacking prior knowledge of the target project. Across the updated tasks, cross-project variance remains substantial for dependency- and pointer-oriented analyses, with performance often polarizing by project context. This pattern suggests that retrieval or lightweight adaptation to project conventions could mitigate data-shift effects in practice.

The magnitude of this variance carries two broader implications. First, it speaks against training data memorization as the primary driver of performance: if models were recalling stored analysis outputs, scores would not polarize so sharply across projects from the same public repositories. The evaluation targets themselves (ASTs, CFGs, call graphs, taint flows) are compositional artifacts derived through structural reasoning, not verbatim content likely to appear in training corpora; this observation, combined with the systematic error patterns documented earlier (fabricated edges, incorrect alias sets), suggests that results reflect genuine capability and its limits. Second, the variance indicates that the surrounding code context substantially influences analysis quality. Solidity projects with complex inter-contract call flows score notably lower than Java libraries dominated by shallow getter/setter patterns, with per-project F1 spanning 0--1.0 for data dependence and 0--0.8 for taint analysis. Because our evaluation uses a modular, function/module-level granularity without a controlled experiment that varies the amount of context provided to the model, this remains an observational finding. Future work could address this through (1)~a context-sensitivity ablation that progressively varies input scope (isolated function $\rightarrow$ in-file context $\rightarrow$ cross-file dependencies $\rightarrow$ full project context), and (2)~a context-window saturation study examining the performance degradation curve as input token count increases.

\subsection{Prompt Strategy Comparison (within-model)}
\label{sec:prompt_strategy}
\begin{figure}[H]
	\centering
	\begin{subfigure}[b]{0.48\textwidth}%
		\centering
		\includegraphics[width=\textwidth]{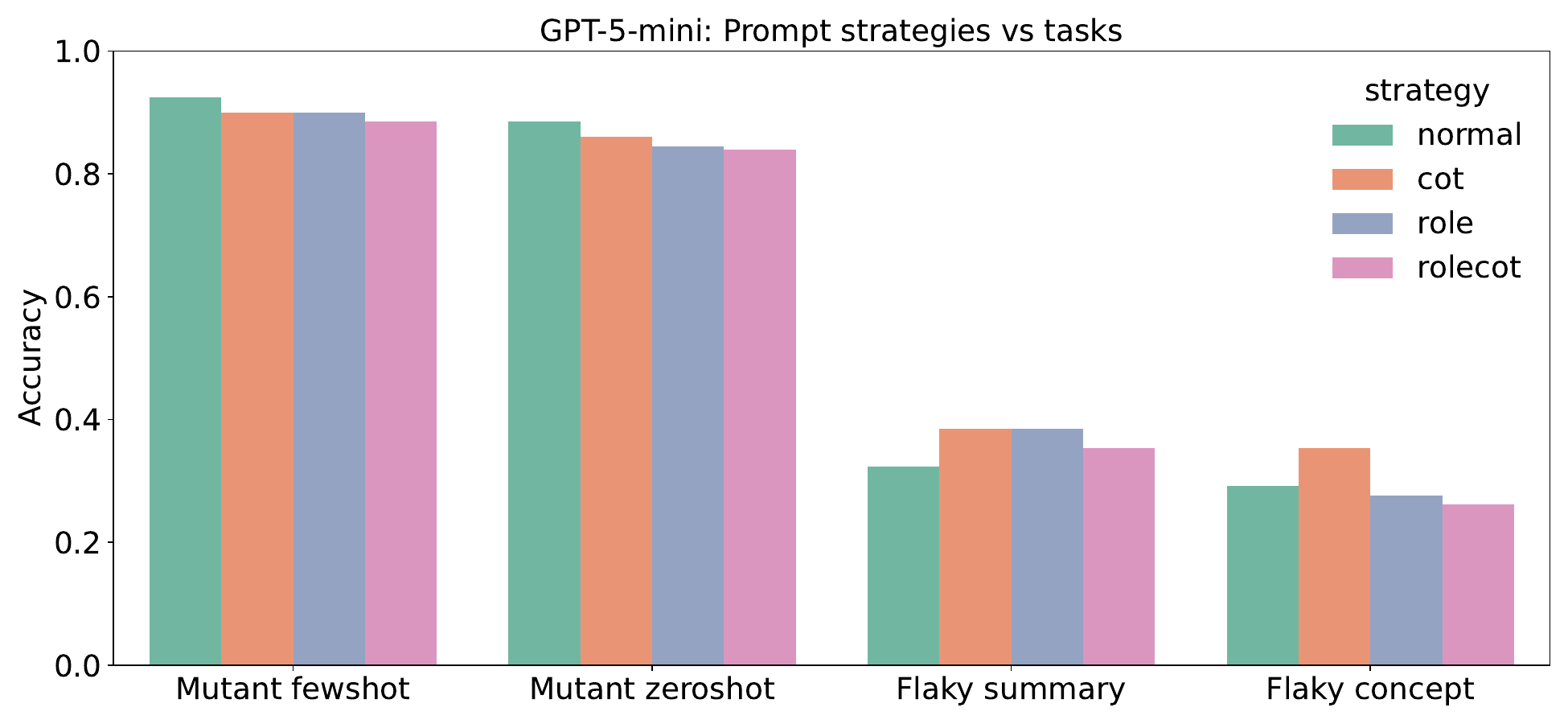}
		\caption*{GPT-5-mini}
		\label{fig:mutant_strategy_gpt5mini}
	\end{subfigure}\hfill
	\begin{subfigure}[b]{0.48\textwidth}%
		\centering
		\includegraphics[width=\textwidth]{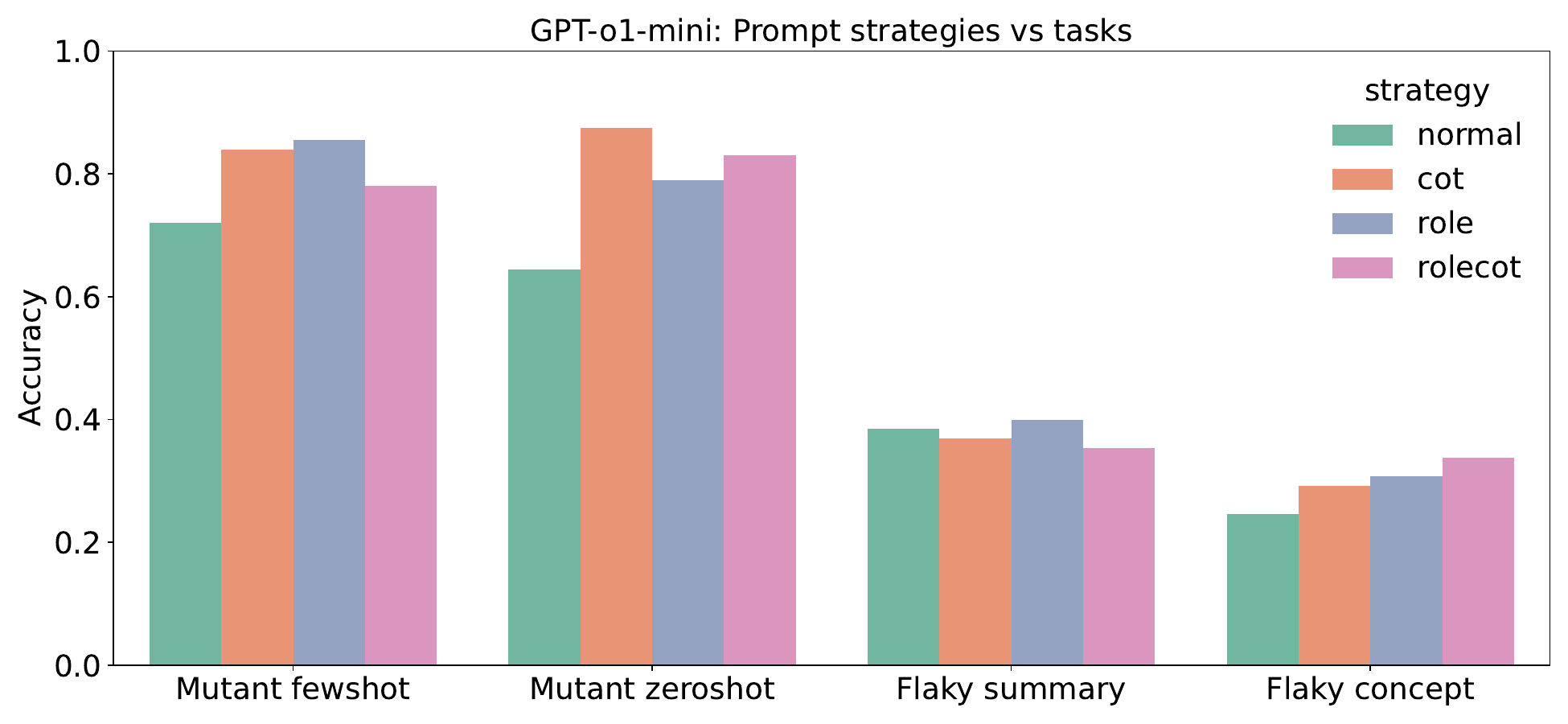}
		\caption*{GPT-o1-mini}
		\label{fig:mutant_strategy_gpto1}
	\end{subfigure}
	\caption{Within-model prompt strategies vs tasks (accuracy).}
	\label{fig:strategy_within_model}
\end{figure}

We systematically compare prompt strategies (normal, CoT, role, role+CoT) within the same model on dynamic tasks (RQ3: Equivalent Mutant Detection, Flaky Test Reasoning) to understand whether reasoning scaffolds improve performance uniformly across models. Figure~\ref{fig:strategy_within_model} reports per-task accuracies for two representative models. Statistical significance is assessed via McNemar's test for binary Mutant classification (paired contingency tables) and paired bootstrap for multi-class Flaky accuracy/F1, with 95\% confidence intervals reported. Table~\ref{tab:strategy_significance} summarizes key comparisons.

\paragraph{\textbf{GPT-5-mini.}}
\begin{itemize}
	\item Flaky (concept): CoT shows modest improvements in accuracy ($\Delta acc=+0.062$) and macro-F1 ($\Delta F1=+0.073$), though not statistically significant. Confusion-matrix analysis reveals systematic shifts: \texttt{floating point operations}$\rightarrow$\texttt{async wait} increases by 2 instances, while \texttt{concurrency}$\rightarrow$\texttt{async wait} decreases by 1, suggesting CoT clarifies temporal boundaries but struggles with arithmetic semantics.
	\item Mutant (fewshot): CoT significantly degrades accuracy ($\Delta acc=-0.065$, $p<0.001$), with confusion dominated by 1$\rightarrow$0 flips (12 instances). This indicates CoT's reasoning scaffolds induce over-conservativeness on this model, rejecting true equivalents.
\end{itemize}

\paragraph{\textbf{GPT-o1-mini.}}
\begin{itemize}
	\item Flaky (concept): role+CoT shows notable improvements in accuracy ($\Delta acc=+0.092$) and F1 ($\Delta F1=+0.085$), though with high uncertainty. Confusion analysis reveals \texttt{async wait}$\rightarrow$\texttt{test case timeout} decreases by 2 instances and \texttt{floating point operations}$\rightarrow$\texttt{time} increases by 2, reflecting finer-grained temporal attribution.
	\item Mutant (fewshot): role achieves the largest gain ($\Delta acc=+0.255$, $p<10^{-12}$), with 1$\rightarrow$0 errors dropping substantially (52 instances) and 0$\rightarrow$1 rising slightly (1 instance). This suggests role-based scaffolding aligns with GPT-o1-mini's reasoning architecture, unlike GPT-5-mini.
\end{itemize}

Strategy--model interactions are significant. CoT/role benefit attribution tasks (Flaky) across models but have opposite effects on binary equivalence detection (Mutant): harmful for GPT-5-mini, beneficial for GPT-o1-mini. This heterogeneity underscores the need for model-specific prompt tuning and cautions against universal prompting recipes in dynamic reasoning tasks.

\begin{table*}[h]
	\centering
	\caption{Prompt strategy effects within models (relative to normal baseline). Mutant: accuracy change with McNemar test; Flaky: accuracy and macro-F1 change with bootstrap 95\% confidence intervals.}
	\label{tab:strategy_significance}
	\scalebox{0.75}{
		\begin{tabular}{ll|cc|cc}
			\toprule
			Model & Strategy & Mutant $\Delta$acc & $p$-value & Flaky $\Delta$acc [95\%CI] & Flaky $\Delta$F1 [95\%CI] \\
			\midrule
			\multirow{3}{*}{GPT-5-mini} & CoT & $-0.065$ & $<0.001$ & $+0.062$ [$-0.046$, $+0.169$] & $+0.073$ [$-0.019$, $+0.171$] \\
			& role & $-0.025$ & $0.063$ & $-0.015$ [$-0.123$, $+0.092$] & $+0.005$ [$-0.093$, $+0.088$] \\
			& role+CoT & $-0.040$ & $0.008$ & $-0.031$ [$-0.108$, $+0.062$] & $-0.040$ [$-0.115$, $+0.032$] \\
			\midrule
			\multirow{3}{*}{GPT-o1-mini} & CoT & $+0.230$ & $<10^{-7}$ & $+0.046$ [$-0.062$, $+0.154$] & $+0.034$ [$-0.069$, $+0.142$] \\
			& role & $+0.255$ & $<10^{-12}$ & $+0.062$ [$-0.062$, $+0.169$] & $+0.062$ [$-0.065$, $+0.173$] \\
			& role+CoT & $+0.180$ & $<10^{-6}$ & $+0.092$ [$-0.015$, $+0.200$] & $+0.085$ [$-0.035$, $+0.192$] \\
			\bottomrule
	\end{tabular}}
\end{table*}

\subsection{Agreement and Validity of Assessments}
\label{sec:agreement}
To validate our evaluation methodology, we conduct comprehensive reliability analysis at two levels: (i)~human--human consistency when labeling pass/fail for structural tasks (AST/CFG/CG), and (ii)~human--machine consistency between expert consensus and automated indicators. This analysis demonstrates that our formal evaluation rubrics (detailed in Section~\ref{sec:evaluation}) achieve high human agreement and quantifies the extent to which automated metrics approximate expert judgment, thereby establishing the reliability bounds of our assessment framework.

\paragraph{\textbf{Inter-Rater Reliability (Human--Human Agreement).}}
We compute agreement between two independent reviewers (r1, r2) using a suite of coefficients to ensure robustness against class imbalance and marginal distribution artifacts. 
We report observed agreement ($P_o$, the proportion of samples where reviewers agree), Cohen's $\kappa$, Gwet's AC1, and PABAK. As shown in Table~\ref{tab:consistency_struct}, the three structural tasks 
achieve high agreement under multiple indices, with observed agreement 
$P_o$ = 0.944--0.982, Cohen's $\kappa$ ranging 0.844-0.936, and Gwet's AC1 
ranging 0.914-0.975. These values exceed the standard threshold ($\kappa \geq 0.81$ per Landis \& Koch) and demonstrate that our formal rubrics (Section~\ref{sec:evaluation}) operationalize semantic reconstructability into testable, reliable criteria. Gwet's AC1, which is robust to prevalence effects, corroborates Cohen's $\kappa$ by showing even higher agreement, confirming that the evaluation protocol is stable across varying pass/fail distributions.

\begin{table*}[h]
\centering
\caption{Structural-task agreement (human--human and human--machine) under multiple reliability indices. Human--human columns establish evaluation reliability bounds; human--machine columns validate automated metrics against expert consensus.}
\label{tab:consistency_struct}
\scalebox{0.95}{
\begin{tabular}{l|cccc|ccccc}
\toprule
\multirow{2}{*}{Task} & \multicolumn{4}{c|}{\textbf{Human--Human (r1 vs r2)}} & \multicolumn{5}{c}{\textbf{Human--Machine (consensus vs auto)}}\\
 & $P_o$ & $\kappa$ & AC1 & PABAK & $P_o$ & $\kappa$ & AC1 & PABAK & F1 \\
\midrule
AST & 0.982 & 0.936 & 0.975 & 0.964 & 0.804 & 0.311 & 0.727 & 0.609 & 0.882 \\
CFG & 0.949 & 0.858 & 0.920 & 0.898 & 0.780 & 0.376 & 0.661 & 0.560 & 0.858 \\
CG  & 0.944 & 0.844 & 0.914 & 0.889 & 0.698 & 0.253 & 0.497 & 0.396 & 0.791 \\
\bottomrule
\end{tabular}}
\end{table*}

\paragraph{\textbf{Human--Machine Agreement: Multi-Metric Validation.}}

We compare automated Layer~1 judgments against human consensus (formed via majority vote from r1, r2; disagreements adjudicated by r3). Table~\ref{tab:consistency_struct} reveals a critical methodological insight: Cohen's $\kappa$ alone provides incomplete assessment of human-machine agreement. The observed $\kappa = 0.25$--$0.38$ reflects prevalence effects inherent in highly imbalanced data (70--80\% pass rate), a well-documented phenomenon where Cohen's $\kappa$ is systematically suppressed despite high observed agreement~\cite{gwet2008computing,feinstein1990high}. This underscores the necessity of prevalence-corrected metrics for accurate agreement assessment.

Prevalence-corrected metrics reveal the true agreement: Gwet's AC1 ($0.50$--$0.73$) and PABAK ($0.40$--$0.61$) demonstrate \textit{substantial} concordance for AST/CFG and \textit{moderate} concordance for CG. Combined with high observed accuracy ($P_o = 0.70$--$0.80$) and F1 ($0.79$--$0.88$), this validates that automated metrics approximate expert judgment with meaningful accuracy. The AST/CFG results (AC1 $> 0.65$) indicate \textit{substantial} agreement per Landis \& Koch standards; CG's moderate AC1 ($0.497$) reflects genuine task complexity where multiple valid factorizations exist, not metric failure.

This multi-metric validation exemplifies statistical best practices in agreement analysis: relying solely on Cohen's $\kappa$ would yield misleading conclusions in imbalanced settings; the combination of $\kappa$, AC1, PABAK, and $P_o$ provides robust assessment across varying prevalence~\cite{gwet2008computing}. The pattern, high human--human agreement ($\kappa \geq 0.84$) versus substantial human--machine AC1 ($\approx 0.50$--$0.73$), quantifies the gap between automated approximations and expert judgment, validating our three-layer protocol: automated metrics enable efficient screening at scale (Layer 1), while human adjudication captures semantic nuances (Layer 2).

\paragraph{\textbf{Model Pass Rates Under Human Consensus.}}
Table~\ref{tab:model_pass_consensus} reports pass rates for six representative models under human consensus labels (r1, r2 majority; disagreements excluded). Top commercial models demonstrate strong structural generation: GPT-5-codex achieves 97.3\% on AST/CFG and 96.0\% on CG; Gemini-2.5-Pro reaches 94.7--96.0\% on AST/CFG and 92.0\% on CG; Claude-Sonnet-4 attains 92.0--96.0\% on AST/CFG and 85.3\% on CG. Mid-tier models (CodeLlama-70b: 73.3--82.7\%; GPT-4o-mini: 68.0--77.3\%) show competent but less consistent performance. Earlier baselines (CodeLlama-13b: 21.3--49.3\%) reveal substantial capability gaps. These results, validated through high inter-rater reliability ($\kappa \geq 0.84$), establish empirical performance bounds free from subjective bias.

\begin{table*}[h]
\centering
\caption{Model pass rates under human consensus (r1, r2 majority; disagreements adjudicated by r3). Only samples with definitive consensus are included.}
\label{tab:model_pass_consensus}
\scalebox{0.8}{
\begin{tabular}{l|cc|cc|cc}
\toprule
Model & AST Pass/Total & AST Rate & CFG Pass/Total & CFG Rate & CG Pass/Total & CG Rate \\
\midrule
Claude-sonnet-4 & 72/75 & 0.960 & 69/75 & 0.920 & 64/75 & 0.853 \\
CodeLlama-13b & 37/75 & 0.493 & 16/75 & 0.213 & 37/75 & 0.493 \\
CodeLlama-70b & 62/75 & 0.827 & 57/75 & 0.760 & 54/75 & 0.733 \\
Gemini-2.5-Pro & 71/75 & 0.947 & 72/75 & 0.960 & 69/75 & 0.920 \\
GPT-4o-mini & 58/75 & 0.773 & 51/75 & 0.680 & 52/75 & 0.693 \\
GPT-5-codex & 73/75 & 0.973 & 73/75 & 0.973 & 72/75 & 0.960 \\
\bottomrule
\end{tabular}}
\end{table*}

\subsection{Positioning LLMs in the Analysis Toolchain}
\label{sec:positioning_llms}

\paragraph{\textbf{Task-Specific Applicability and Deployment Guidance}}
Our evaluation reveals three distinct applicability tiers based on error characteristics and consistency. Unlike traditional analyzers that provide formal guarantees, LLMs operate heuristically with systematic error patterns: fabricated edges in graph generation, high cross-project variance in dependency analysis, and semantic confusion in execution reasoning. These characteristics necessitate task-specific deployment strategies:
\begin{itemize}
\item \textit{Suitable for assisted workflows with validation}: Syntax parsing tasks (AST, expression matching) achieve high pass rates (best-performing models: AST 90\%+, expression matching 84-100\% Hit@10) with predominantly local and recoverable errors, requiring spot-checking.
\item \textit{Experimental/exploratory use only}: Structural graph generation (CFG 80-90\%, CG 68-72\% for best-performing models) requires case-by-case validation due to non-local fabrication errors. Suitable for research prototypes but not production pipelines without human oversight.
\item \textit{Not yet viable}: Dependency analysis (pointer Jaccard 0.29-0.55, taint F1 0.28-0.76) and flaky test reasoning (accuracy 0.25-0.43) fall below practical utility thresholds due to high cross-project variance and semantic confusion. Equivalent mutant detection shows promise for best-performing models (F1 0.90-0.93) but remains inconsistent across model families.
\end{itemize}

\paragraph{\textbf{Hybrid Deployment Patterns}}
Given these applicability tiers, we identify three effective hybrid patterns that leverage LLM strengths while mitigating limitations:
\begin{enumerate}
\item \textit{Triage-and-validation}: LLM pre-screens code for suspicious patterns; traditional analyzer performs expensive inter-procedural analysis only on flagged regions, reducing analysis cost while maintaining precision. Our results suggest this pattern is better suited for local, intra-procedural pattern screening, where LLMs show stronger performance (Section~\ref{sec:capability_hierarchy}). For vulnerabilities that span multiple files or procedures (e.g., a taint source in one module reaching a sink in another), the LLM-based triage step alone is unlikely to capture the full cross-file context needed. In such cases, the pattern would need to be augmented with retrieval-based context assembly or iterative multi-step analysis to gather relevant cross-module information before the LLM screening step.
\item \textit{Explanation synthesis}: Static analyzer extracts structures (e.g., call graphs, taint flows); LLM generates human-readable summaries for code reviews; tool re-validates critical findings.
\item \textit{Incomplete-code handling}: For syntactically invalid code (IDE suggestions, documentation snippets), LLM infers intended structure; once code compiles, switch to deterministic tools for correctness checks.
\end{enumerate}

These patterns recognize that LLMs are better suited for scenarios requiring flexibility and context-awareness (explaining results, handling malformed inputs) rather than as replacements for deterministic analyzers in safety-critical contexts.

\subsection{Implications for Research and Practice}
\label{sec:implications}

\noindent{\textbf{Practice.}}
Deploy LLMs based on error characteristics: syntax parsing (local, recoverable errors) suits assisted workflows with spot-checking; structural generation (fabrication errors) requires case-by-case validation; dependency analysis and flaky test reasoning (high cross-project variance, semantic confusion) need deterministic verification. For safety-critical applications, pair LLM outputs with formal tools (Section~\ref{sec:positioning_llms}).

Our within-model ablation (Section~\ref{sec:prompt_strategy}) reveals model-dependent prompt effects: Chain-of-Thought degrades GPT-5-mini ($\Delta$acc=$-6.5\%$, $p<0.001$) but improves GPT-o1-mini ($\Delta$acc=$+23.0\%$, $p<10^{-7}$) on mutant detection. Practitioners must validate prompt strategies per model-task combination rather than applying universal recipes.

\noindent\textbf{Code Generation Limitations.}
While our work evaluates code \textit{analysis}, the capability hierarchy (strong syntax, weak execution semantics) suggests caution for LLM-generated code involving concurrency, state management, or security boundaries. The execution-semantics reasoning gaps observed in flaky test diagnosis may manifest as subtle bugs in generated control flow or data handling. Code review and testing remain essential for safety-critical components.

\noindent\textbf{Research Directions.}
Flaky test reasoning remains challenging across all models, revealing fundamental gaps in non-determinism attribution. Promising approaches: (1)~Hybrid training with execution traces or symbolic constraints to ground runtime behavior; (2)~LLM-tool co-reasoning where models generate hypotheses and symbolic engines verify them; (3)~Domain-specific reasoning protocols decomposing complex tasks into verifiable sub-steps.

\section{Threats to Validity}
\label{sec:threats_to_validity}

\paragraph{Internal Validity}
Our three-layer evaluation protocol (automated metrics, expert adjudication, consistency validation; Section~\ref{sec:evaluation}) achieved high inter-rater reliability for structural tasks. However, subjective judgment in edge cases (e.g., semantically valid alternative graph factorizations) remains a potential source of bias despite formal rubrics and adjudication protocols.

Manually designed prompts may not fully exploit models' potential compared to automated prompt tuning~\cite{shin2020autoprompt} or retrieval-augmented generation~\cite{lewis2020retrieval}. We optimized prompts using established practices~\cite{awesome} and validated through within-model strategy comparison (Section~\ref{sec:prompt_strategy}) that core findings persist across prompt variations (normal, CoT, role-based).

LLM outputs exhibit non-unified formatting (node naming, indentation, structural representations). We applied normalization pipelines (Section~\ref{sec:evaluation}) and standardized rubrics to ensure fair comparison, with remaining ambiguities resolved through expert consensus.

\paragraph{External Validity}
Our dataset comprises 3,124 samples across nine tasks and four languages. While manual validation constrains absolute scale, task diversity (syntax, static semantics, dynamic reasoning) and tool-generated references (Tree-sitter, Soot, Joern, Slither, Frama-C) provide systematic coverage. However, advanced analysis tasks (e.g., inter-procedural symbolic execution, shape analysis) remain uncovered, limiting generalizability.

Analysis tasks employ 38 real-world projects (100,947--78,244 LOC; Table~\ref{tab:task_datasets}), while structural tasks use compact programs (1,059--2,356 LOC) constrained by expert validation feasibility and contamination control requirements. This limits insights about performance on extremely large functions ($>$1,000 LOC with deep nesting), though the modular design reflects practical piecewise analysis workflows. While the modular evaluation granularity reflects certain real-world SE practices such as per-function analysis in IDEs and file-level static checks, it does not validate LLM behavior under larger analysis scopes such as whole-program taint tracking or cross-module call graph construction. Our study does not include a controlled experiment that systematically varies the code context provided to the model; we discuss indirect evidence and future directions in Section~\ref{sec:discussion}.

LLMs trained on public repositories may have encountered similar code patterns, potentially inflating performance. We mitigated this through tool-generated datasets (MutantBench~\cite{9440157}, FlakyCat~\cite{akli2022predicting}) not directly in training corpora and cross-project generalization evaluation. Error pattern analysis (hallucinations, fabrications) further validates capability gaps beyond memorization. Nevertheless, complete elimination is infeasible for black-box models. Evaluating on proprietary or previously non-public codebases, combined with membership inference techniques~\cite{shi2024detecting}, would strengthen contamination control and is an important direction for future work.

\paragraph{Construct Validity}
Our metrics emphasize practical usability (e.g., reconstructability for AST/CFG/CG targets structural completeness for downstream analyses, not surface-level formatting). Validated through inter-rater reliability (Section~\ref{sec:agreement}), metrics align with expert judgment. However, alternative designs (e.g., graph edit distance, execution trace similarity) might yield different rankings, particularly for dynamic tasks with inherently approximate ground truth.

Proprietary models' inaccessibility precludes weight-based or attention-pattern analyses. Our behavioral evaluation (inferring capabilities from input-output mappings) limits mechanistic insights (e.g., whether CFG errors stem from training data bias or architectural constraints). Open-weight models would enable deeper causal analysis.

\section{Related Work}
\label{sec:related_work}

\subsection{Code Model Capabilities}

\paragraph{Probing Code Model Internals}

Probing analysis~\cite{rogers2021primer} examines what linguistic or structural knowledge neural models encode internally. For natural language, Clark et al.~\cite{clark2019does} found that BERT~\cite{devlin2018bert} encodes syntactic structure in attention heads; Hewitt \& Manning~\cite{hewitt-manning-2019-structural} proposed structural probing, showing embeddings represent parse trees. For code, Wan et al.~\cite{wan2022they} and Hern{\'a}ndez et al.~\cite{hernandez2022ast} probe how CodeBERT~\cite{feng2020codebert}, CodeT5~\cite{wang2021codet5}, and UnixCoder~\cite{guo2022unixcoder} learn syntax; Troshin \& Chirkova~\cite{troshin-chirkova-2022-probing} and Wei et al.~\cite{10.1145/3664606} examine semantic representations. Kou et al.~\cite{kou2024large} study attention pattern similarities between programmers and LLMs during code reading. These studies focus on internal representations. In contrast, we evaluate end-to-end generation capabilities: whether LLMs can produce correct ASTs, CFGs, and analysis results in zero-shot scenarios, spanning syntax, static semantics, and dynamic reasoning.

\paragraph{LLM Capabilities on SE Tasks}

LLMs have been evaluated on diverse SE tasks. Chen et al.~\cite{chen2021evaluating} introduced HumanEval for code generation; Austin et al.~\cite{austin2021program} studied program synthesis via verification-guided decoding. Du et al.~\cite{du2024classeval} proposed ClassEval for class-level generation, finding models perform worse on interdependent methods. Izadi et al.~\cite{izadi2024practical} evaluated code completion using real IDE usage data, finding offline benchmarks poorly correlate with production performance. For code repair, Xia \& Zhang~\cite{xia2023keep} proposed dialogue-based APR; Sobania et al.~\cite{sobania2023analysis} analyzed ChatGPT's bug-fixing strategies. For code understanding, Tian et al.~\cite{tian2023chatgpt} evaluated ChatGPT on generation, repair, and summarization. Jin et al.~\cite{jin2024llms}, Hou et al.~\cite{10.1145/3695988}, Fan et al.~\cite{fan2023large}, and Nguyen \& Nguyen~\cite{nguyen2023generative} survey LLMs' SE applications and impacts.

These studies measure downstream task performance. We focus on foundational program analysis capabilities, including AST/CFG/CG generation, data flow, and dynamic reasoning, which underpin such tasks, revealing a capability hierarchy across model families.

\subsection{Traditional Program Analysis and Hybrid Approaches}

Static analysis has a rich history in PL/SE research. Symbolic analyzers~\cite{cousot1977abstract} provide formal guarantees but require whole-program visibility and struggle with dynamic features. Flow-sensitive analyses~\cite{smaragdakis2015pointer} achieve high precision at exponential cost; flow-insensitive variants~\cite{steensgaard1996points} scale better but sacrifice accuracy. Taint analysis~\cite{kim2014survey} tracks data-flow security but faces path-explosion problems. Dynamic analysis~\cite{ERNST200735} observes runtime behavior but requires test suites. Hybrid approaches~\cite{godefroid2008automated} combine symbolic execution with concrete testing.
Recent work combines LLMs with traditional tools. Wadhwa et al.~\cite{wadhwa2024core} propose CORE, partnering LLMs with static analyzers (CodeQL, SonarQube) to fix code quality issues. Li et al.~\cite{li2025llm-symexec} integrate LLMs with symbolic execution using path-based decomposition. Zhang et al.~\cite{zhang2025dfuzz} propose DFUZZ, an LLM-driven fuzzing approach for bug detection. These hybrid approaches leverage complementary strengths: traditional tools provide formal guarantees but require language-specific configuration; LLMs offer cross-language generalization but produce heuristic outputs requiring validation.

Our evaluation characterizes which tasks suit LLM standalone use, which require validation, and which necessitate hybrid approaches, informing deployment strategies in Section~\ref{sec:positioning_llms}.

\subsection{Evaluation Methodologies}

Empirical studies have assessed code models on various tasks. Lu et al.~\cite{lu2021codexglue} introduced CodeXGLUE for code completion and translation; Ren et al.~\cite{ren2020codebleu} proposed CodeBLEU for code quality evaluation. Chen et al.~\cite{chen2021evaluating} benchmarked Codex on HumanEval; Li et al.~\cite{li2022competition} extended evaluation to competitive programming (APPS).
Recent work emphasizes rigorous testing. Liu et al.~\cite{liuxw2023evalplus} augmented HumanEval with more test cases, revealing accuracy drops and changing model rankings. Du et al.~\cite{du2024classeval} showed function-level evaluations may not reflect real-world complexity involving interdependent methods. Steenhoek et al.~\cite{steenhoek2024dataflow} study dataflow analysis; VulnRepairEval~\cite{wang2025vulnrepaireval} evaluates vulnerability detection for LLM.

We evaluate nine program analysis tasks across four languages with a three-layer framework combining automated metrics, expert adjudication, and consistency validation using prevalence-corrected metrics. This reveals a capability hierarchy across model families that informs hybrid deployment strategies.

\section{Conclusion}
\label{sec:conclusion}

We evaluated 21 LLMs on nine program analysis tasks spanning syntax parsing, static semantics, and dynamic reasoning, using 3,124 code samples across four languages. Our three-layer evaluation protocol achieves high inter-rater reliability (Cohen's $\kappa = 0.844$--$0.936$) and substantial human-machine agreement (AC1 = 0.500--0.727). The evaluation framework, prompts, and datasets are released for reproducible assessment.
Our findings reveal a consistent capability hierarchy across model families and scales: (1)~Strong syntax parsing: best-performing models achieve 90\%+ pass rates (AST) and 84-100\% Hit@10 (expression matching); (2)~Moderate static analysis competence: CFG/CG generation reaches 80-90\% and 68-72\% respectively, but with fabrication errors requiring validation; (3)~Mixed dynamic reasoning: equivalent mutant detection achieves F1 0.90-0.93 for best-performing models, while flaky test reasoning remains challenging (accuracy 0.25-0.43). 
Within-model ablation studies on dynamic reasoning tasks reveal significant model-strategy interactions: Chain-of-Thought degrades GPT-5-mini on mutant detection ($\Delta$acc$=-0.065$, $p<0.001$) but improves GPT-o1-mini ($\Delta$acc$=+0.230$, $p<10^{-7}$), indicating that prompt strategy must be tuned per model-task combination.
These findings inform deployment strategies: syntax parsing achieves high accuracy suitable for assisted workflows with spot-checking; structural analysis suits research prototypes with validation; dynamic reasoning and dependency analysis require hybrid approaches combining LLM flexibility with deterministic tool verification. Three effective patterns are triage-and-validation, explanation synthesis, and incomplete-code handling.
Future work should address: (1)~Bridging the execution-semantics gap through hybrid training with execution traces or LLM-tool co-reasoning; (2)~Mitigating cross-project variance via retrieval or project-specific adaptation; (3)~Investigating whether architectural changes (e.g., explicit state tracking) are necessary beyond scaling.

\begin{acks}
This research is supported by the Ministry of Education, Singapore under its Academic Research Fund Tier 3 (Award ID: MOET32020-0004). Any opinions, findings and conclusions or recommendations expressed in this material are those of the author(s) and do not reflect the views of the Ministry of Education, Singapore.
Shangqing Liu and Qiang Hu are the co-corresponding authors of this paper.
\end{acks}

\bibliographystyle{ACM-Reference-Format}
\bibliography{sample-base}

\appendix

\section{Complete Prompt Templates}
\label{appendix:prompts}

This appendix provides the complete prompt templates used in our experiments, as described in Section~\ref{sec:prompt}. All prompts follow either role-based or instruction-based patterns with explicit placeholder definitions. These templates enable exact reproduction of our experimental setup.

\subsection{AST Generation (Role-based)}
\label{appendix:prompt_ast}

\begin{tcolorbox}[colback=gray!5,colframe=gray!40,breakable]
\small
\textbf{Prompt:}

You are a [LANG] Abstract Syntax Tree (AST) parser. I will give you a [LANG] code file. You give me its AST in Json format. Each AST node only has three attributes, children, type and value.

The input file is
```
[INPUT\_CODE]
```
\end{tcolorbox}

\noindent Instantiation: [LANG] $\in$ \{C, Python, Java, Solidity\}; [INPUT\_CODE] is the source code to analyze.

\subsection{Expression Matching (Role-based)}
\label{appendix:prompt_expression}

\begin{tcolorbox}[colback=gray!5,colframe=gray!40,breakable]
\small
\textbf{Prompt:}

You are an AI trained to detect similar code expressions. Given a Smart Contract code and a specific target code expression, your task is to find and list the most similar expressions within the provided Smart Contract code. I will show you the answer format and then please analyze the new input following code file and search for expressions that closely resemble the target code piece provided.

```
\{``Answer'':``Yes'' or ``No'', ``similar\_expressions'': [
   \{
     ``function\_name'': the matched function name,
     ``line\_number'': line\_number,
     ``expression'': the similar code
   \}
 ]
 ``Reason'': your reason
 \}
```

Input Smart Contract Code:
```Solidity
[INPUT\_CODE]
```

Input Specific Target Code Expression:
```Target Expression
[INPUT\_EXPRESSION] 
```

Please identify the similar expressions, their corresponding function name and their corresponding line numbers in the code file. You also need to replace the function calls ``add'', ``sub'', ``div'', ``mul'', ``divCeil'' in the found similar expressions with ``+'', ``-'', ``/'', and ``*''. Put your results in JSON format at the beginning.
\end{tcolorbox}

\noindent Instantiation: [INPUT\_CODE] is the Solidity smart contract; [INPUT\_EXPRESSION] is the target expression from whitepaper.

\subsection{Control Flow Graph (Role-based)}
\label{appendix:prompt_cfg}

\begin{tcolorbox}[colback=gray!5,colframe=gray!40,breakable]
\small
\textbf{Prompt:}

You are a control flow graph analyzer for [LANG]. I will give you a [LANG] program and you tell me its control flow graph. The output format is json file, including nodes and edges.

The input file is 
```
[INPUT\_CODE]
```
\end{tcolorbox}

\noindent Instantiation: [LANG] $\in$ \{C, Python, Java, Solidity\}; [INPUT\_CODE] is the source code to analyze.

\subsection{Call Graph (Role-based)}
\label{appendix:prompt_cg}

\begin{tcolorbox}[colback=gray!5,colframe=gray!40,breakable]
\small
\textbf{Prompt:}

You are a call graph analyzer for [LANG]. I will give you a [LANG] program and you tell me its call graph. The output format is json file, including nodes and edges.

The input file is 
```
[INPUT\_CODE]
```
\end{tcolorbox}

\noindent Instantiation: [LANG] $\in$ \{C, Python, Java, Solidity\}; [INPUT\_CODE] is the source code to analyze.

\subsection{Data Dependency Analysis (Role-based)}
\label{appendix:prompt_data_dep}

\begin{tcolorbox}[colback=gray!5,colframe=gray!40,breakable]
\small
\textbf{Prompt:}

You are a helpful code program analysis tool for Smart Contract. You analyze the Solidity contract code and classify if two variables or contract states have a data dependency relationship. The labels you use are `yes', `no' and `unknown'. `yes' means they are data dependent. `no' means they are not data dependent. Otherwise, they are labelled `unknown'. You first give the label and then explain the reason.

The code is 
```
[INPUT\_CODE]
```.

You first give the label and then explain the reason. Please answer the following question: is the variable [VAR\_NAME] in the function [FUNCTION\_NAME] data depended on the variable [VAR\_NAME] in the function [FUNCTION\_NAME]?
\end{tcolorbox}

\noindent Instantiation: [INPUT\_CODE] is the Solidity or Java code; [VAR\_NAME] and [FUNCTION\_NAME] are the specific variables and functions to analyze.

\subsection{Taint Analysis (Role-based)}
\label{appendix:prompt_taint}

\begin{tcolorbox}[colback=gray!5,colframe=gray!40,breakable]
\small
\textbf{Prompt:}

You are a helpful code program analysis tool for Smart Contract. You analyze the Solidity contract code and classify if the variable or contract state is controlled by the user. The labels you use are `yes', `no' and `unknown'. `yes' means it is controlled by the user. `no' means it is not controlled by the user. Otherwise, it is labelled `unknown'. You first give the label and then explain the reason.

The code is 
```
[INPUT\_CODE]
```

You first give the label and then explain the reason. Please answer the following question: is the variable [VAR\_NAME] in the function [FUNCTION\_NAME] controlled by the user?
\end{tcolorbox}

\noindent Instantiation: [INPUT\_CODE] is the Solidity or Java code; [VAR\_NAME] and [FUNCTION\_NAME] specify the target variable.

\subsection{Pointer Analysis (Role-based)}
\label{appendix:prompt_pointer}

\begin{tcolorbox}[colback=gray!5,colframe=gray!40,breakable]
\small
\textbf{Prompt:}

You are a pointer analysis tool for C programs. I will provide a C file to you and you do the pointer analysis about it. You analyze what variables the pointers points to in the provided code. The code is
```
[INPUT\_CODE]
```

Please provide you answer in Json format that includes the list of the variable names each pointer points to:
\end{tcolorbox}

\noindent Instantiation: [INPUT\_CODE] is the C source code to analyze.

\subsection{Equivalent Mutant Detection (Instruction-based)}
\label{appendix:prompt_mutant}

\begin{tcolorbox}[colback=gray!5,colframe=gray!40,breakable]
\small
\textbf{Prompt:}

Please analyze the two following provided code files in C or Java. Identify if they are semantically equal. `Semantically equal' means two codes have the same meaning, that they have the same output given the same input. Here are three semantically equal examples:

The first example pair is
``` Code 1
double f(double M, double x) \{
x = (M + x) / 2;
return x;
\}
```
``` Mutant Code 1
double f(double M, double x) \{
x = (M + x++ ) / 2;
return x;
\}
```

Yes. The two codes are semantically equal because `M + x++' first does `M + x' and then `x++'. Therefore, `(M + x) / 2' is the same with `(M + x++) / 2'.

Please identify if the two following codes are semantically equal. Please only answer `yes' or `no'. `yes' means they are semantically equal. `no' means they are not.

Input:
```Code[INPUT\_CODE]```
\end{tcolorbox}

\noindent Instantiation: [INPUT\_CODE] contains both the original code and mutant code.

\subsection{Flaky Test Reasoning (Instruction-based)}
\label{appendix:prompt_flaky}

\begin{tcolorbox}[colback=gray!5,colframe=gray!40,breakable]
\small
\textbf{Prompt:}

Please analyze the following provided test code in Java. Identify the reason why it is flaky test. `Flaky test' means one test sometimes pass and sometimes fails. There are 13 reasons about the flaky test.

Here are the definitions of 13 flaky test reasons: 

Reason 1, async wait. We classify a commit into the Async Wait category when the test execution makes an asynchronous call and does not properly wait for the result of the call to become available before using it.

Reason 2, test order dependency. We classify a commit into this category when the test outcome depends on the order in which the tests are run.

Reason 3, time. Relying on the system time introduces non-deterministic failures, e.g., a test may fail when the midnight changes in the UTC time zone. Some tests also fail due to the precision by which time is reported as it can vary from one platform to another.

Reason 4, IO. I/O operations (in addition to those for networks) may also cause flakiness.

Reason 5, concurrency. We classify a commit in this category when the test non-determinism is due to different threads interacting in a non-desirable manner (but not due to asynchronous calls from the Async Wait category), e.g., due to data races, atomicity violations, or deadlocks.

Reason 6, network. Tests whose execution depends on network can be flaky because the network is a resource that is hard to control. In such cases, the test failure does not necessarily mean that the CUT itself is buggy, but rather the developer does not account for network uncertainties.

Reason 7, resource leak. A resource leak occurs whenever the application does not properly manage (acquire or release) one or more of its resources, e.g., memory allocations or database connections, leading to intermittent test failures.

Reason 8, randomness. The use of random numbers can also make some tests flaky. In the cases that we analyzed, tests are flaky because they use a random number generator without accounting for all the possible values that may be generated.

Reason 9, unordered collections. In general, when iterating over unordered collections (e.g., sets), the code should not assume that the elements are returned in a particular order. If it does assume, the test outcome can become non-deterministic as different executions may have a different order.

Reason 10, test case timeout. Flaky tests experiencing non-deterministic timeouts related to a single test belong to this category. It is comparable to the Test Suite Timeout (reported later), with the difference that the size of a single test grew over time without adjusting the max runtime value.

Reason 11, too restrictive range. In this category, some of the valid output values are outside the assertion range considered at test design time, so the test fails when they show up. In other words, such test cases have a range of predefined values for which the test is allowed to pass; if this range is defined too restrictively, tests may start failing in a not deterministic way.

Reason 12, floating point operations. Dealing with floating point operations is known to lead to tricky non-deterministic cases, especially in the high-performance computing community. Even simple operations like calculating the average of an array require thorough coding to avoid overflows, underflows, problems with non-associative addition, etc. Such problems can also be the root cause of flaky tests.

Reason 13, platform dependency. In many ways, it is possible for an execution to differ because of platform dependencies, for example because the size of an object (which is accessible via sys.getsizeof) differs between 32-bit and 64-bit systems.

Please identify the reason why the following code is flaky. Your answers are from `async wait', `test order dependency', `time', `IO', `concurrency', `network', `resource leak', `randomness', `unordered collections', `test case timeout', `too restrictive range', `floating point operations' and `platform dependency'.

Input:
```Code
[INPUT\_CODE]
```
\end{tcolorbox}

\noindent Instantiation: [INPUT\_CODE] is the Java test code to analyze.

\vspace{0.5cm}

\noindent\textbf{Note:} All prompts are available in our online repository~\cite{mywebsite} with complete instantiation examples.

\section{Detailed Normalization Pipelines}
\label{appendix:normalization}

This appendix provides implementation details for the Layer~1 automated comparison normalization pipelines described in Section~\ref{sec:eval_protocol}. These specifications enable reproducibility and support researchers implementing similar evaluation frameworks.

\subsection{AST Normalization Pipeline}
\label{appendix:ast_norm}

We build gold-standard ASTs using Tree-sitter, a production-grade incremental parser supporting multiple languages through unified grammar specifications, and compare LLM outputs against these references. This comparison encounters three fundamental challenges reflecting the tension between compiler-focused and human-focused representations.

\subsubsection{Challenges}
1). Internal Node Proliferation. Tree-sitter emits numerous nodes serving compilation purposes (type qualifiers, implicit casts, translation-unit wrappers) that carry no reconstructable semantic information.
2). Nesting Depth Inconsistency. Tree-sitter produces deeply nested trees (seven to eight levels) with intermediate packaging layers, while LLMs generate flatter structures (three to four levels) mirroring human code comprehension patterns.
3). Semantic Granularity Mismatch. Node type naming varies across Tree-sitter language grammars (\texttt{if\_statement} in Python grammar, \texttt{IfStatement} in Clang AST dumps for C/C++, \texttt{BranchNode} in LLM outputs) reflecting different syntactic abstractions of identical control constructs.

\subsubsection{Normalization Strategies}

Our normalization pipeline addresses these challenges through three strategies grounded in cross-tool empirical analysis:

\begin{enumerate}[label=(\roman*), leftmargin=*, itemsep=3pt]
\item \textit{Node filtering} via pattern-based classification derived from systematic examination of Tree-sitter outputs across our four evaluation languages (C, Java, Python, Solidity). The classification operates through keyword-based matching identifying three categories:
\begin{itemize}[leftmargin=*, itemsep=1pt]
    \item \textit{Drop rules}: compiler artifacts (e.g., \texttt{TypedefDecl}, \texttt{\_\_builtin} functions, namespace declarations, pragma directives) consistently present in parser outputs yet absent in human code descriptions are removed.
    \item \textit{Flatten rules}: non-semantic wrappers (e.g., \texttt{TranslationUnitDecl}, \texttt{ImplicitCastExpr}, type qualifiers, parenthesization layers) preserving no control-flow or data-flow semantics are collapsed by promoting children to parent level.
    \item \textit{Keep rules}: semantically critical nodes (function declarations with names or parameters, control structures, expressions) are preserved.
\end{itemize}
This filtering empirically reduces tree depth from typical Tree-sitter output (seven to eight levels) to human-comprehensible ranges (three to five levels).

\item \textit{Type coarsening} by mapping fine-grained node types to 15~unified semantic categories selected to span fundamental programming constructs across imperative and object-oriented paradigms. Control flow and statement categories include: \texttt{FUNCTION}, \texttt{BRANCH}, \texttt{LOOP}, \texttt{CALL}, \texttt{RETURN}, \texttt{ASSIGN}, \texttt{TRY}, \texttt{EXCEPT}. Declaration and data categories include: \texttt{TYPE\_DECL}, \texttt{INTERFACE}, \texttt{ENUM}, \texttt{BLOCK}, \texttt{PARAM}, \texttt{LITERAL}, \texttt{IDENT}. 

The mapping operates via language-agnostic tokenization (splitting \texttt{camelCase} and \texttt{snake\_case}, lowercasing) and keyword matching with disambiguation rules prioritizing control-flow semantics over syntactic packaging (e.g., \texttt{RETURN} takes precedence over \texttt{FUNCTION} to avoid misclassifying \texttt{Return(lambda)} nodes).

\item \textit{Backbone extraction} retains eight construct types forming program control-flow skeletons: \texttt{FUNCTION}, \texttt{BRANCH}, \texttt{LOOP}, \texttt{CALL}, \texttt{RETURN}, \texttt{ASSIGN}, \texttt{TRY}, \texttt{EXCEPT}, along with their parent-child relationships. This focuses evaluation on structural patterns developers prioritize when reasoning about code behavior rather than complete syntactic trees.
\end{enumerate}

\subsection{CFG Normalization Pipeline}
\label{appendix:cfg_norm}

We construct gold-standard CFGs using language-specific static analyzers: Joern~\cite{joern} for C (producing Code Property Graph representations), tree-sitter-graph for Python (AST-derived control flow), and Soot for Java (Jimple intermediate representation).

\subsubsection{Challenges}
1). Basic-Block Granularity Mismatch. Static analyzers decompose statements into micro-operation sequences (temporary variable assignments, load/store operations, arithmetic steps) producing node counts three to five times larger than LLM outputs.
2). Node Type Naming Variation. Naming manifests across tool ecosystems (\texttt{if\_stmt} in tree-sitter-graph, \texttt{IfStatement} in Soot, \texttt{ConditionalBlock} in LLM outputs).
3). Abstraction Level Differences. Tools include implementation artifacts (SSA phi nodes in Joern representing data-flow merge points, polymorphic dispatch details like \texttt{specialinvoke} and \texttt{virtualinvoke} in Soot Jimple, control-flow join markers) that LLMs abstract away.

\subsubsection{Normalization Strategies}

We apply three transformations calibrated to preserve control-flow semantics while tolerating micro-operation variance:

\begin{enumerate}[label=(\roman*), leftmargin=*, itemsep=3pt]
\item \textit{Node-type normalization} maps fine-grained labels to ten coarse categories via token-based keyword matching (keywords like \texttt{if}, \texttt{loop}, \texttt{return}, \texttt{call}) augmented with regex heuristics for language-specific call patterns (e.g., \texttt{obj.method(...)}, \texttt{new~Type(...)}). This yields seven backbone types (\texttt{ENTRY}, \texttt{EXIT}, \texttt{BRANCH}, \texttt{LOOP}, \texttt{CALL}, \texttt{TRY}, \texttt{EXCEPT}) plus three auxiliary categories (\texttt{JUMP} for break/continue, \texttt{MERGE} for phi/join nodes, \texttt{NORMAL} for other statements).

\item \textit{Graph canonicalization} collapses linear chains of auxiliary nodes (\texttt{NORMAL}, \texttt{MERGE}, \texttt{JUMP}) by rewiring edges to connect nearest backbone nodes via breadth-first traversal. Adaptive triggering is calibrated through empirical analysis of our evaluation dataset containing 75~gold CFGs. We observed that code examples in our dataset typically contain fewer than 15~nodes with sparse auxiliary-node ratios (below 50\%), while production static-analyzer IRs consistently exceed 15~nodes with auxiliary-node densities above 80\% reflecting their micro-operation decomposition. 

Coarsening activates only when both conditions hold (graph size $\geq$15~nodes \textit{and} auxiliary-node ratio $\geq$80\%) to preserve interpretable small-scale graphs unchanged while aligning heavily instrumented outputs. This design empirically reduces Java/Soot CFG sizes by factors of three to six (from 30--80~nodes down to 5--20~backbone nodes) and C/Python graphs by 1.5--2$\times$, aligning tool granularities with LLM abstraction levels.

\item \textit{Backbone-focused metrics} extract node/edge category distributions and five binary existence flags (\texttt{has\_branch}, \texttt{has\_loop}, \texttt{has\_call}, \texttt{has\_try}, \texttt{has\_except} defined at the graph level), computing coverage over the seven backbone types rather than all nodes to reduce sensitivity to micro-operation counting artifacts. 

For Java, where Soot's Jimple representation introduces pervasive micro-operations (temporary variable manipulations, explicit type conversions) even for simple methods, we enforce canonicalization and compute backbone-restricted recall focusing only on control-significant nodes. Fallback lightweight skeletons (\texttt{ENTRY-EXIT-CALL} triplets) are accepted when full backbone coverage is incomplete, reflecting human tolerance for interface-style outputs lacking full method bodies.
\end{enumerate}

\subsection{Call Graph Normalization Pipeline}
\label{appendix:cg_norm}

We generate gold-standard call graphs using Joern for C, tree-sitter-based static analysis for Python, Soot for Java, and Slither~\cite{slither} for Solidity.

\subsubsection{Challenges}

1). Qualifier Inconsistency. Tools emit fully qualified names (\texttt{com.example.MyClass.method} in Soot Java bytecode analysis, \texttt{module.Class.method} in Python static analysis) while LLMs use context-dependent abbreviations (\texttt{method}, \texttt{MyClass.method}).
2). Synonym Variation. Equivalent operations receive diverse names across implementations and language idioms (\texttt{println} vs. \texttt{print} vs. \texttt{printf} for output, \texttt{init} vs. \texttt{initialize} vs. \texttt{constructor} for initialization).
3). Optional Library Granularity. Static tools enumerate comprehensive library scaffolding (IO streams, logging frameworks, collection utilities, concurrency primitives, system calls) that LLMs and human reviewers treat as implementation details when summarizing core application logic.

\textbf{Anonymous and Compiler-Generated Variants.} Lambdas and inner classes produce ordinal names (\texttt{Lambda\$1.apply}, \texttt{Lambda\$2.apply} in Soot bytecode; \texttt{MyClass\$1.<init>}, \texttt{MyClass\$2.<init>} for anonymous inner classes) differing across compilations yet representing semantically equivalent constructs.

\subsubsection{Normalization Strategies}

Our pipeline employs five strategies calibrated against human call-graph judgment patterns observed in manual validation pilots (20~sample cases per language, annotated independently by two domain experts to establish inter-rater agreement patterns):

\begin{enumerate}[label=(\roman*), leftmargin=*]
\item \textit{Identifier tokenization} normalizes naming variations by lowercasing, splitting compound names (handling \texttt{camelCase}, \texttt{snake\_case}, \texttt{package.qualifiers} uniformly), and removing language-specific artifacts (generics like \texttt{<T>}, parameter signatures like \texttt{(int,String)}), reducing heterogeneous qualified paths to comparable token sequences.

\item \textit{Synonym expansion} incorporates domain knowledge through explicit mapping rules covering operations where human annotators consistently accepted lexical variants as semantically equivalent. Rules derived from cross-language API analysis include: \texttt{print} $\leftrightarrow$ \texttt{println} $\leftrightarrow$ \texttt{printf} $\leftrightarrow$ \texttt{format} (IO family); \texttt{init} $\leftrightarrow$ \texttt{initialize} $\leftrightarrow$ \texttt{setup} $\leftrightarrow$ \texttt{constructor} (initialization family); \texttt{length} $\leftrightarrow$ \texttt{size} (collection queries).

\item \textit{Fuzzy matching} computes token-set Jaccard similarity with suffix-emphasis heuristics calibrated to observed inter-annotator agreement patterns. When human experts agreed on node equivalence despite qualifier mismatches (e.g., accepting \texttt{A.B.method} as equivalent to \texttt{method}), we found method-name (last-token) matches in 95\% of agreed pairs and class-plus-method (last-two-token) matches in 88\% of agreed pairs, leading to programmatic similarity lifts of 0.8~and~0.9~respectively. 

Category alignment (both nodes classified into the same functional category via keyword patterns covering eight domains: io, logging, error handling, events, system calls, concurrency, collections, utilities) lifts weak lexical matches (initial Jaccard below 0.30) to acceptance threshold when human annotators marked such pairs acceptable despite low string overlap.

\item \textit{Wrapper collapse} iteratively removes thin auxiliary nodes (in-degree and out-degree both unity, relaxed to two for dynamic-dispatch scaffolding like function-pointer indirection) belonging to eight optional categories identified through annotation analysis: library/system calls that both expert annotators consistently excluded when asked to summarize core application call structure. This empirically reduces Java CG sizes by 40--60\% (from 80--100~nodes including library scaffolding to 30--50~essential application nodes) and C/Python graphs by 20--30\%.

\item \textit{Variant-key normalization} groups ordinal variants (lambda instances, anonymous inner classes) into semantic equivalence classes using composite keys combining functional category, method-name token, and class-context anchor. Examples: \texttt{Lambda\$1.apply} and \texttt{Lambda\$2.apply} both map to \texttt{other:apply}; \texttt{MyClass\$1.<init>} and \texttt{MyClass\$2.<init>} both map to \texttt{other:ctor:myclass}. This deduplicates 15--25\% of lambda/inner-class nodes observed in typical Java and Python graphs.
\end{enumerate}

\subsubsection{Evaluation Strategy}

Evaluation restricts to essential edges connecting code-defined core application functions (excluding optional library categories) while tolerating resolution patterns observed as acceptable in human validation: indirect calls via one-hop intermediates, dynamic dispatch resolving to alternative concrete implementations, shared-caller scenarios where multiple callees represent polymorphic dispatch targets.

\subsection{Threshold Calibration Methodology}
\label{appendix:threshold_calibration}

Quantitative thresholds across all normalization pipelines (CFG coarsening at 80\% auxiliary-node density with 15-node minimum, CG suffix-matching at 0.8--0.9~similarity, category-alignment threshold at 0.30) were calibrated through iterative refinement: 

\begin{enumerate}[leftmargin=*]
\item Initial values derived from dataset statistics (e.g., median auxiliary-node density separating pedagogical examples from production tool IRs, observed inter-rater agreement rates on name-mismatch cases)
\item Adjusted via systematic comparison against manual validation labels on a held-out calibration subset (15~cases per task)
\item Refined until automated pass/fail decisions achieved 85\%+~agreement with expert judgments
\item Balanced precision (avoiding false alignments across semantically distinct constructs) with recall (accepting representational variations within semantic equivalence classes)
\end{enumerate}

Semantic equivalence is complex (particularly for higher-level abstractions like call graphs where human judgment weighs application logic structure over library plumbing details), necessitating expert validation (Layer~2) to adjudicate cases where structural metrics provide incomplete assessment of correctness.

\section{Complete Experimental Results}
\label{appendix:complete_results}

This appendix provides complete experimental results for all 21 evaluated models, supplementing the representative subsets shown in the main text.

\subsection{Data Dependency and Taint Analysis: Complete Results}
\label{appendix:dp_taint_full}

Figure~\ref{fig:data_dep_taint_complete} presents per-project F1 scores on Data Dependency (blue) and Taint Analysis (orange) for all 21 evaluated models. The main text (Section~\ref{sec:result}, Figure~\ref{fig:data_dep_taint}) shows six representative models selected to cover top commercial systems, mid-tier open-source models, and baselines.

\section{Complete Results}
\begin{figure*}[t]
	\centering
	
	\begin{subfigure}[b]{0.3\textwidth}\centering
		\includegraphics[width=\textwidth]{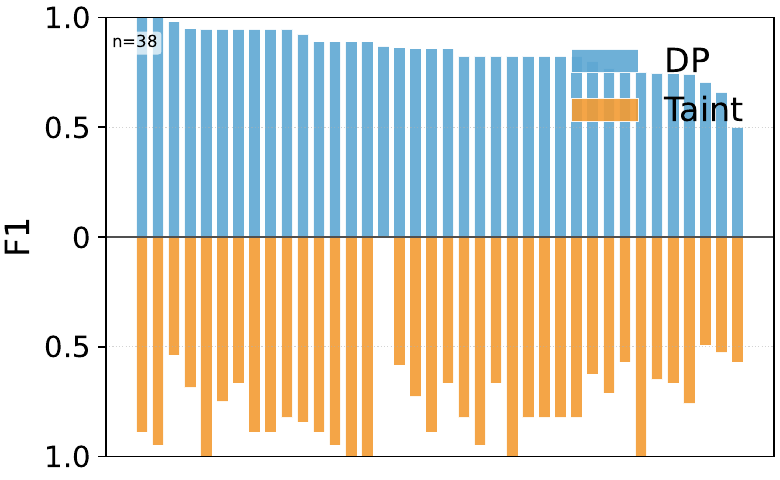}
		\caption*{\small GPT-5-nano}
	\end{subfigure}%
	\begin{subfigure}[b]{0.3\textwidth}\centering
		\includegraphics[width=\textwidth]{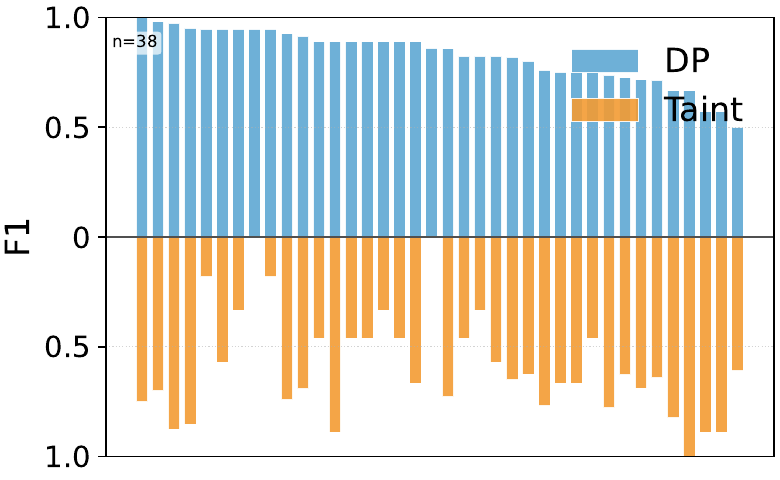}
		\caption*{\small GPT-5-mini}
	\end{subfigure}%
	\begin{subfigure}[b]{0.3\textwidth}\centering
		\includegraphics[width=\textwidth]{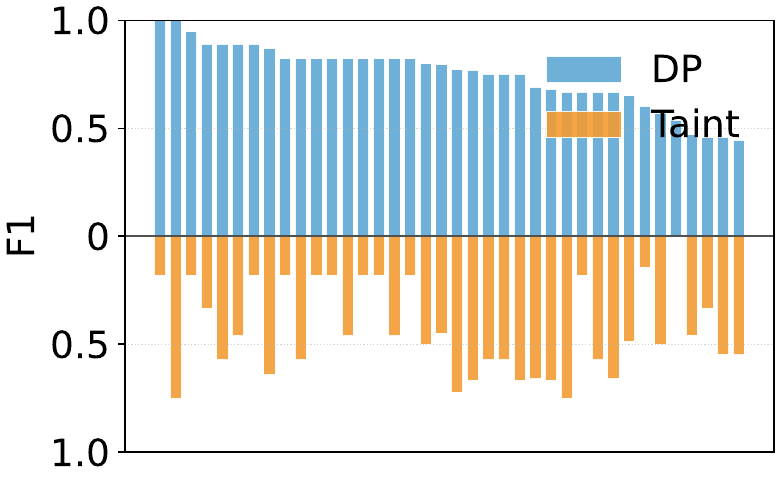}
		\caption*{\small GPT-o1-mini}
	\end{subfigure}%
	
	\begin{subfigure}[b]{0.3\textwidth}\centering
		\includegraphics[width=\textwidth]{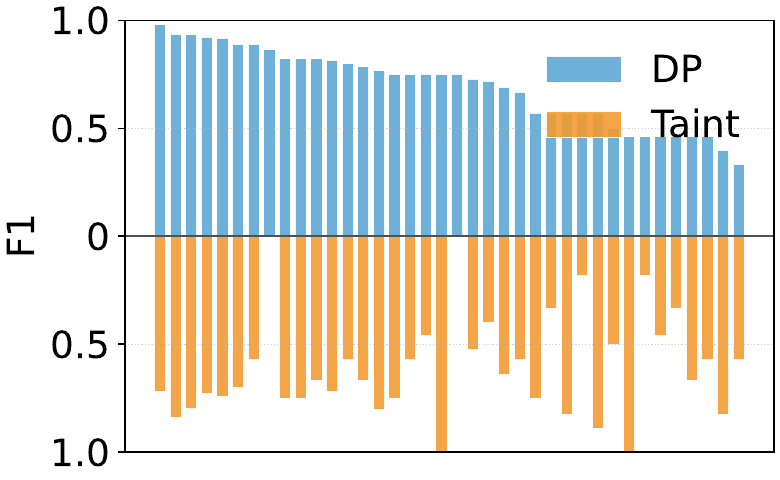}
		\caption*{\small GPT-5-codex}
	\end{subfigure}%
	\begin{subfigure}[b]{0.3\textwidth}\centering
		\includegraphics[width=\textwidth]{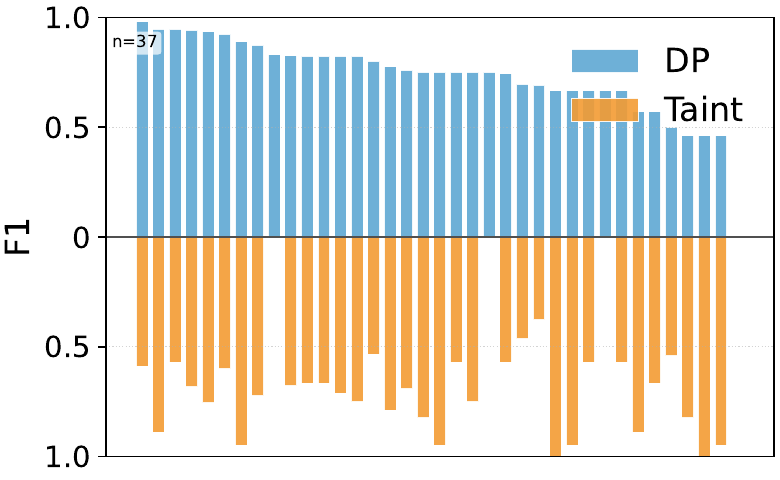}
		\caption*{\small GPT-OSS-20B}
	\end{subfigure}%
	\begin{subfigure}[b]{0.3\textwidth}\centering
		\includegraphics[width=\textwidth]{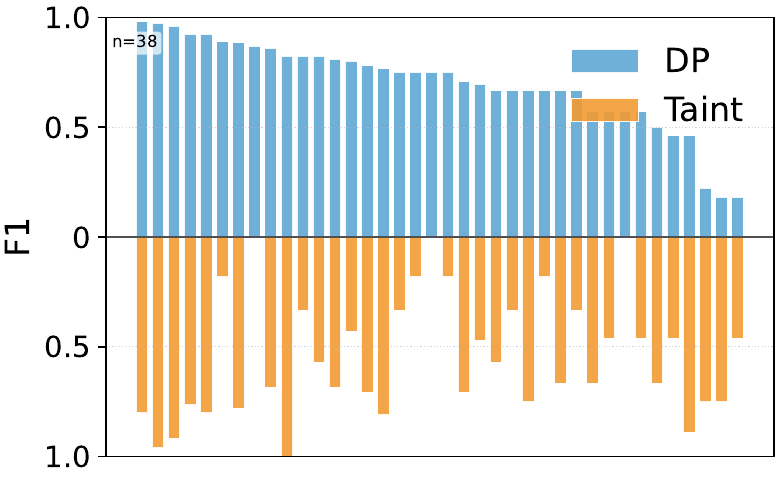}
		\caption*{\small GPT-5}
	\end{subfigure}%
	
	\begin{subfigure}[b]{0.3\textwidth}\centering
		\includegraphics[width=\textwidth]{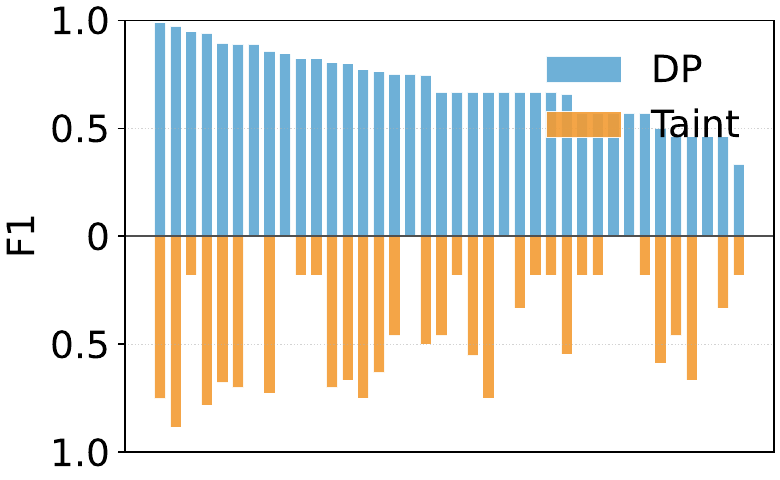}
		\caption*{\small GPT-o4-mini}
	\end{subfigure}%
	\begin{subfigure}[b]{0.3\textwidth}\centering
		\includegraphics[width=\textwidth]{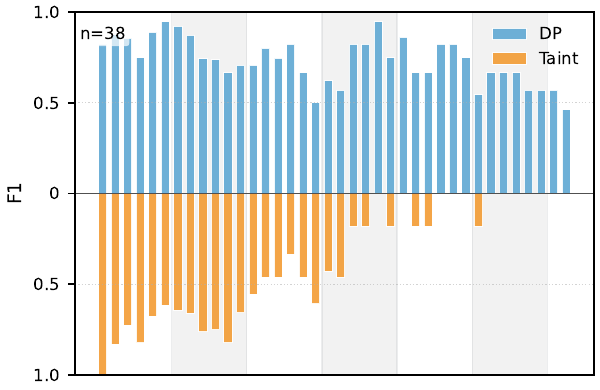}
		\caption*{\small GPT-o3-mini}
	\end{subfigure}%
	\begin{subfigure}[b]{0.3\textwidth}\centering
		\includegraphics[width=\textwidth]{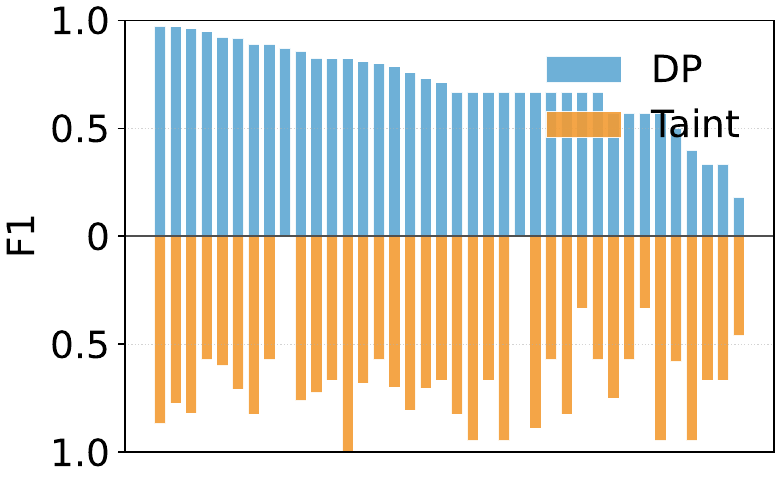}
		\caption*{\small GPT-o3}
	\end{subfigure}%
	
	\begin{subfigure}[b]{0.3\textwidth}\centering
		\includegraphics[width=\textwidth]{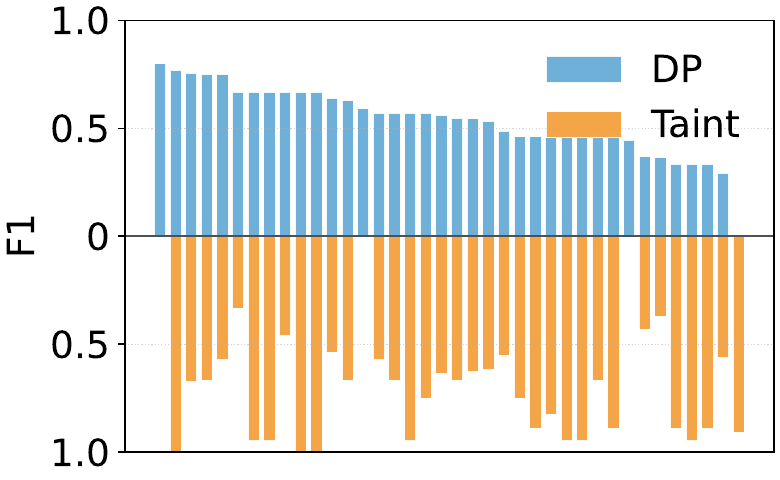}
		\caption*{\small GPT-4o-mini}
	\end{subfigure}%
	\begin{subfigure}[b]{0.3\textwidth}\centering
		\includegraphics[width=\textwidth]{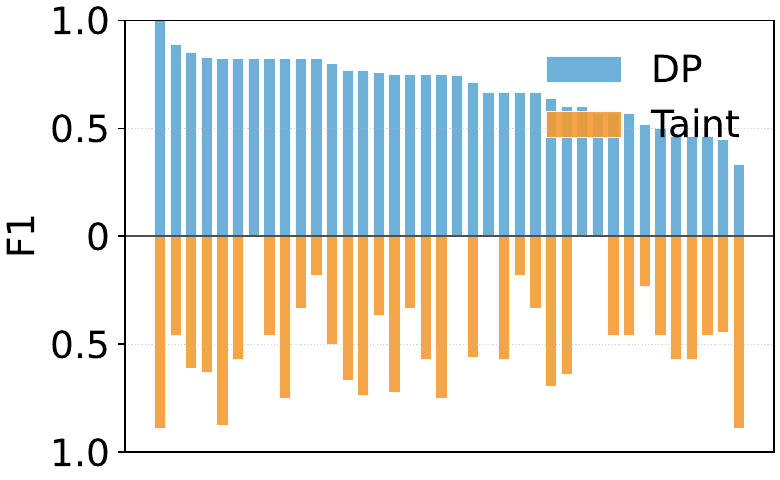}
		\caption*{\small GPT-4o}
	\end{subfigure}%
	
	\caption{Complete F1 results on Data Dependency (blue) and Taint Analysis (orange) for all 21 evaluated models. Each subplot shows per-project F1 scores, with data dependency in the upper section (0–1.0) and taint analysis in the lower section (0–\(\sim\)0.8). (Part 1 of 2)}
	\label{fig:data_dep_taint_complete}
\end{figure*}

\begin{figure*}[t]\ContinuedFloat
	\centering
	
	\begin{subfigure}[b]{0.3\textwidth}\centering
		\includegraphics[width=\textwidth]{sections/figures/dp_taint/Claude-sonnet-4__dp_taint_f1_dual_centered_paper.pdf}
		\caption*{\small Claude-sonnet-4}
	\end{subfigure}%
	\begin{subfigure}[b]{0.3\textwidth}\centering
		\includegraphics[width=\textwidth]{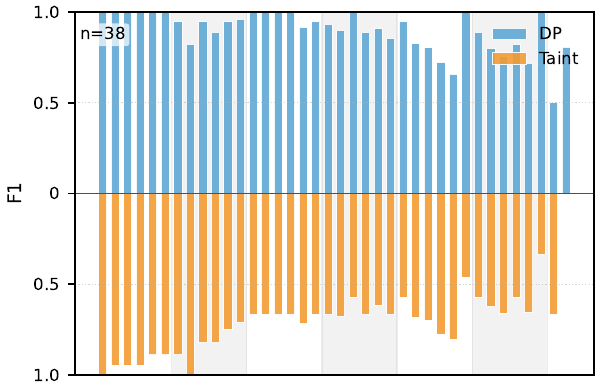}
		\caption*{\small Gemini-2.5-Flash}
	\end{subfigure}%
	\begin{subfigure}[b]{0.3\textwidth}\centering
		\includegraphics[width=\textwidth]{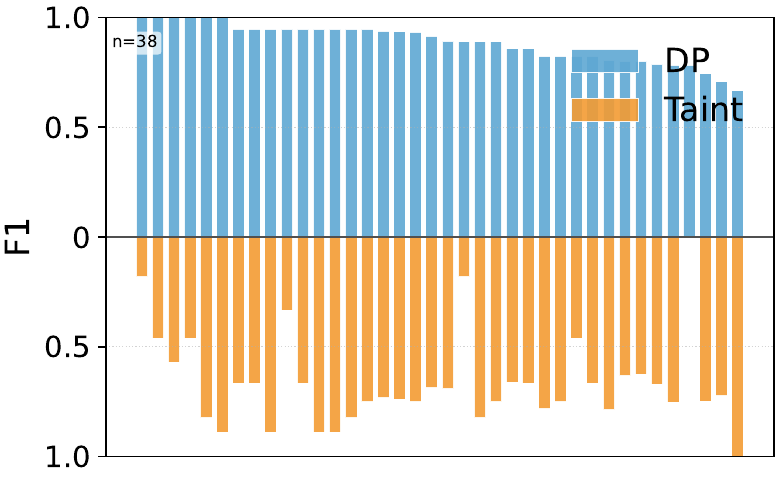}
		\caption*{\small Gemini-2.5-Pro}
	\end{subfigure}%
	
	\begin{subfigure}[b]{0.3\textwidth}\centering
		\includegraphics[width=\textwidth]{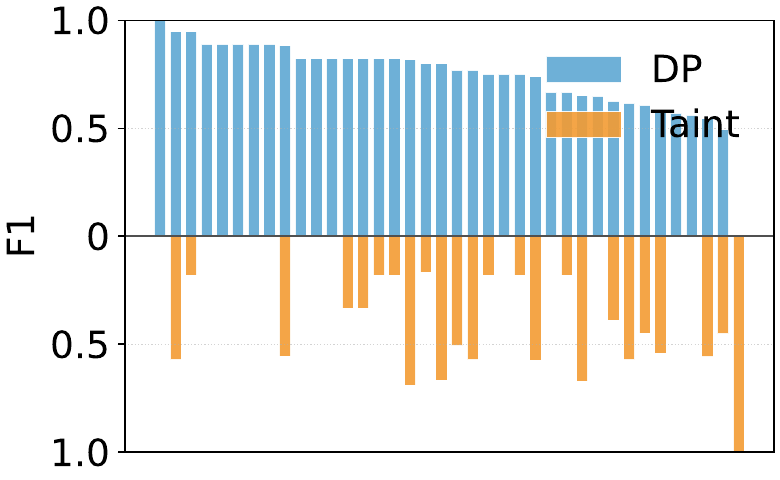}
		\caption*{\small Deepseekchat-v3}
	\end{subfigure}%
	\begin{subfigure}[b]{0.3\textwidth}\centering
		\includegraphics[width=\textwidth]{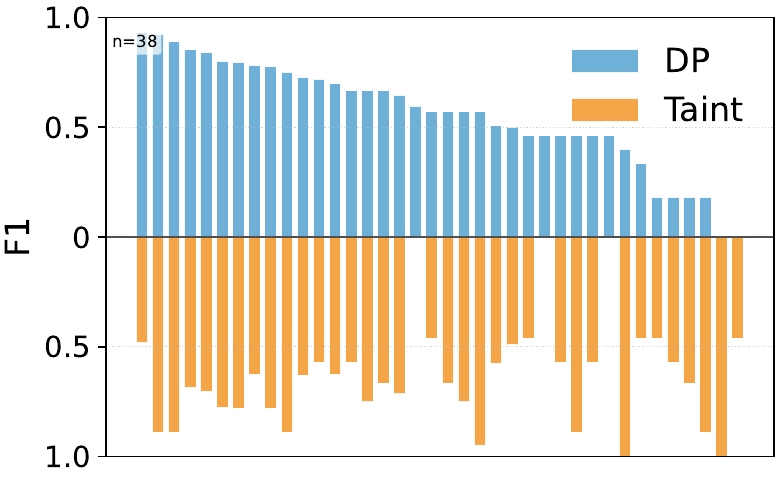}
		\caption*{\small Qwen3-next-80b-a3b}
	\end{subfigure}%
	\begin{subfigure}[b]{0.3\textwidth}\centering
		\includegraphics[width=\textwidth]{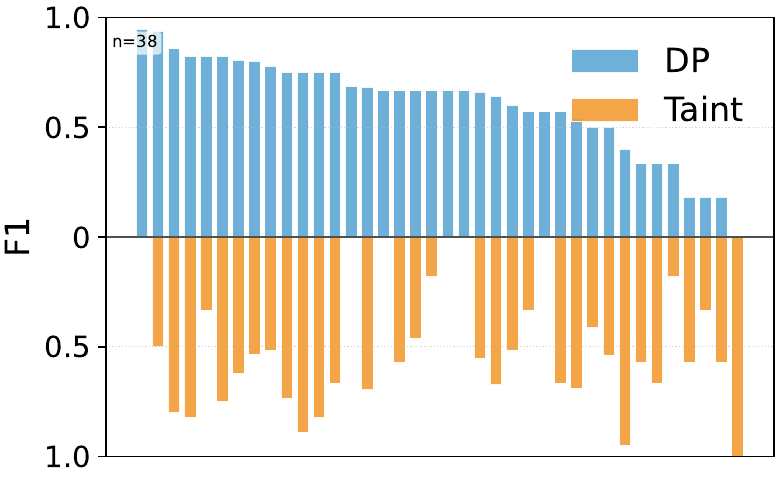}
		\caption*{\small Qwen3-coder-plus}
	\end{subfigure}%
	
	\begin{subfigure}[b]{0.3\textwidth}\centering
		\includegraphics[width=\textwidth]{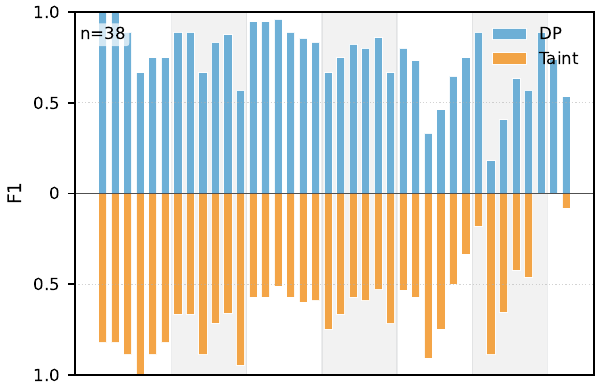}
		\caption*{\small Kimi-K2}
	\end{subfigure}%
	\begin{subfigure}[b]{0.3\textwidth}\centering
		\includegraphics[width=\textwidth]{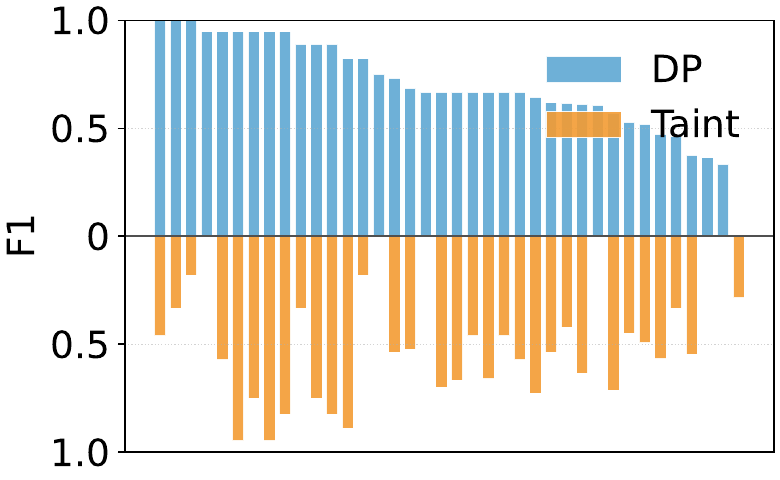}
		\caption*{\small StarChat}
	\end{subfigure}%
	\begin{subfigure}[b]{0.3\textwidth}\centering
		\includegraphics[width=\textwidth]{sections/figures/dp_taint/CodeLlama-70b__dp_taint_f1_dual_centered_paper.pdf}
		\caption*{\small CodeLlama-70b}
	\end{subfigure}%
	
	\begin{subfigure}[b]{0.3\textwidth}\centering
		\includegraphics[width=\textwidth]{sections/figures/dp_taint/CodeLlama-13b__dp_taint_f1_dual_centered_paper.pdf}
		\caption*{\small CodeLlama-13b}
	\end{subfigure}%
	
	\caption{Complete F1 results on Data Dependency and Taint Analysis. (Part 2 of 2)}
\end{figure*}

\subsection{Flaky Test Reasoning: Complete Results}
\label{appendix:flaky_full}

\begin{figure}[!htbp]
   \centering
   \begin{subfigure}[b]{0.3\textwidth}\centering\includegraphics[width=\textwidth]{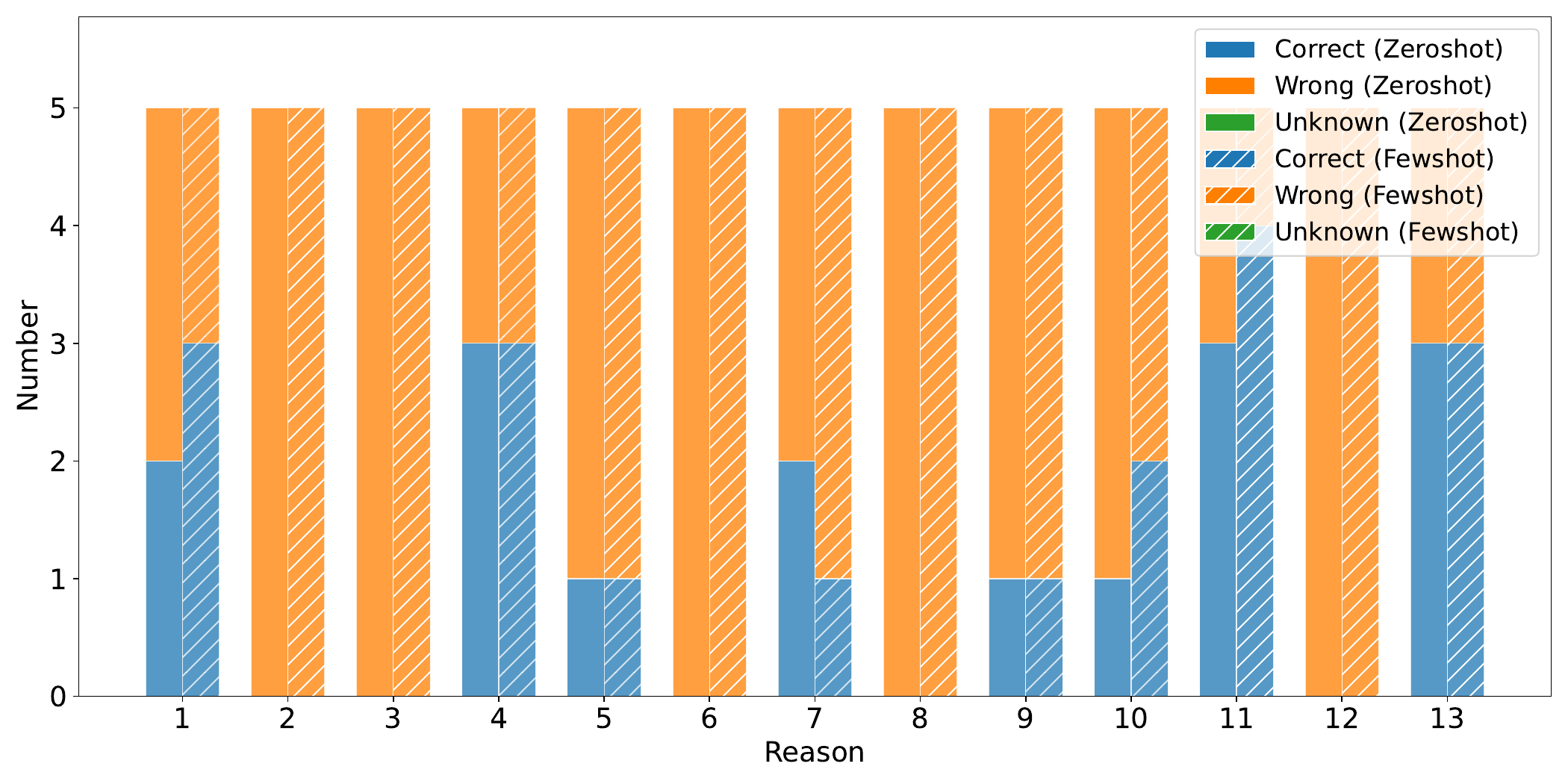}\caption*{\small{GPT-5-nano}}\end{subfigure}%
   \begin{subfigure}[b]{0.3\textwidth}\centering\includegraphics[width=\textwidth]{sections/figures/flaky/fig12/flaky_gpt-5-mini_fewshot_zeroshot.pdf}\caption*{\small{GPT-5-mini}}\end{subfigure}%
   \begin{subfigure}[b]{0.3\textwidth}\centering\includegraphics[width=\textwidth]{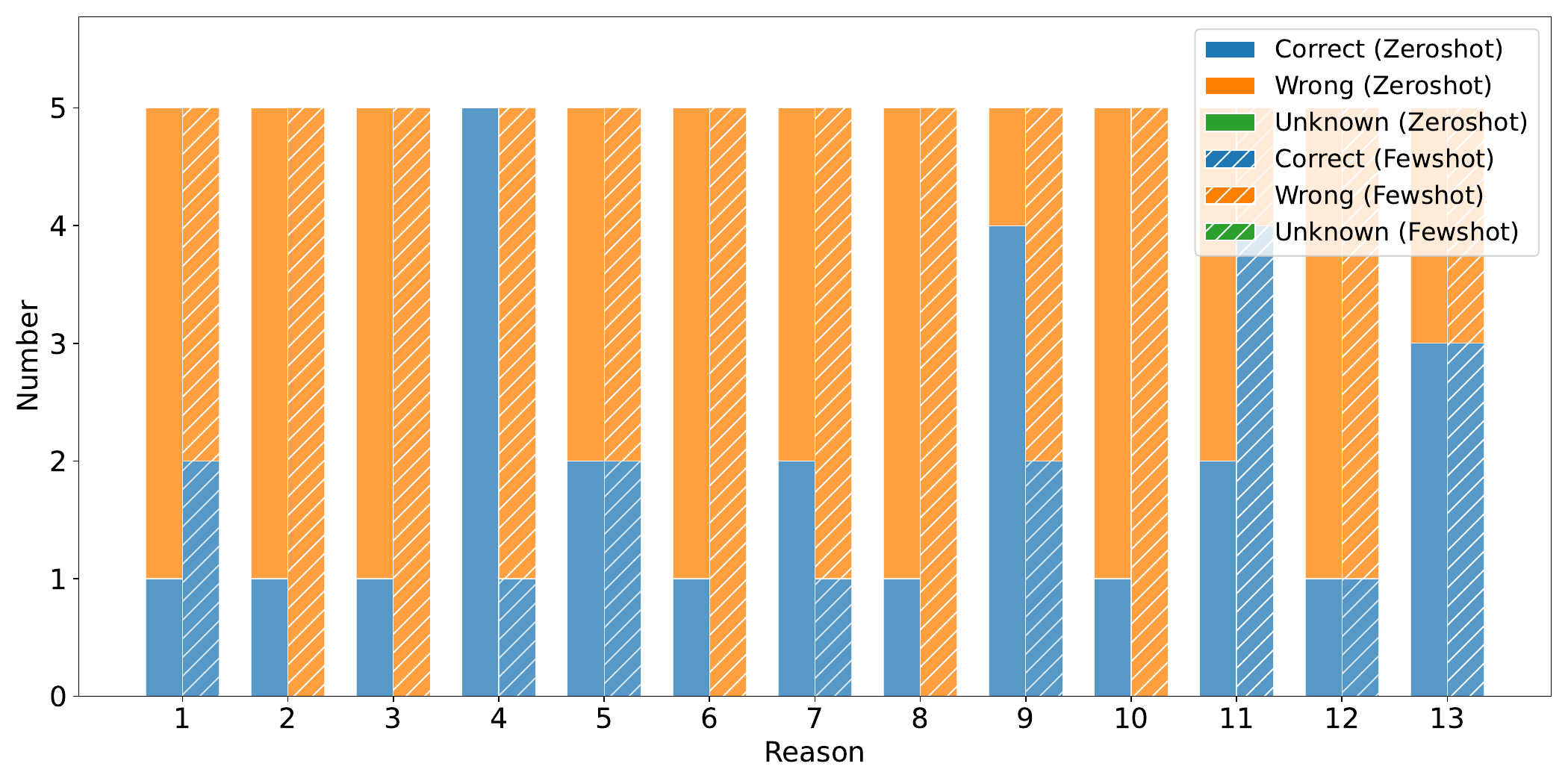}\caption*{\small{GPT-o1-mini}}\end{subfigure}
   
   \begin{subfigure}[b]{0.3\textwidth}\centering\includegraphics[width=\textwidth]{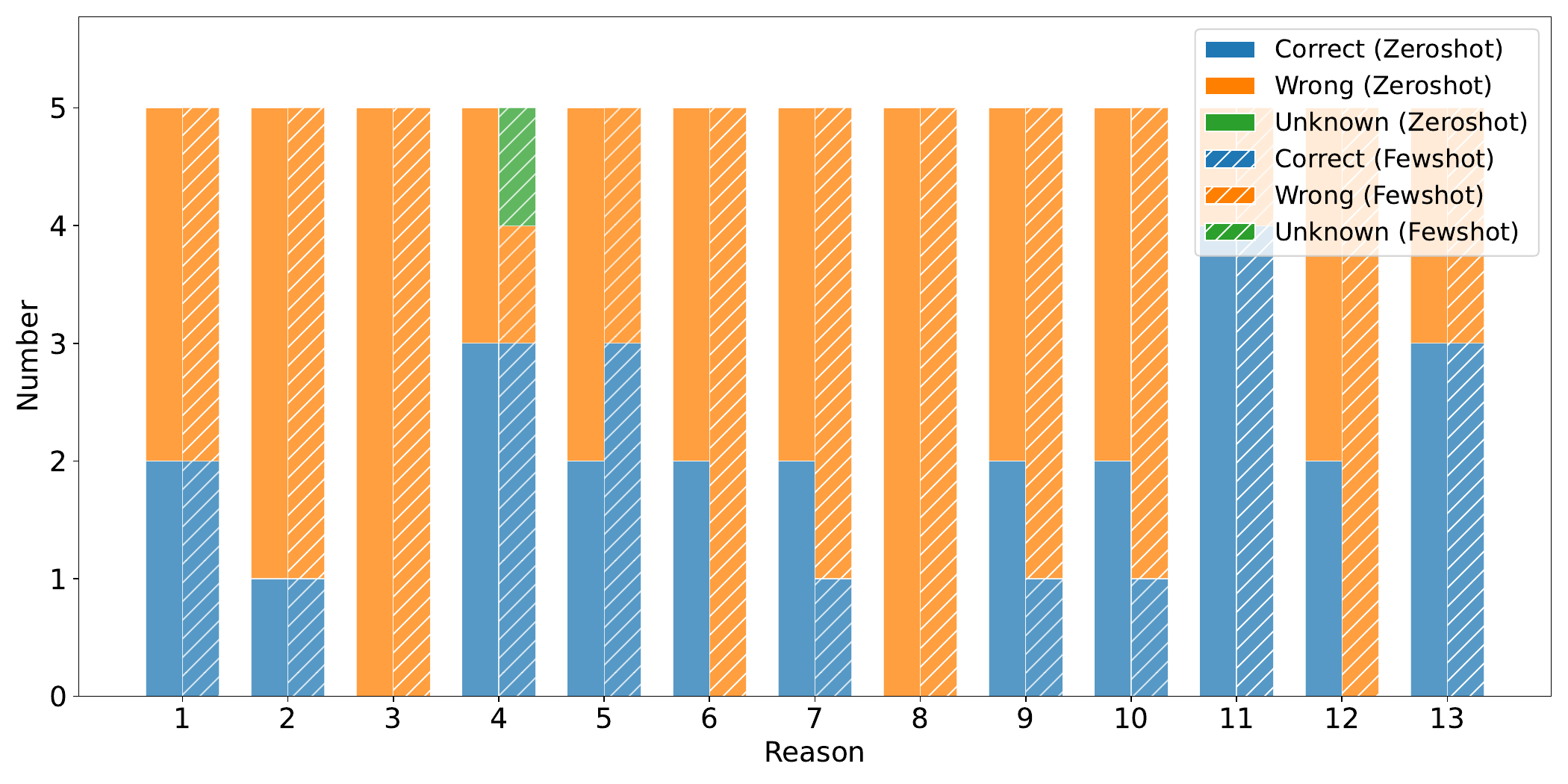}\caption*{\small{GPT-5-codex}}\end{subfigure}%
   \begin{subfigure}[b]{0.3\textwidth}\centering\includegraphics[width=\textwidth]{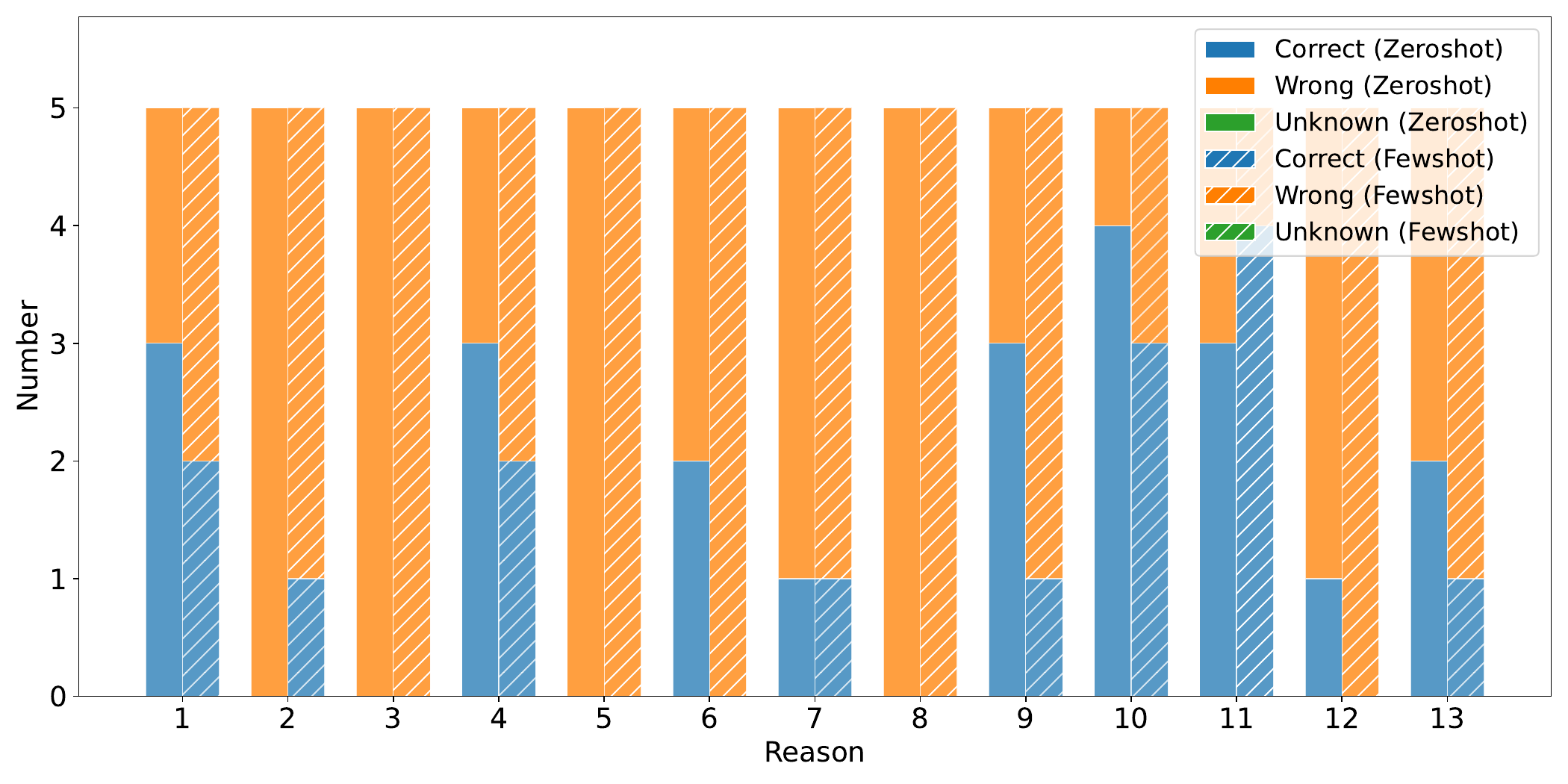}\caption*{\small{GPT-OSS-20B}}\end{subfigure}%
   \begin{subfigure}[b]{0.3\textwidth}\centering\includegraphics[width=\textwidth]{sections/figures/flaky/fig12/flaky_gpt-5_fewshot_zeroshot.pdf}\caption*{\small{GPT-5}}\end{subfigure}
   
   \begin{subfigure}[b]{0.3\textwidth}\centering\includegraphics[width=\textwidth]{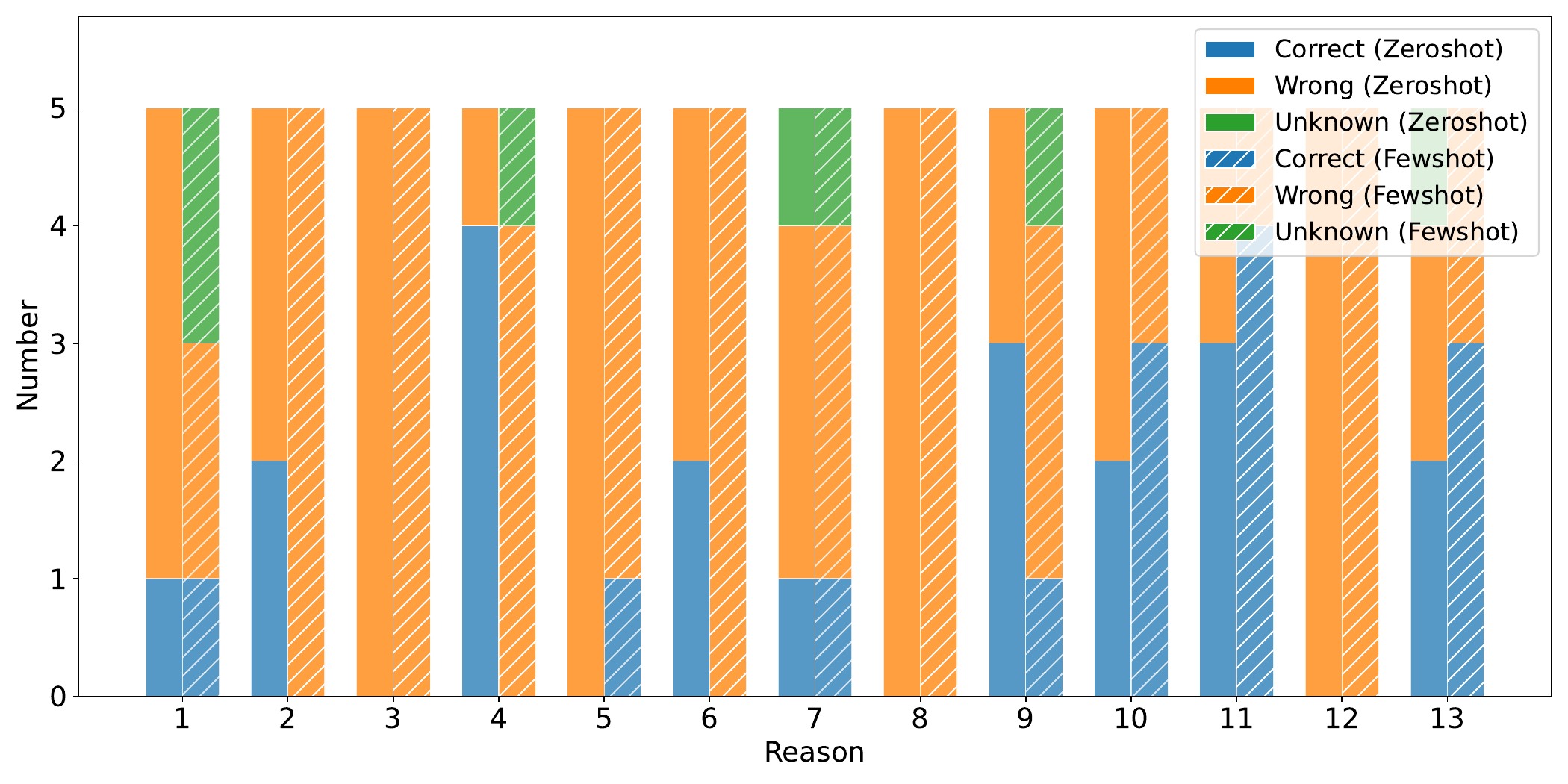}\caption*{\small{GPT-o4-mini}}\end{subfigure}%
   \begin{subfigure}[b]{0.3\textwidth}\centering\includegraphics[width=\textwidth]{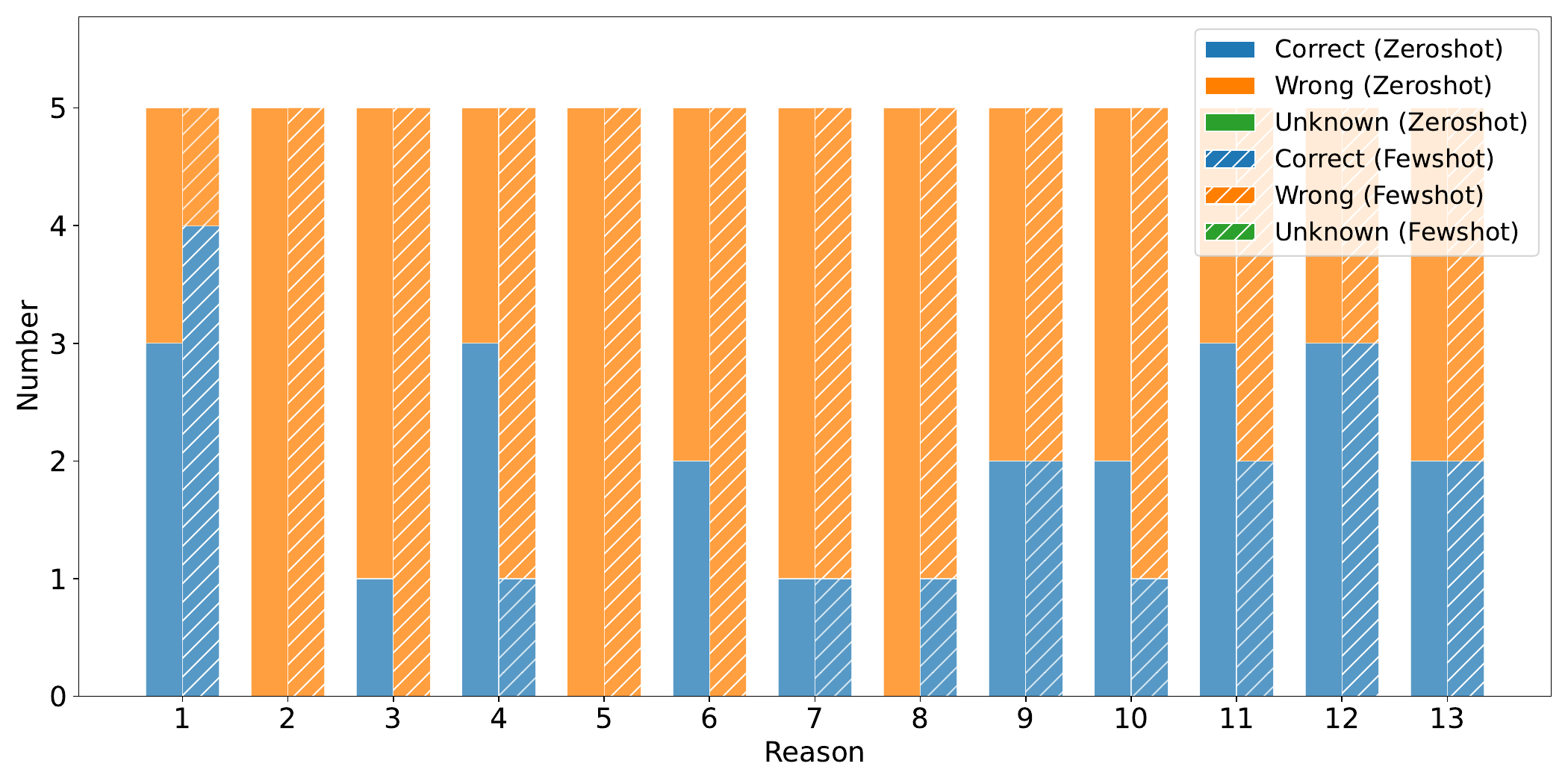}\caption*{\small{GPT-o3-mini}}\end{subfigure}%
   \begin{subfigure}[b]{0.3\textwidth}\centering\includegraphics[width=\textwidth]{sections/figures/flaky/fig12/flaky_o3_fewshot_zeroshot.pdf}\caption*{\small{GPT-o3}}\end{subfigure}
   
   \begin{subfigure}[b]{0.3\textwidth}\centering\includegraphics[width=\textwidth]{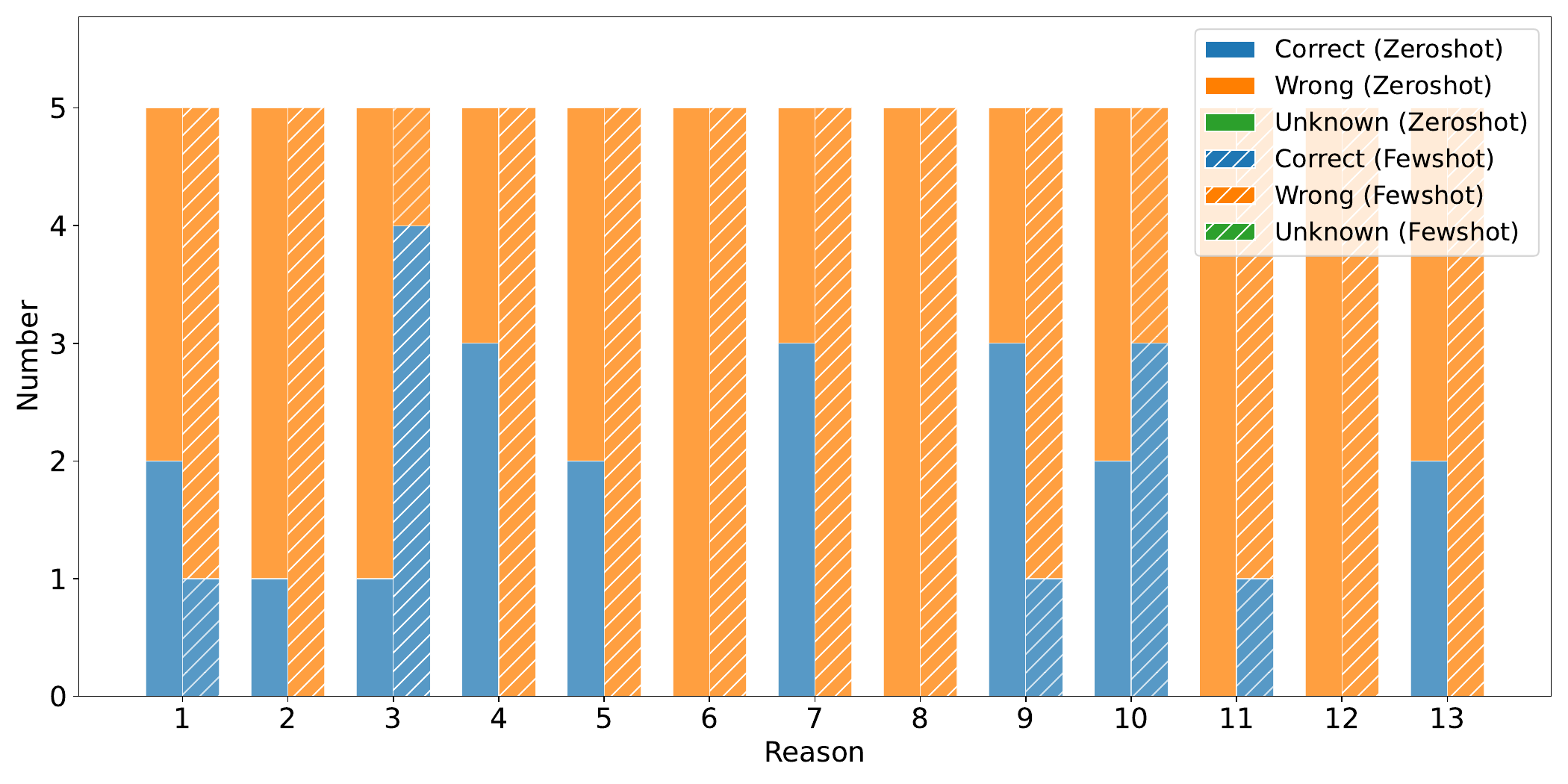}\caption*{\small{GPT-4o-mini}}\end{subfigure}%
   \begin{subfigure}[b]{0.3\textwidth}\centering\includegraphics[width=\textwidth]{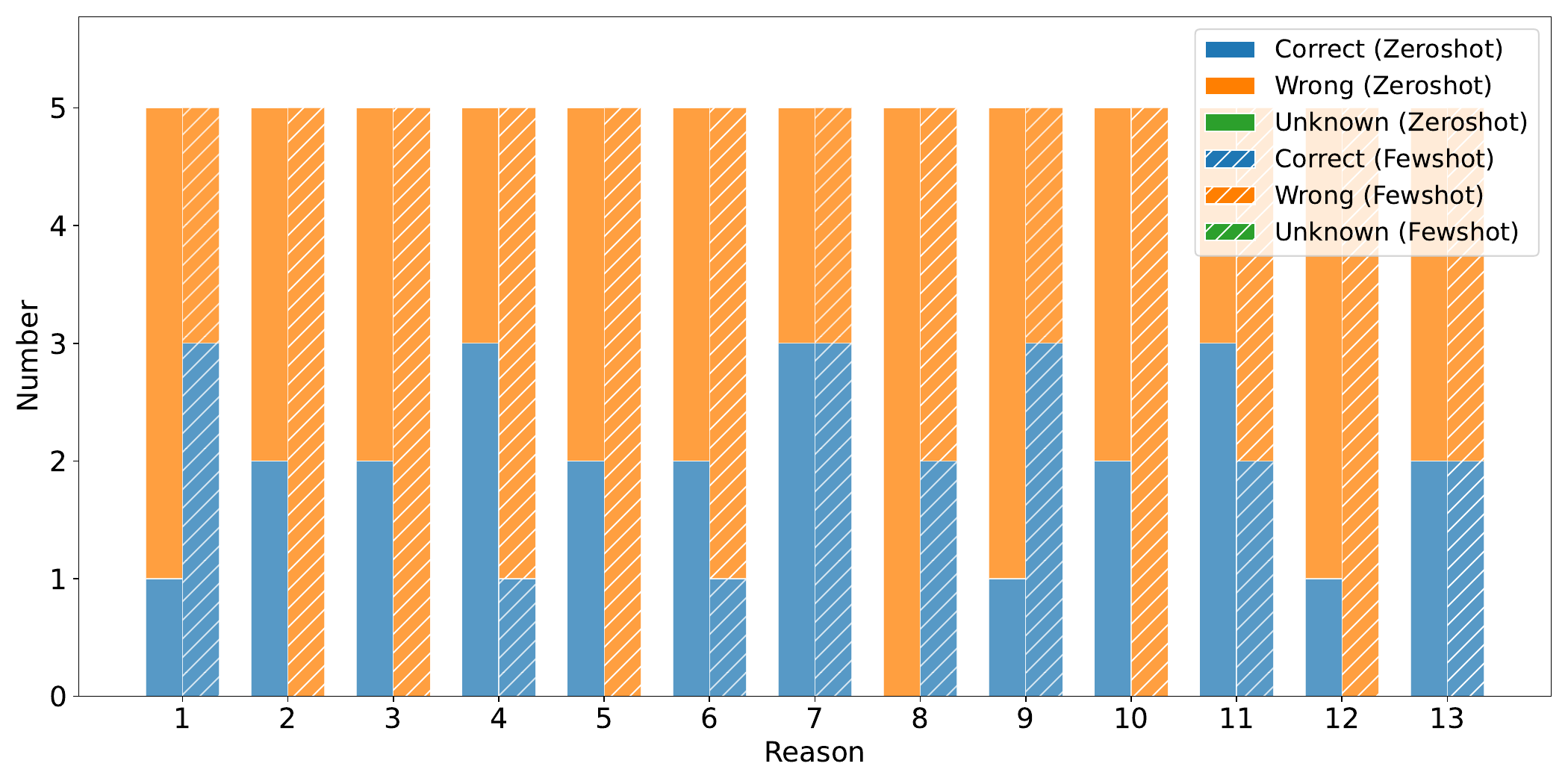}\caption*{\small{GPT-4o}}\end{subfigure}%
   \begin{subfigure}[b]{0.3\textwidth}\centering\includegraphics[width=\textwidth]{sections/figures/flaky/fig12/flaky_Claude-sonnet-4_fewshot_zeroshot.pdf}\caption*{\small{Claude-sonnet-4}}\end{subfigure}
   \begin{subfigure}[b]{0.3\textwidth}\centering\includegraphics[width=\textwidth]{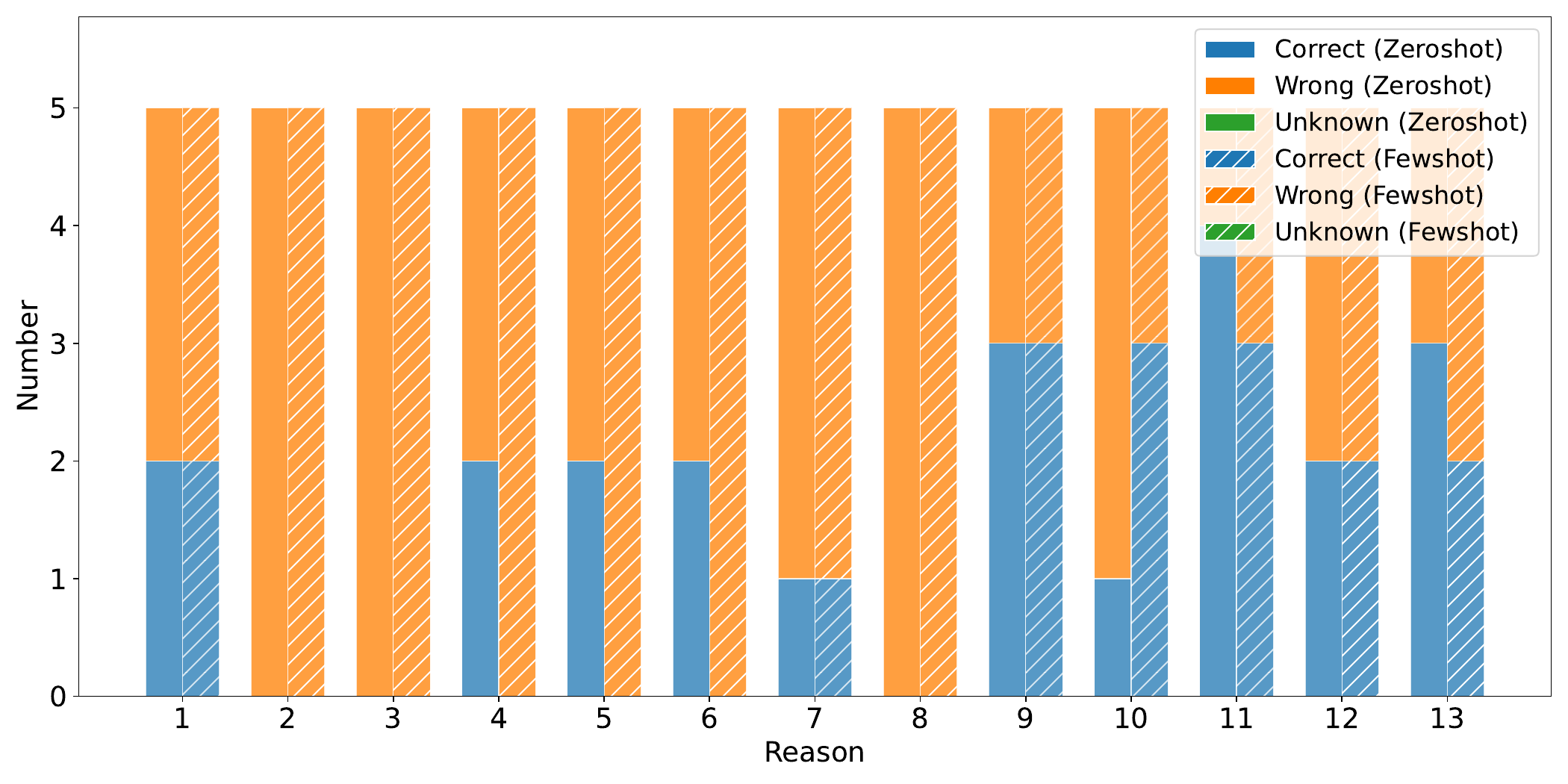}\caption*{\small{Gemini-2.5-Flash}}\end{subfigure}%
   \begin{subfigure}[b]{0.3\textwidth}\centering\includegraphics[width=\textwidth]{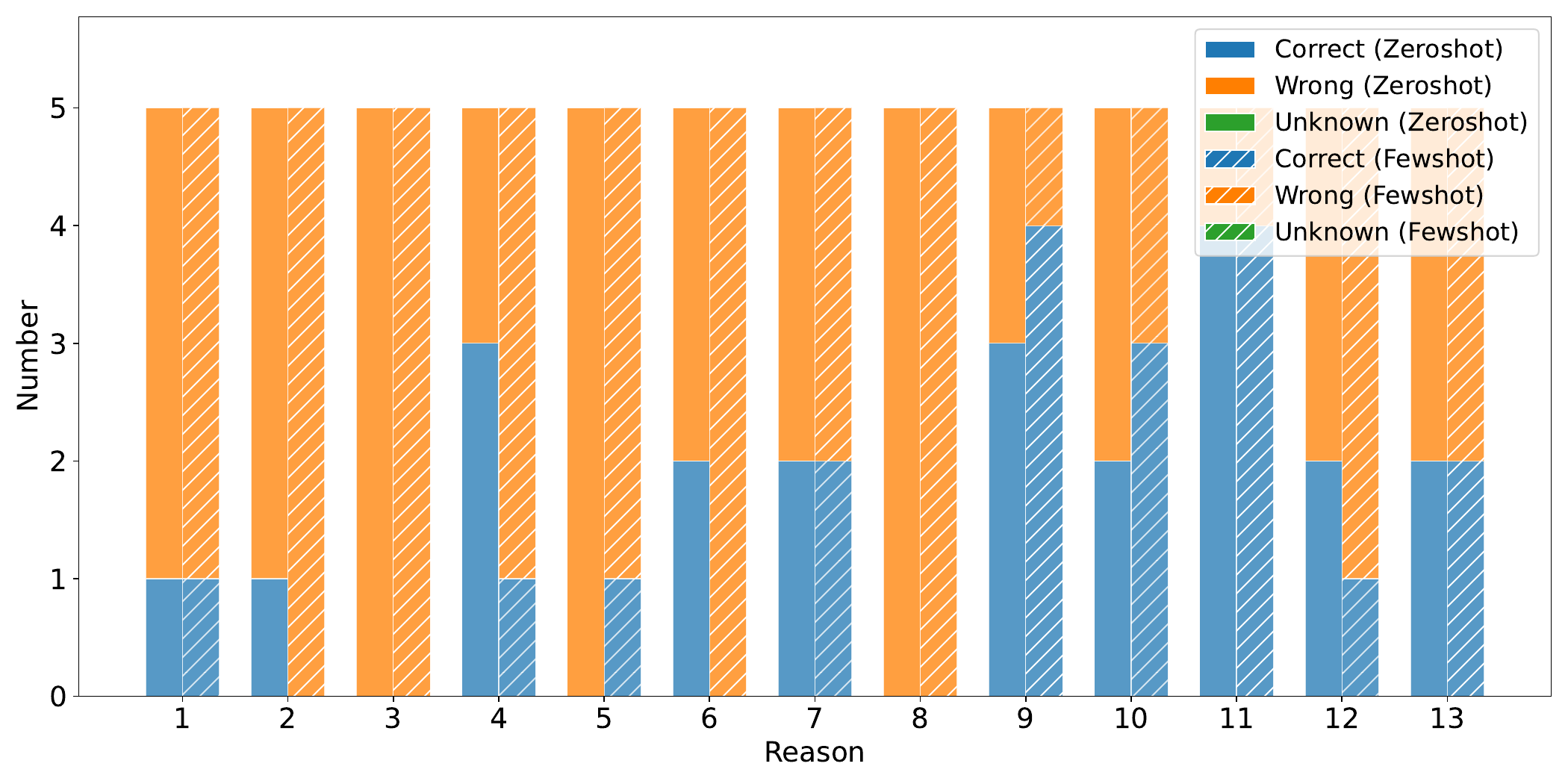}\caption*{\small{Gemini-2.5-Pro}}\end{subfigure}%
   \begin{subfigure}[b]{0.3\textwidth}\centering\includegraphics[width=\textwidth]{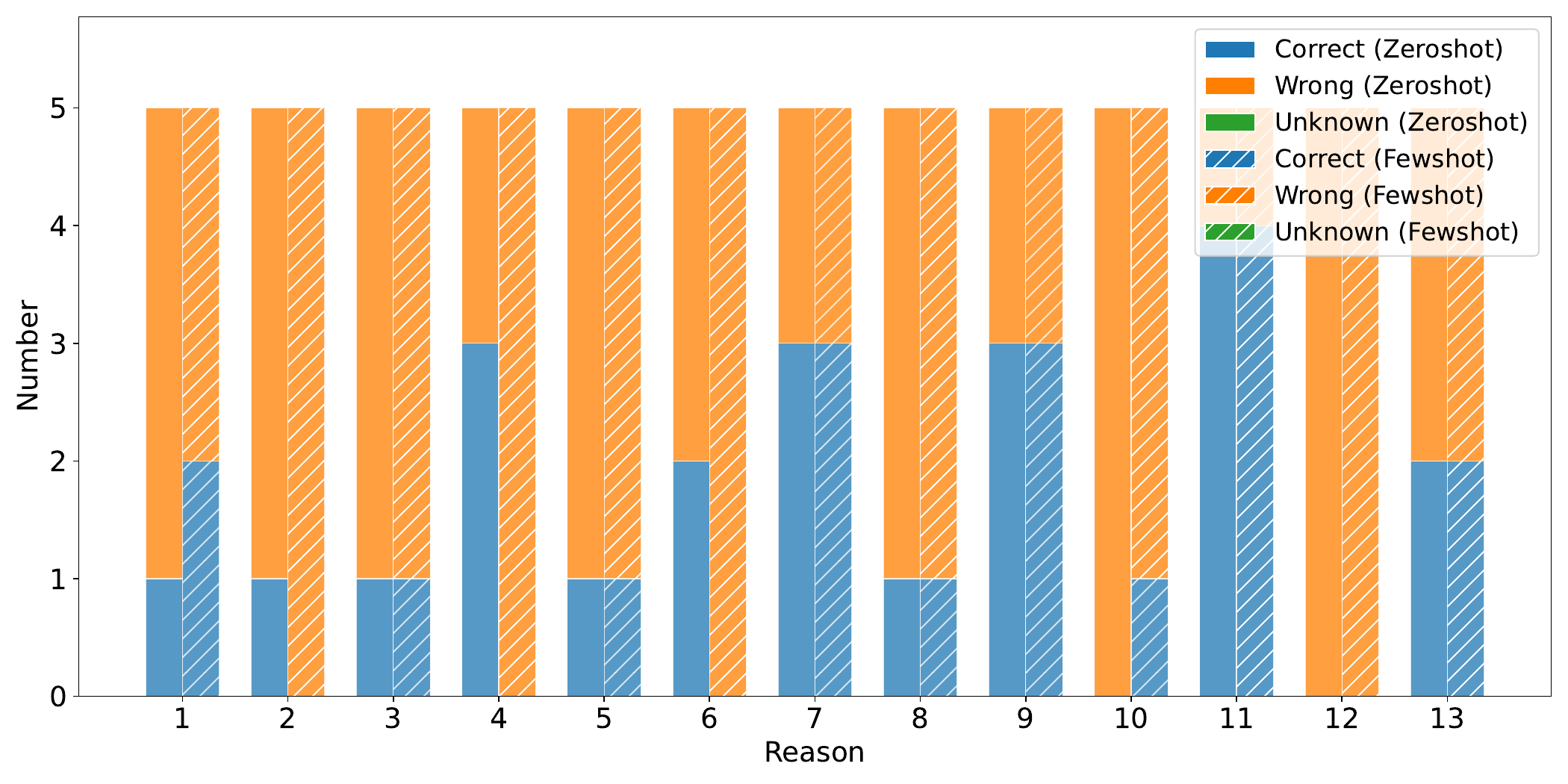}\caption*{\small{Deepseekchat-v3}}\end{subfigure}
   
   \begin{subfigure}[b]{0.3\textwidth}\centering\includegraphics[width=\textwidth]{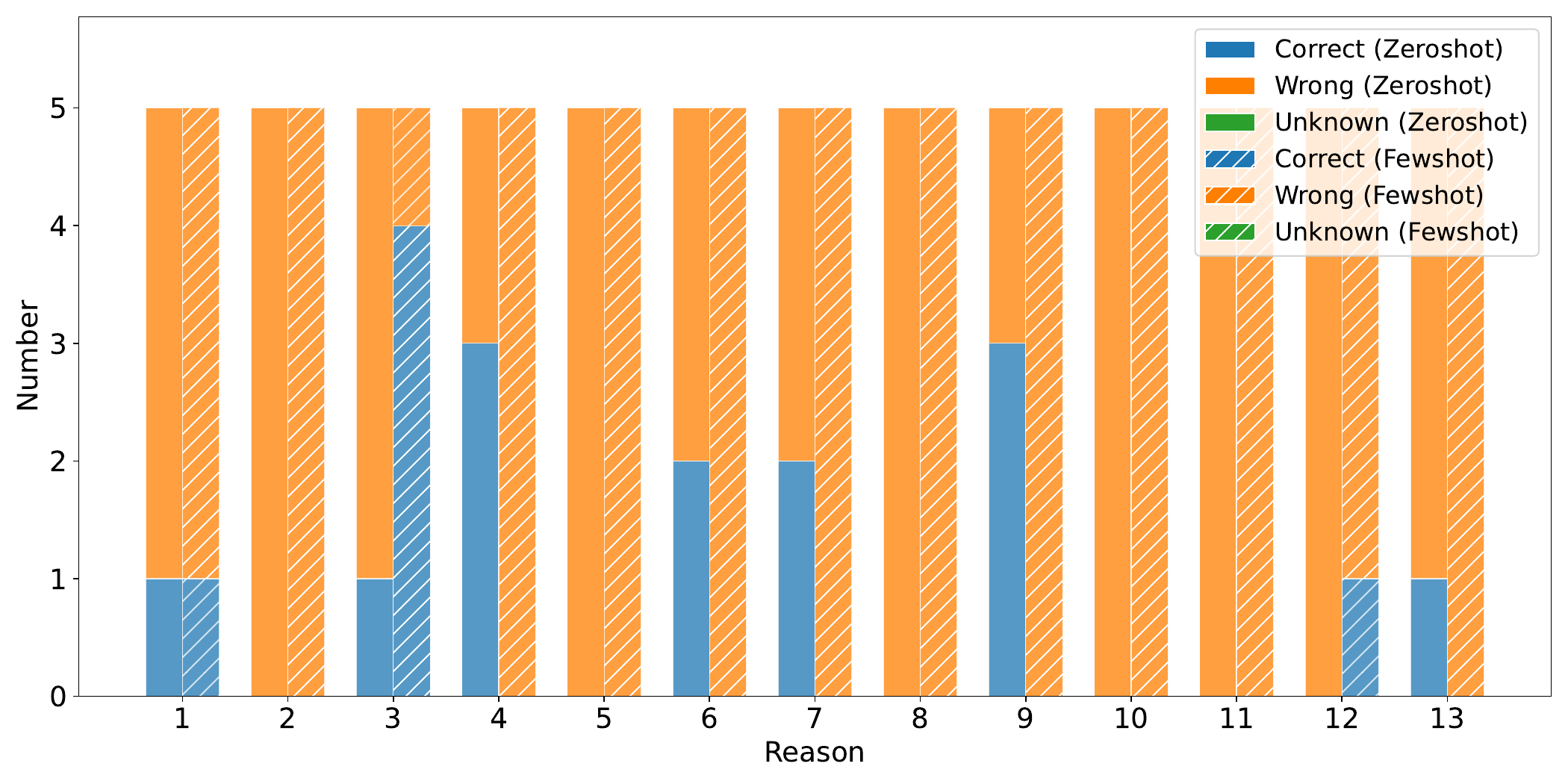}\caption*{\small{Qwen3-next-80b-a3b}}\end{subfigure}%
   \begin{subfigure}[b]{0.3\textwidth}\centering\includegraphics[width=\textwidth]{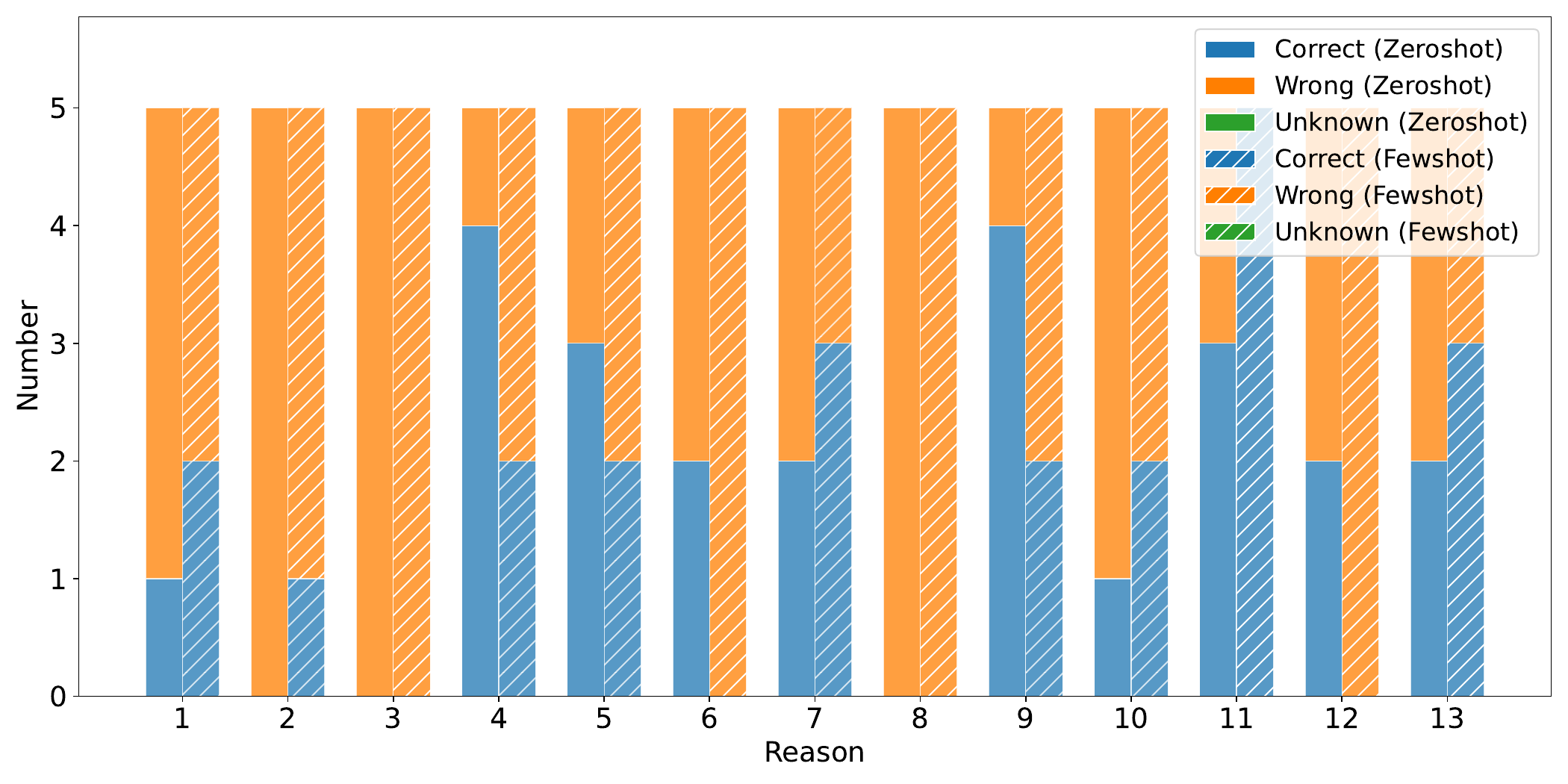}\caption*{\small{Qwen3-coder-plus}}\end{subfigure}%
   \begin{subfigure}[b]{0.3\textwidth}\centering\includegraphics[width=\textwidth]{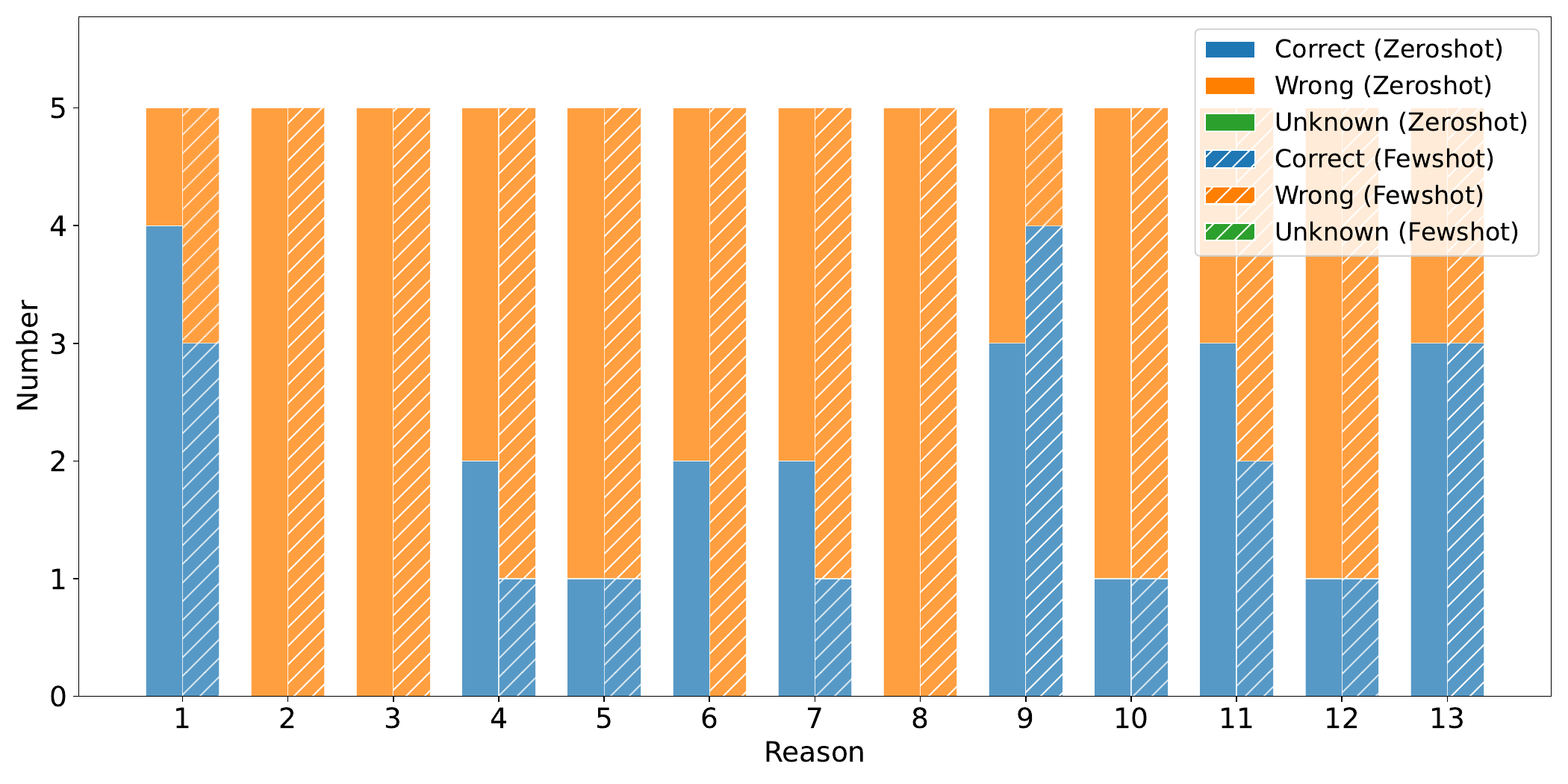}\caption*{\small{Kimi-K2}}\end{subfigure}
   
   \begin{subfigure}[b]{0.3\textwidth}\centering\includegraphics[width=\textwidth]{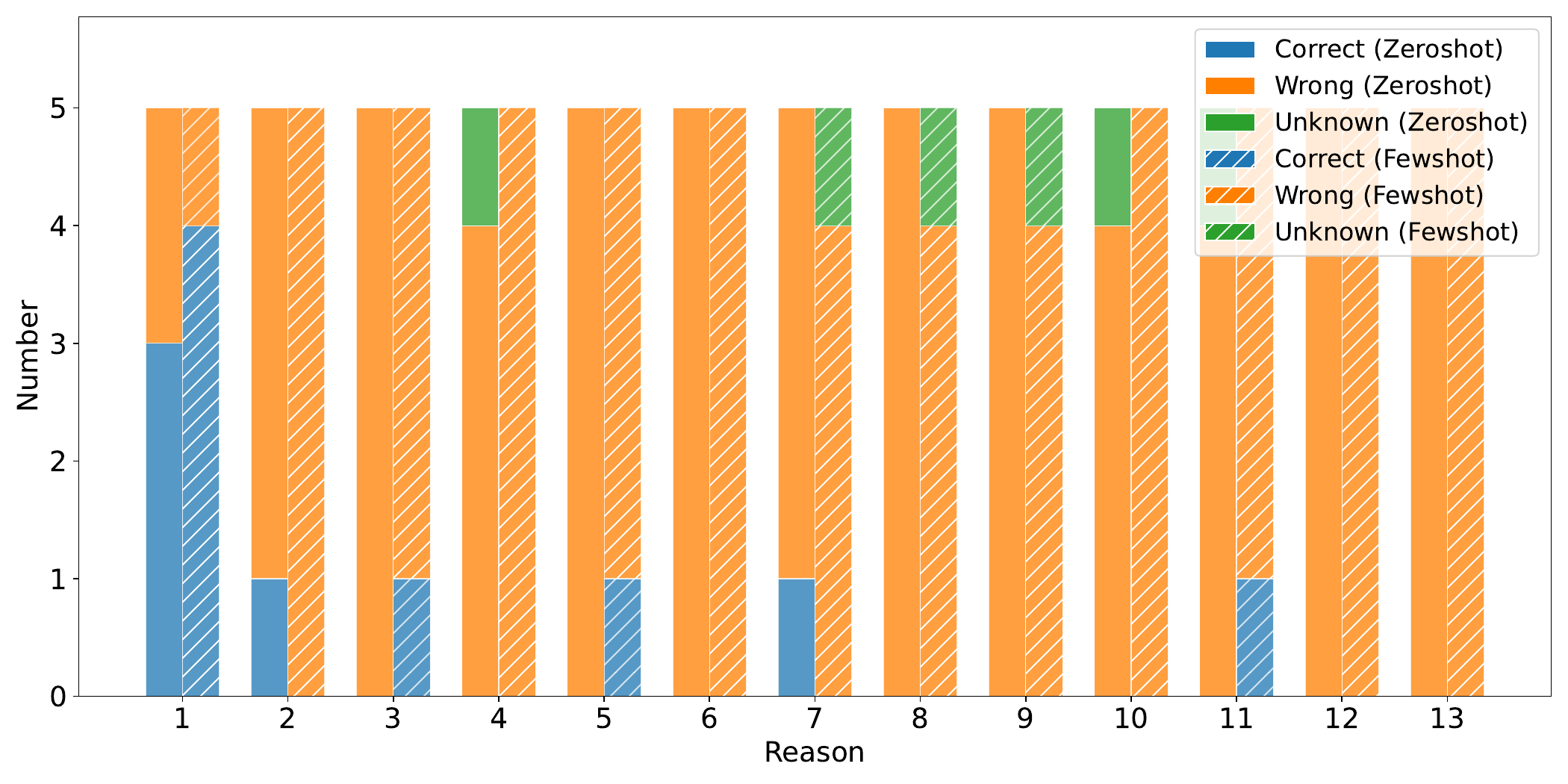}\caption*{\small{StarChat}}\end{subfigure}%
   \begin{subfigure}[b]{0.3\textwidth}\centering\includegraphics[width=\textwidth]{sections/figures/flaky/fig12/flaky_CodeLlama-70b_fewshot_zeroshot.pdf}\caption*{\small{CodeLlama-70b}}\end{subfigure}%
   \begin{subfigure}[b]{0.3\textwidth}\centering\includegraphics[width=\textwidth]{sections/figures/flaky/fig12/flaky_CodeLlama-13b_fewshot_zeroshot.pdf}\caption*{\small{CodeLlama-13b}}\end{subfigure}
   
   \caption{Complete predictions of all 21 LLMs for Flaky Test Reasoning across 13 test categories. Each subplot compares few-shot (left bar) and zero-shot (right bar) performance. Representative models are highlighted in the main text (Figure~\ref{fig:flaky_test}).}
   \label{fig:flaky_test_complete}
\end{figure}

Figure~\ref{fig:flaky_test_complete} presents per-class predictions for Flaky Test Reasoning across all 21 evaluated models under both few-shot and zero-shot settings. The main text (Section~\ref{sec:result}, Figure~\ref{fig:flaky_test}) shows six representative models covering best-performing commercial systems and open-source baselines.

\subsubsection{Confusion Matrices for Representative Models}

To provide deeper insights into error patterns, we present confusion matrices for representative models. The main text (Section~\ref{sec:result}, Figure~\ref{fig:flaky_confmat_gpt5mini_concept}) shows GPT-5-mini as a representative case. Here we provide additional confusion matrices for models spanning different performance tiers and architectures.

\begin{figure}[]
   \centering
   \begin{subfigure}[b]{0.48\textwidth}
      \centering
      \includegraphics[width=\textwidth]{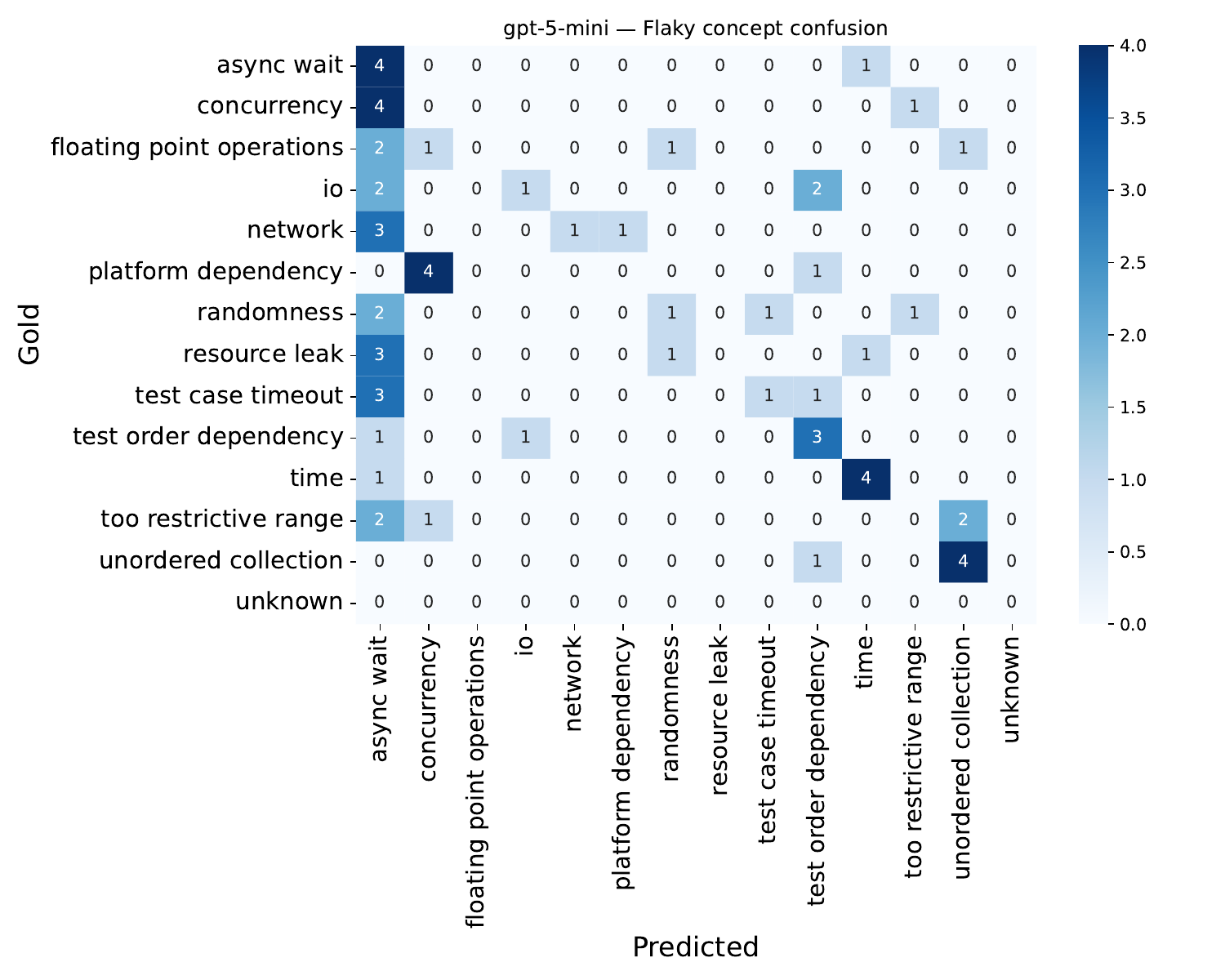}
      \caption{GPT-5-mini (concept split)}
      \label{fig:confmat_gpt5mini_concept_appendix}
   \end{subfigure}\hfill
   \begin{subfigure}[b]{0.48\textwidth}
      \centering
      \includegraphics[width=\textwidth]{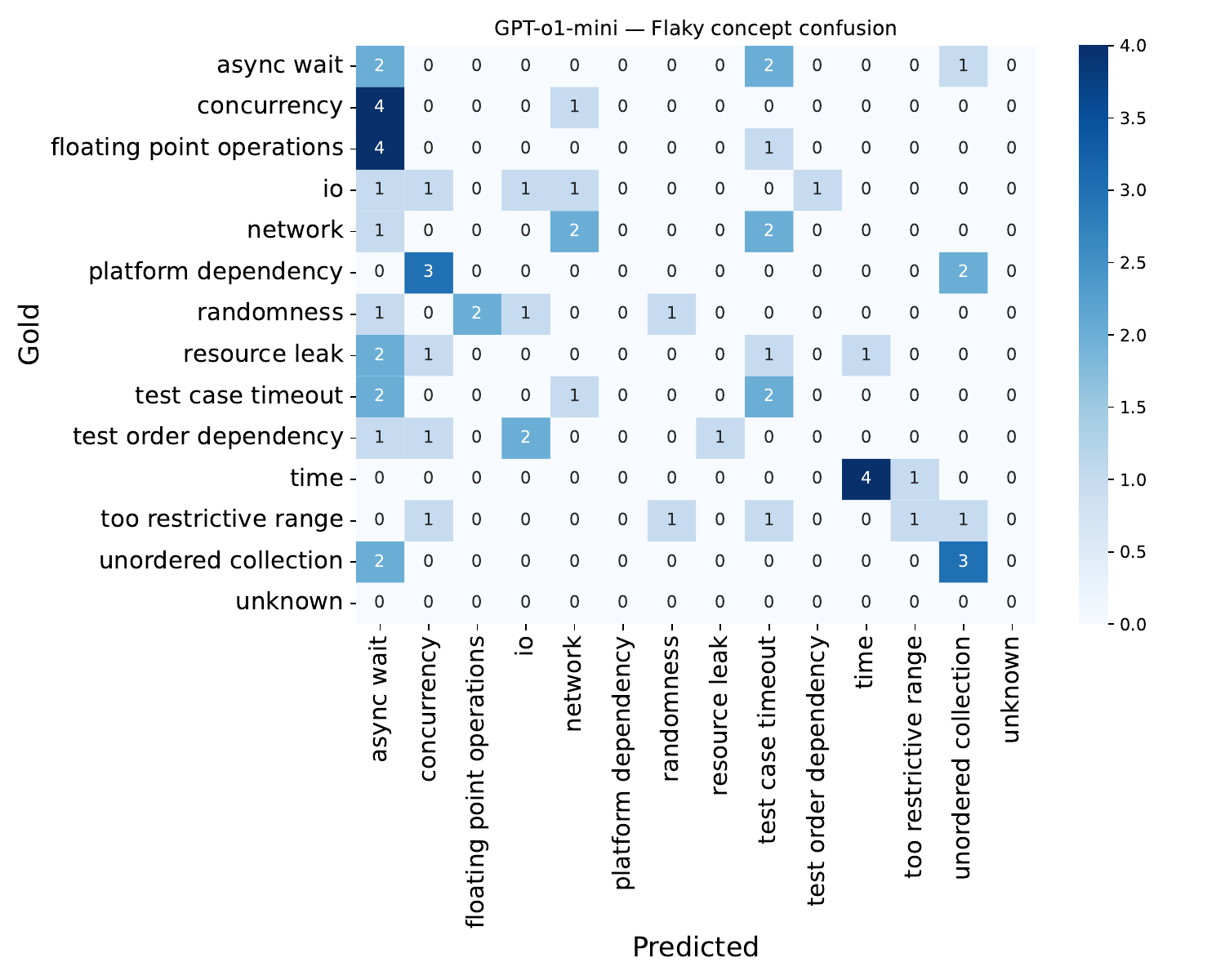}
      \caption{GPT-o1-mini (concept split)}
      \label{fig:confmat_gpto1mini_concept}
   \end{subfigure}
   
   \vspace{0.2cm}
   
   \begin{subfigure}[b]{0.48\textwidth}
      \centering
      \includegraphics[width=\textwidth]{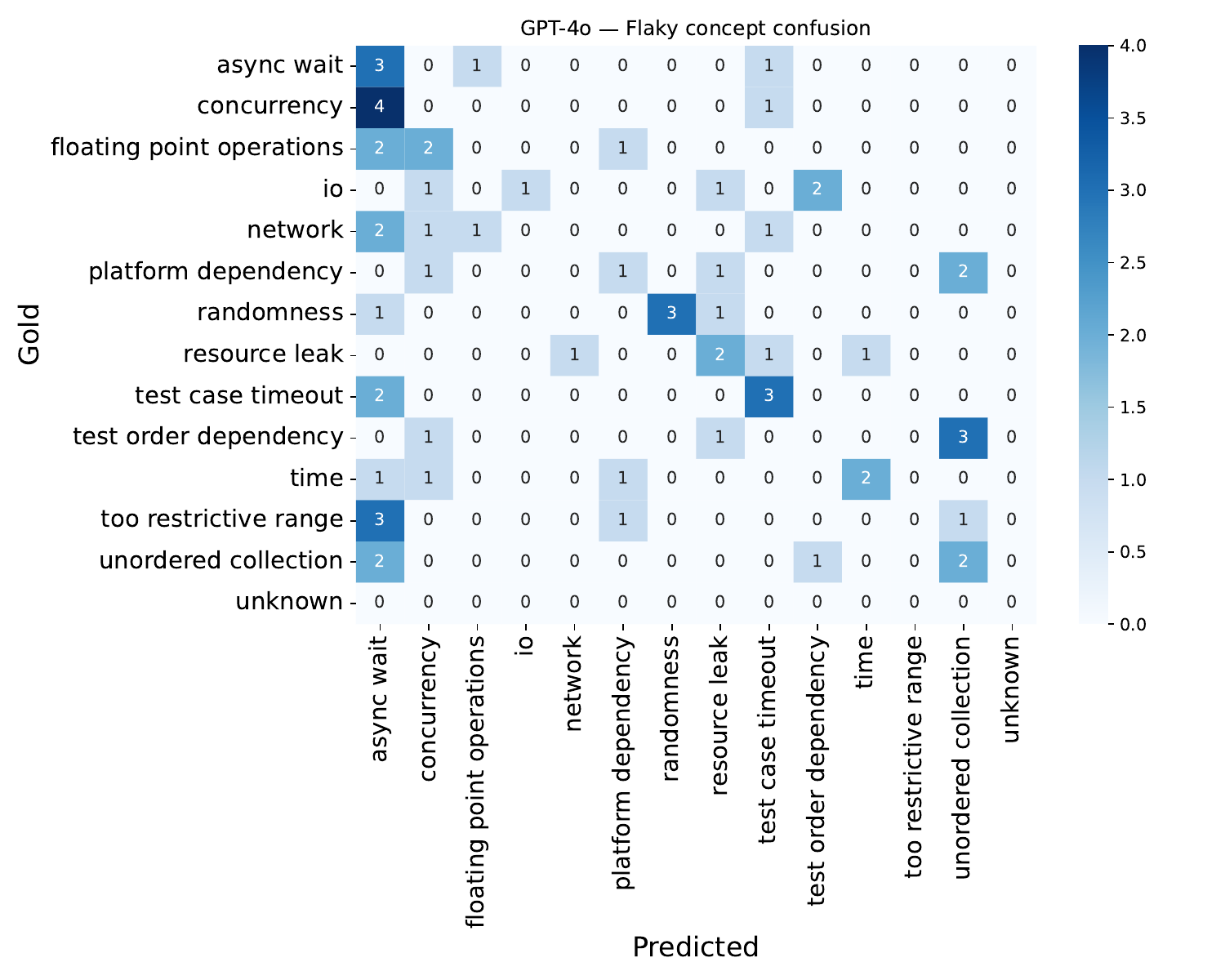}
      \caption{GPT-4o (concept split)}
      \label{fig:confmat_gpt4o_concept}
   \end{subfigure}\hfill
   \begin{subfigure}[b]{0.48\textwidth}
      \centering
      \includegraphics[width=\textwidth]{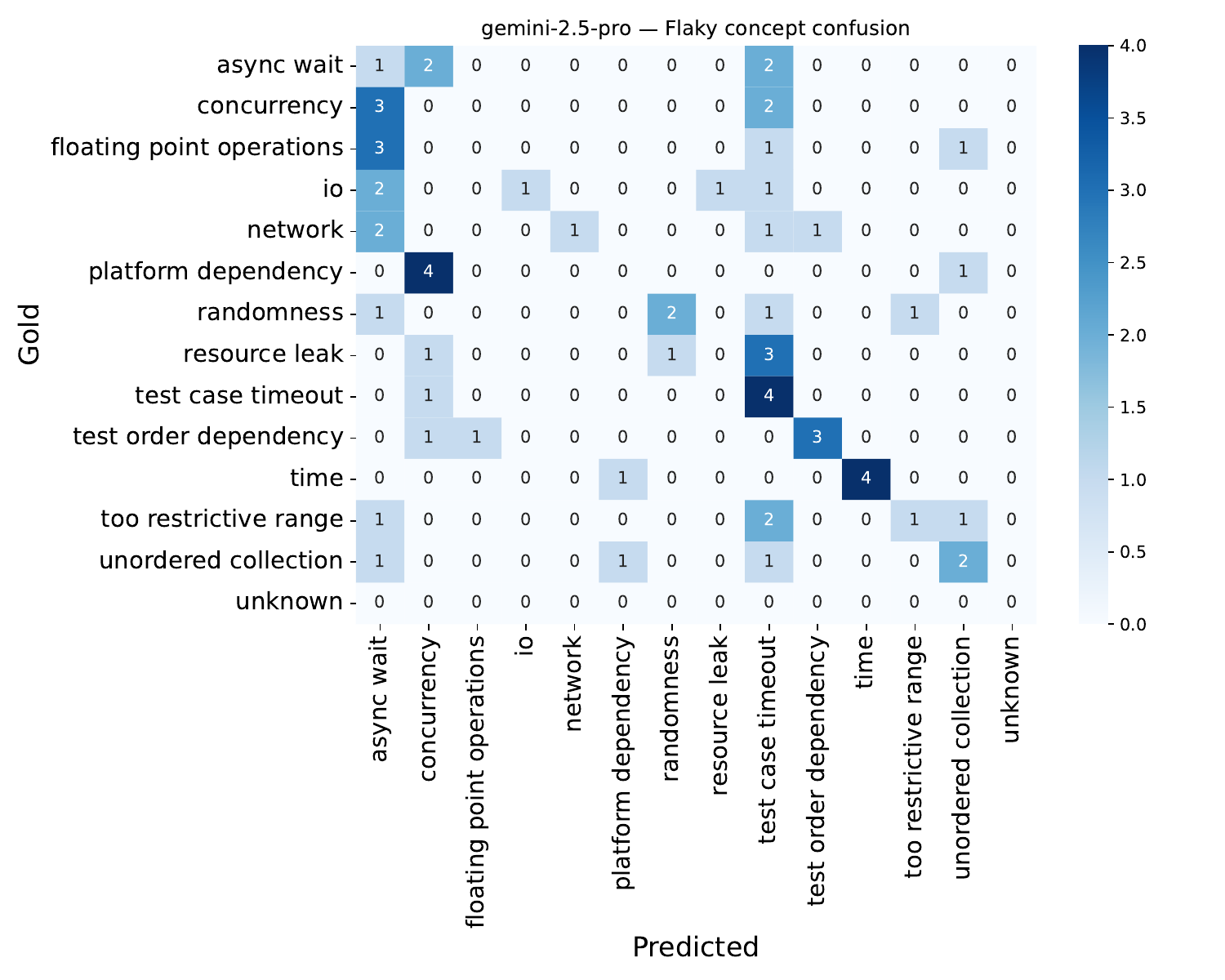}
      \caption{Gemini-2.5-Pro (concept split)}
      \label{fig:confmat_gemini25pro_concept}
   \end{subfigure}
   
   \vspace{0.2cm}
   
   \begin{subfigure}[b]{0.48\textwidth}
      \centering
      \includegraphics[width=\textwidth]{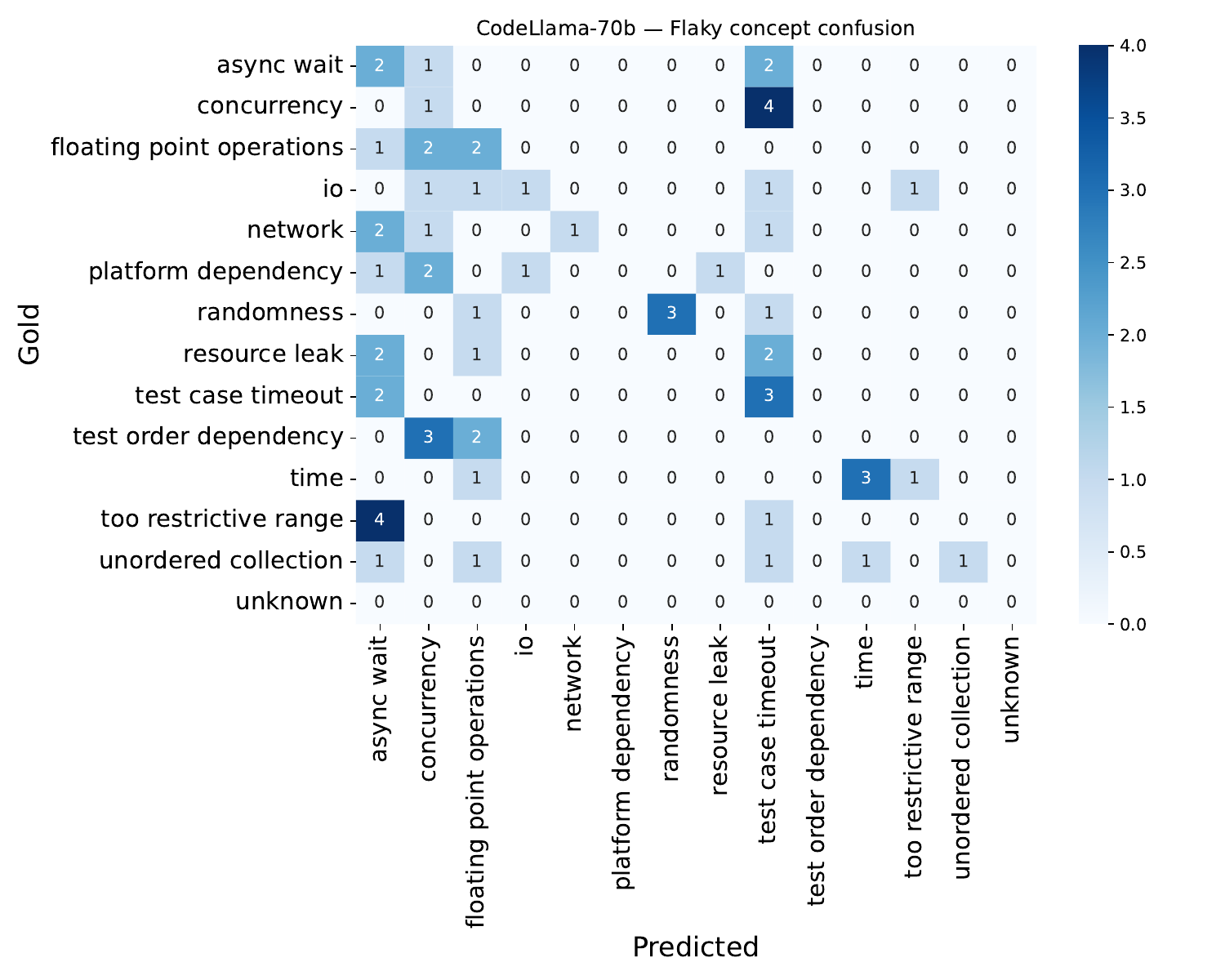}
      \caption{CodeLlama-70b (concept split)}
      \label{fig:confmat_codellama70b_concept}
   \end{subfigure}\hfill
   \begin{subfigure}[b]{0.48\textwidth}
      \centering
      \includegraphics[width=\textwidth]{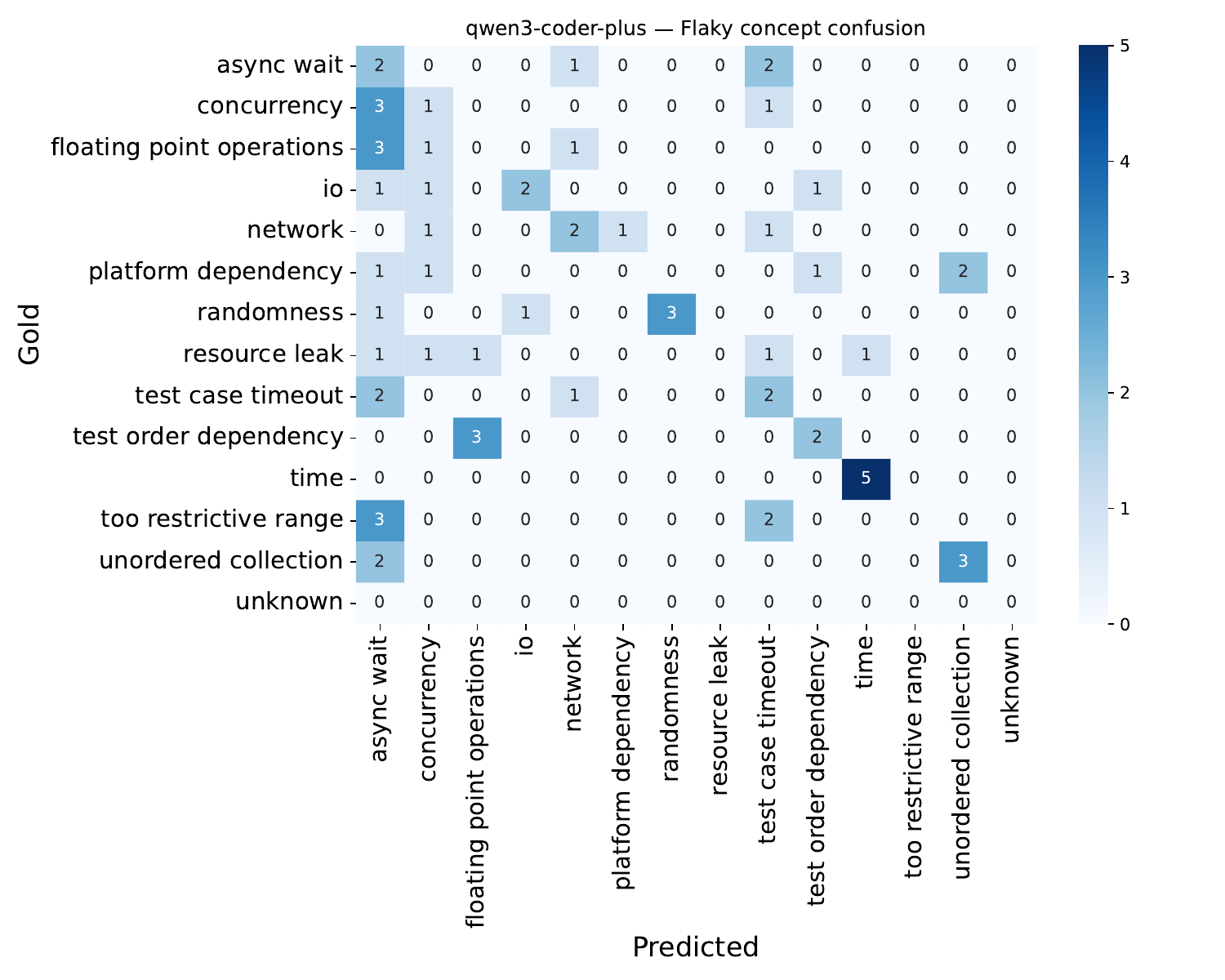}
      \caption{Qwen3-coder-plus (concept split)}
      \label{fig:confmat_qwen3coderplus_concept}
   \end{subfigure}
   
   \caption{Confusion matrices for Flaky Test Reasoning (concept split) across representative models. Rows represent true categories, columns represent predicted categories. Darker colors indicate higher confusion frequencies. These matrices reveal systematic error patterns: (1) over-attribution to ``async wait'', (2) time-adjacent confusion clusters (time, test\_case\_timeout, async\_wait), and (3) low recall for concurrency, io, and floating\_point categories.}
   \label{fig:flaky_confusion_matrices_concept}
\end{figure}

\begin{figure}[]
   \centering
   \begin{subfigure}[b]{0.48\textwidth}
      \centering
      \includegraphics[width=\textwidth]{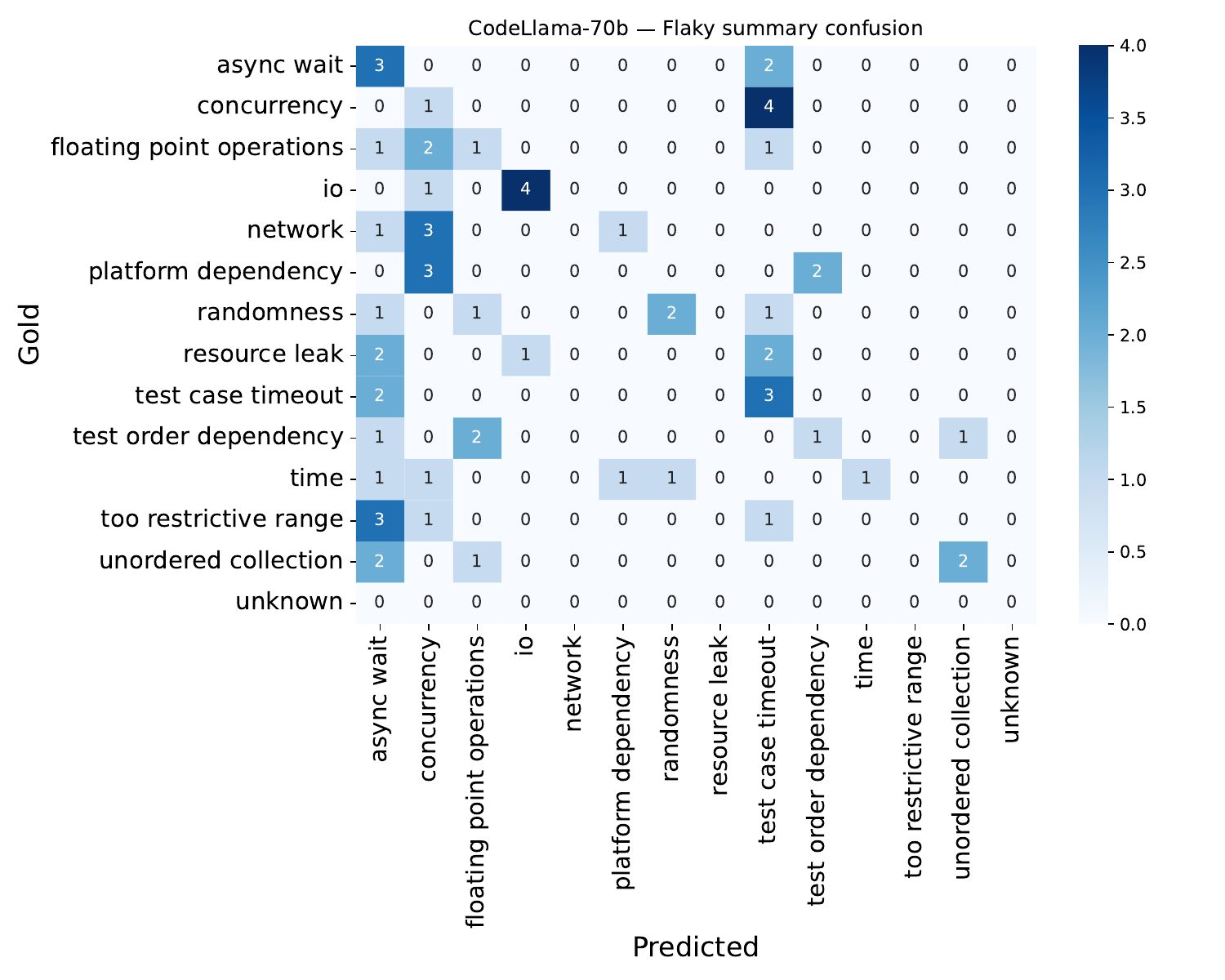}
      \caption{CodeLlama-70b (summary split)}
      \label{fig:confmat_codellama70b_summary}
   \end{subfigure}\hfill
   \begin{subfigure}[b]{0.48\textwidth}
      \centering
      \includegraphics[width=\textwidth]{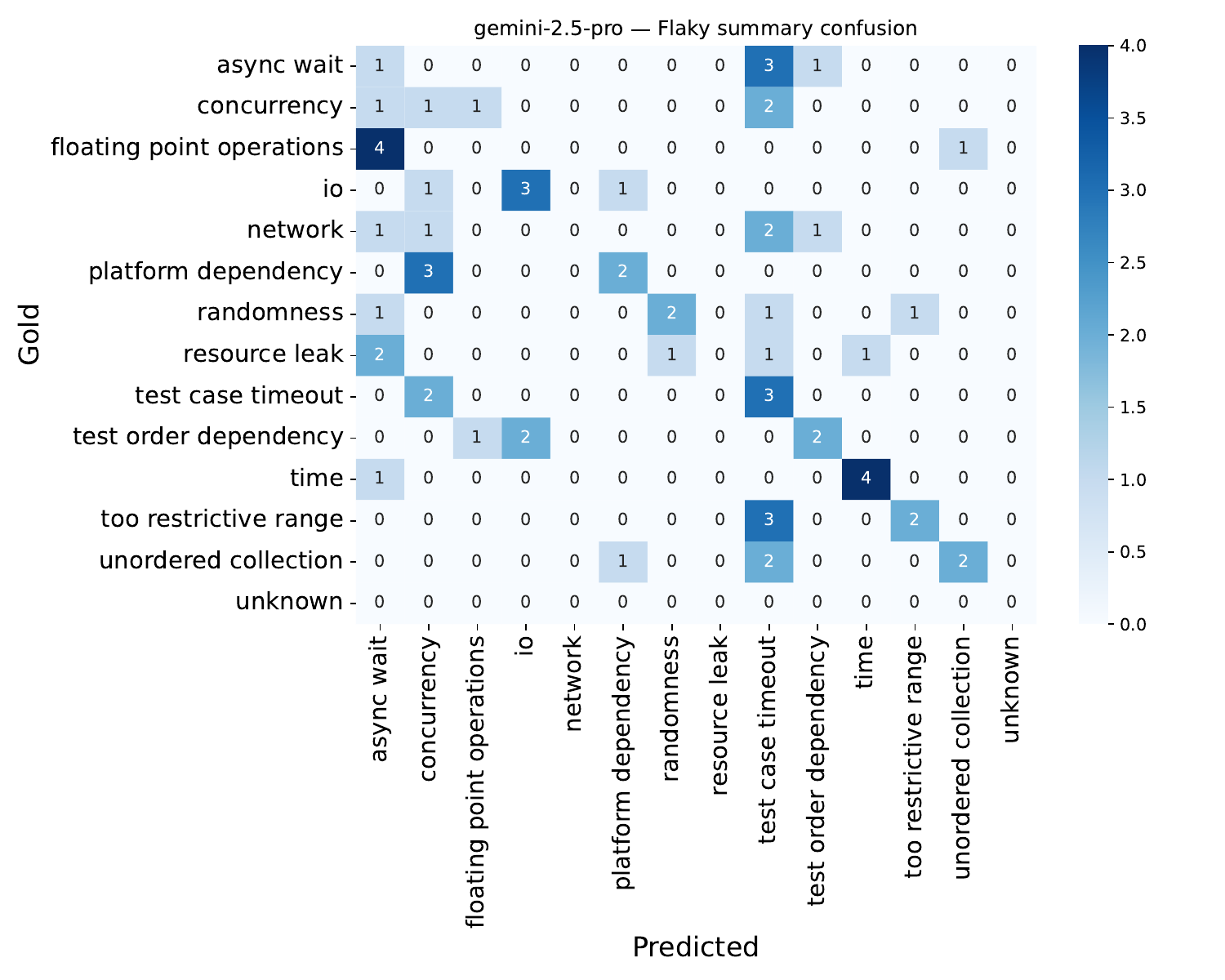}
      \caption{Gemini-2.5-Pro (summary split)}
      \label{fig:confmat_gemini25pro_summary}
   \end{subfigure}
   
   \vspace{0.3cm}
   
   \begin{subfigure}[b]{0.48\textwidth}
      \centering
      \includegraphics[width=\textwidth]{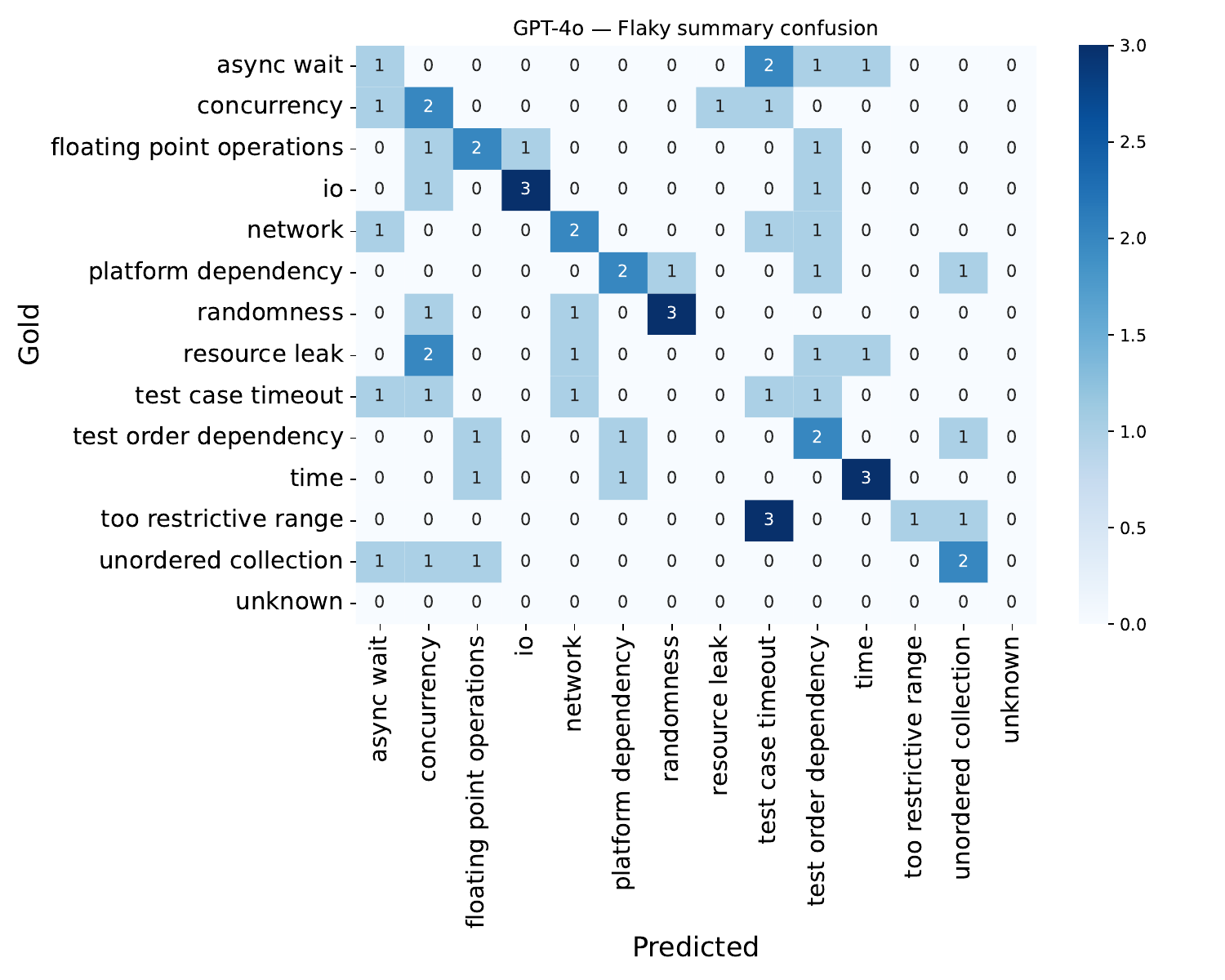}
      \caption{GPT-4o (summary split)}
      \label{fig:confmat_gpt4o_summary}
   \end{subfigure}\hfill
   \begin{subfigure}[b]{0.48\textwidth}
      \centering
      \includegraphics[width=\textwidth]{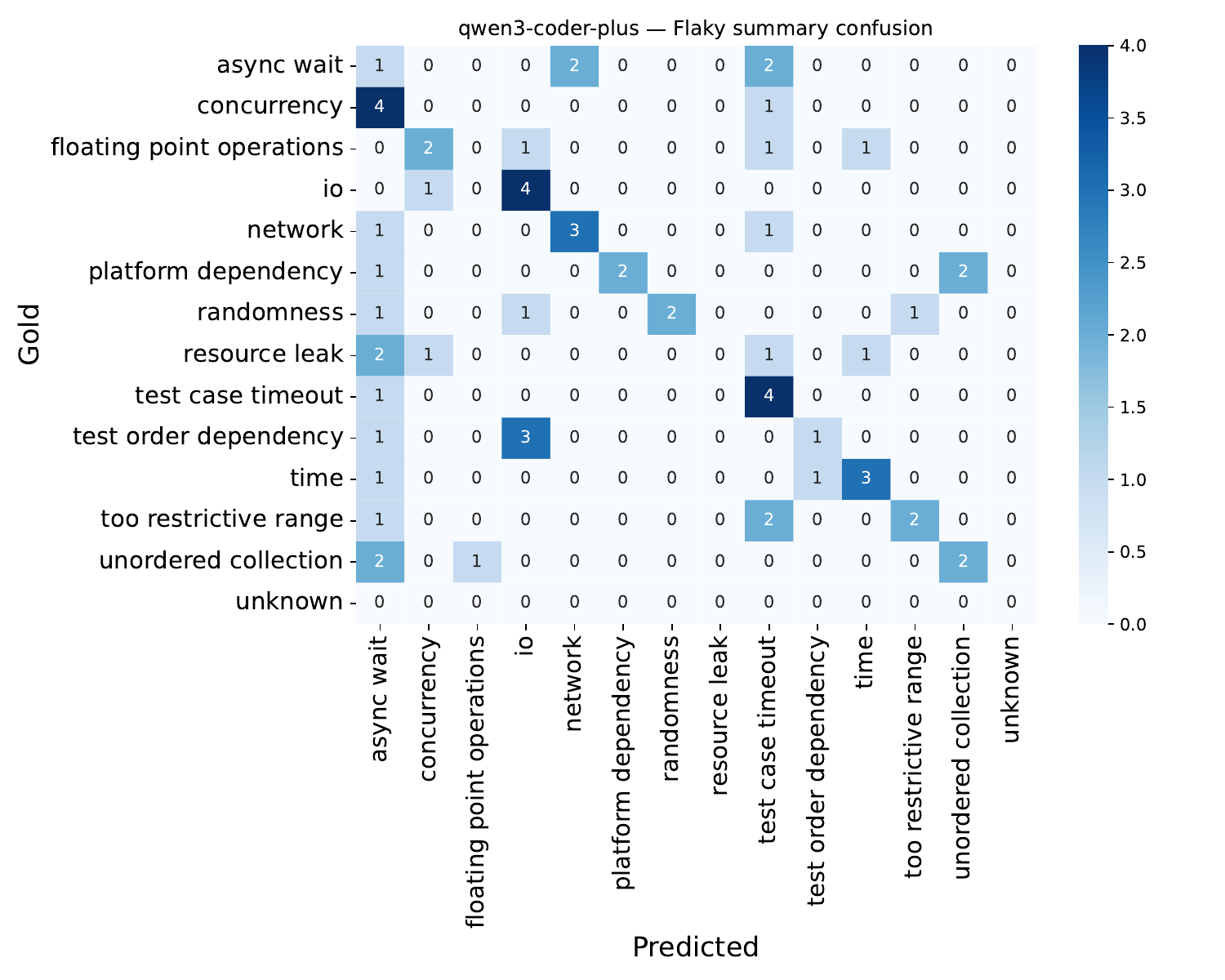}
      \caption{Qwen3-coder-plus (summary split)}
      \label{fig:confmat_qwen3coderplus_summary}
   \end{subfigure}
   
   \caption{Confusion matrices for Flaky Test Reasoning (summary split) across representative models. The summary split uses test descriptions rather than full code, testing whether models can diagnose flaky test causes from high-level summaries. Error patterns remain consistent with the concept split, indicating that confusion stems from fundamental execution-semantics reasoning gaps rather than code comprehension difficulties.}
   \label{fig:flaky_confusion_matrices_summary}
\end{figure}

\end{document}